\documentclass[11pt]{article}

\usepackage[margin=0.8in]{geometry}
\usepackage{amsmath,amssymb,amsthm,mathtools}
\usepackage{enumitem}
\usepackage{microtype}
\usepackage{xcolor}
\usepackage{mathpazo}
\usepackage{titlesec}
\usepackage{etoolbox}
\usepackage{mdframed}
\usepackage{tikz}
\usetikzlibrary{arrows.meta}
\usepackage{pgfplots}
\pgfplotsset{compat=1.18}
\usepackage[hidelinks]{hyperref}

\titlespacing*{\section}{0pt}
  {2.3ex plus .7ex minus .2ex}{1ex plus .2ex}
\titlespacing*{\subsection}{0pt}
  {1.8ex plus .5ex minus .2ex}{.7ex plus .15ex}
\titlespacing*{\subsubsection}{0pt}
  {1.5ex plus .4ex minus .2ex}{.6ex plus .15ex}
\setlist{
  topsep=4pt plus 1pt minus 1pt,
  itemsep=2pt plus 1pt,
  parsep=2pt,
  partopsep=0pt
}
\newcommand{\manuscriptdisplayspacing}{%
  \setlength{\abovedisplayskip}{8pt plus 2pt minus 3pt}%
  \setlength{\belowdisplayskip}{8pt plus 2pt minus 3pt}%
  \setlength{\abovedisplayshortskip}{0pt plus 2pt}%
  \setlength{\belowdisplayshortskip}{5pt plus 2pt minus 2pt}%
}
\AddToHook{cmd/normalsize/after}{\manuscriptdisplayspacing}
\AtBeginDocument{\manuscriptdisplayspacing}

\makeatletter
\patchcmd{\@maketitle}{\vskip 2em}{\vskip .8em}{}
  {\PackageWarning{manuscript}{Title top spacing patch failed}}
\patchcmd{\@maketitle}{\vskip 1.5em}{\vskip .8em}{}
  {\PackageWarning{manuscript}{Title-author spacing patch failed}}
\patchcmd{\@maketitle}{\vskip 1em}{\vskip .25em}{}
  {\PackageWarning{manuscript}{Title-date spacing patch failed}}
\patchcmd{\@maketitle}{\vskip 1.5em}{\vskip .7em}{}
  {\PackageWarning{manuscript}{Title bottom spacing patch failed}}
\makeatother

\newenvironment{revision}{\color{black}}{}
\newenvironment{newrevision}{\color{black}}{}

\newtheorem{theorem}{Theorem}[section]
\newtheorem*{informaltheorem}{Theorem}
\newtheorem{proposition}[theorem]{Proposition}
\newtheorem{lemma}[theorem]{Lemma}
\newtheorem{corollary}[theorem]{Corollary}
\newtheorem{conjecture}[theorem]{Conjecture}
\theoremstyle{definition}
\newtheorem{definition}[theorem]{Definition}
\newtheorem{protocol}{Protocol}
\theoremstyle{remark}
\newtheorem{remark}[theorem]{Remark}

\mdfdefinestyle{protocolshade}{
  backgroundcolor=black!4,
  hidealllines=true,
  fontcolor=.,
  innerleftmargin=10pt,
  innerrightmargin=10pt,
  innertopmargin=3pt,
  innerbottommargin=4pt,
  skipabove=0.4\baselineskip,
  skipbelow=0.4\baselineskip,
  nobreak=false
}

\newcommand{\Bdd}{B_2^d}
\newcommand{\D}{\mathcal D}
\newcommand{\J}{\mathcal J}
\newcommand{\T}{\mathcal T}
\newcommand{\E}{\mathbb E}
\newcommand{\Prb}{\mathbb P}
\newcommand{\Cov}{\operatorname{Cov}}
\newcommand{\Var}{\operatorname{Var}}
\newcommand{\sgn}{\operatorname{sgn}}
\newcommand{\proj}{\operatorname{proj}}
\newcommand{\KL}{D_{\mathrm{KL}}}

\title{\textcolor{black}{Query-Dependent Reading for Quantized Inner-Product Estimation: A Minimax Separation}}
\author{\normalsize Krishnakumar Balasubramanian\textsuperscript{1,2},
  Siddharth Bhandari\textsuperscript{1}, and
  Shiva Prasad Kasiviswanathan\textsuperscript{1}\\
  \normalsize \textsuperscript{1}Amazon\thanks{This work does not relate to the authors' positions at Amazon.}\\
  \normalsize \textsuperscript{2}Department of Statistics, University of California, Davis}
\date{}

\begin{document}
\maketitle

\begin{abstract}
\textcolor{black}{We study quantization for inner-product estimation under separate storage and read budgets.}  We ask a fundamental question: how much can accuracy
improve when the query is used to choose which stored bits to read?
We work in the static bit-probe model,
the one-bit-cell version of Yao's static cell-probe model
\cite{yao-cell-probe}.  \textcolor{black}{For \(x,y\) in the Euclidean
unit ball of \(\mathbb R^d\), an encoder stores \(w\) bits about \(x\)
before the query \(y\) is known.}  After \(y\) arrives, a
reader probes at most \(r<w\) bits to estimate \(x^\top y\).  Only storage and probes
are charged; protocols may be randomized and computation is unrestricted.
\textcolor{black}{We measure performance by the normalized risk:
\(d\) times the worst-case mean-squared error over \(x,y\), with
expectation taken over protocol randomness.}

 We first show that extra storage provides no benefit for query-independent
  access.  For every finite \(d\) and \(w\geq r\), a reader with
  query-independent addresses is equivalent to fully reading an
  \(r\)-bit message.  This holds even when later probes depend on previously
  returned bits.
  At fixed rates \(W=w/d\) and \(R=r/d\), the normalized minimax error
  approaches \(1/(4^R-1)\), independent of \(W\).
  \textcolor{black}{The upper bound follows from known full-read
  quantization results \cite{ordentlich-polyanskiy}, while a matching
  lower bound was known for unbiased compressors
  \cite{safaryan-compression}.
  We establish the converse for arbitrary decoders that estimate the
  inner product directly.}
\textcolor{black}{All rate limits are taken after \(d\to\infty\) at
fixed \(W,R\).}

\textcolor{black}{For query-adaptive access, we show how query-dependent selection of quantization bits can substantially improve accuracy beyond conventional full-read quantization. We give an explicit protocol based on Gaussian projections and random thresholds, with two read batches and one adaptive transition. } \textcolor{black}{Analyzing the selected bits' mean response gives
an asymptotic normalized error bound of}
\((16+o(1))/[R\log(W/R)]\) as \(W/R\to\infty\).
Thus, at fixed \(R\), its error vanishes as \(W\) grows, while the
query-independent error remains \(1/(4^R-1)\).
A refined selection rule and decoder improve the coefficient \(16\)
to \(\pi/4\).
\textcolor{black}{We also prove a finite-dimensional lower bound that applies
to arbitrary encoders, decoders, and fully branching readers.}
In the doubly sparse regime \(W/R\to\infty\) and
\(R\log(W/R)\to0\), the converse gives normalized error at least
\((1-o(1))/[2R\log(W/R)]\).  Hence, the refined construction is optimal
up to a factor of \(\pi/2\) in the leading constant.

\end{abstract}

\section{Introduction}
\begin{revision}
Many data structures store an object before knowing which queries they
will receive.  The stored record may be large, even though each query
reads only a small part of it.  We study records that are written once
and then reused, with separate budgets for storage and reading.
Our aim is to understand whether the query can help the reader choose
more informative bits.
\end{revision}

One example is the key--value (KV) cache used by autoregressive
transformers.  A token's key--value pair \((k_i,v_i)\) is stored before
future queries are known.  At decoding step \(t\), attention scores that
entry using \(q_t^\top k_i/\sqrt{d_h}\), where \(q_t\) is the current
query and \(d_h\) is the attention-head dimension
\cite{vaswani-transformer}.  Each entry is written once and reused,
while its relevance changes with the query.
Our companion paper studies query-adaptive KV-cache quantization under
separate storage and read budgets
\cite{bhandari2026readmattersqueryadaptivequantization}.
Here, we isolate the estimation of one query--key inner product.
We write it as \(x^\top y\), with source \(x\) and query \(y\) in the
Euclidean unit ball
\(\Bdd=\{u\in\mathbb R^d:\|u\|_2\le1\}\).

\textcolor{black}{We model this setting using the standard
\emph{static bit-probe model}, the one-bit-cell version of Yao's static
cell-probe model \cite{yao-cell-probe}.}
\textcolor{black}{Static means that the stored record remains fixed across
queries.}  An encoder stores \(w\) bits about \(x\).
After the query \(y\) arrives, a reader probes at most \(r\le w\) cells.  The
returned bits form a transcript, and a decoder uses \(y\) and this transcript
to estimate \(x^\top y\).  Protocols may be randomized.  The model charges
only storage and probes; computation and arithmetic precision are
unrestricted.

We measure the largest mean-squared error over all \(x,y\in\Bdd\), averaged
over protocol randomness, and then minimize this error over protocols.  This
is a distribution-free minimax criterion.  We write \(W=w/d\) and \(R=r/d\)
for the write and read rates per dimension.  These budgets apply to the
entire vector.  The difference \(W-R\) is the
\emph{write surplus}: \textcolor{black}{stored information that cannot all
be read by one query}.  Our asymptotic results concern normalized risk, meaning MSE
multiplied by \(d\).

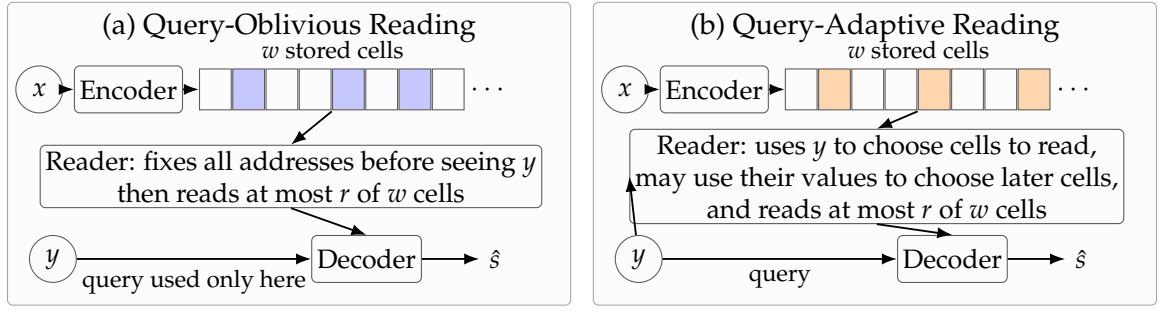
\begin{figure}[!t]
\centering
\begin{tikzpicture}[
  x=1cm,
  y=1cm,
  font=\small,
  panel/.style={
    draw=black!35,
    fill=black!1,
    rounded corners=2pt
  },
  process/.style={
    draw=black!65,
    rounded corners=2pt,
    align=center,
    minimum height=0.62cm,
    inner sep=2pt
  },
  vector/.style={
    circle,
    draw=black!65,
    minimum size=0.62cm,
    inner sep=1pt
  },
  cell/.style={
    draw=black!65,
    minimum width=0.42cm,
    minimum height=0.54cm,
    inner sep=0pt
  },
  fixedcell/.style={cell,fill=blue!22},
  adaptivecell/.style={cell,fill=orange!32},
  flow/.style={
    -{Latex[length=2mm]},
    line width=0.7pt
  },
  note/.style={
    font=\footnotesize,
    align=center
  }
]

\draw[panel] (0,0.35) rectangle (7.45,4.35);
\node[font=\mdseries] at (3.725,4.02)
  {(a) Query-Oblivious Reading};

\node[vector] (lx) at (0.42,3.20) {\(x\)};
\node[process,minimum width=1.05cm] (lw) at (1.60,3.20) {Encoder};
\node[note] at (4.30,3.73) {\(w\) stored cells};
\node[cell]      (l0) at (2.75,3.20) {};
\node[fixedcell] (l1) at (3.19,3.20) {};
\node[cell]      (l2) at (3.63,3.20) {};
\node[cell]      (l3) at (4.07,3.20) {};
\node[fixedcell] (l4) at (4.51,3.20) {};
\node[cell]      (l5) at (4.95,3.20) {};
\node[fixedcell] (l6) at (5.39,3.20) {};
\node[cell]      (l7) at (5.83,3.20) {};
\node at (6.35,3.20) {\(\cdots\)};
\draw[flow] (lx) -- (lw);
\draw[flow] (lw) -- (l0.west);

\node[process,minimum width=3.45cm] (lsel) at (3.75,2.05)
  {Reader: fixes all addresses before seeing \(y\)\\then reads at most \(r\) of \(w\) cells};
\node[vector] (ly) at (0.60,0.96) {\(y\)};
\node[process,minimum width=1.35cm] (ldec) at (4.75,0.96)
  {Decoder};
\node (lout) at (6.45,0.96) {\(\hat{s}\)};
\draw[flow] (4.29,2.91) -- (lsel.north);
\draw[flow] (lsel.south) -- (ldec.north);
\draw[flow] (ly) -- node[below,note] {query used only here} (ldec);
\draw[flow] (ldec) -- (lout);

\draw[panel] (7.75,0.35) rectangle (15.20,4.35);
\node[font=\mdseries] at (11.475,4.02)
  {(b) Query-Adaptive Reading};

\node[vector] (rx) at (8.17,3.20) {\(x\)};
\node[process,minimum width=1.05cm] (rw) at (9.35,3.20) {Encoder};
\node[note] at (12.05,3.73) {\(w\) stored cells};
\node[cell]         (r0) at (10.50,3.20) {};
\node[adaptivecell] (r1) at (10.94,3.20) {};
\node[cell]         (r2) at (11.38,3.20) {};
\node[cell]         (r3) at (11.82,3.20) {};
\node[adaptivecell] (r4) at (12.26,3.20) {};
\node[cell]         (r5) at (12.70,3.20) {};
\node[cell]         (r6) at (13.14,3.20) {};
\node[adaptivecell] (r7) at (13.58,3.20) {};
\node at (14.10,3.20) {\(\cdots\)};
\draw[flow] (rx) -- (rw);
\draw[flow] (rw) -- (r0.west);

\node[process,minimum width=3.55cm] (rsel) at (11.50,2.05)
  {Reader: uses \(y\) to choose cells to read,\\may use their values to choose later cells,\\and reads at most \(r\) of \(w\) cells};
\node[vector] (ry) at (8.35,0.96) {\(y\)};
\node[process,minimum width=1.35cm] (rdec) at (12.50,0.96)
  {Decoder};
\node (rout) at (14.20,0.96) {\(\hat{s}\)};
\draw[flow] (12.04,2.91) -- (rsel.north);
\draw[flow] (rsel.south) -- (rdec.north);
\draw[flow] (ry) -- (rsel.west);
\draw[flow] (ry) -- node[below,note] {query} (rdec);
\draw[flow] (rdec) -- (rout);

\end{tikzpicture}
 \caption{Illustration of finite-write, finite-read inner-product estimation
with query-oblivious and query-adaptive reading.  The encoder stores an
encoding of \(x\) in \(w\) one-bit cells before \(y\) is known.
In both panels, the reader retrieves at most \(r\le w\) of these cells,
shown in color.  The decoder outputs \(\hat{s}\), an estimate of \(x^\top y\).
In the left panel, the reader fixes all addresses before seeing the query,
then reads those cells.  In the right panel, the query guides address
selection, and later choices may use bits already read.  Our
query-independent branching class also permits choices based on bits
already read, while keeping address selection independent of \(y\).}
\label{fig:local-access-setup}
\end{figure}

Figure~\ref{fig:local-access-setup} shows the finite-write, finite-read
problem and contrasts query-oblivious with query-adaptive reading.  The same
\(w\)-bit record must serve every future query, while each query receives
its own \(r\)-probe budget.

The reader's access rule is the central modeling choice.  A
\emph{query-oblivious} reader fixes its full address list before seeing the
query or any returned bit.  Our stronger \emph{query-independent branching}
class \textcolor{black}{may use earlier returned bits to choose later
addresses, while keeping its address rule independent of \(y\).
Its choices can therefore depend on \(x\) through the stored record.}
A \emph{query-dependent}, or
\emph{query-adaptive}, reader may use \(y\) to choose addresses.  A fully
branching reader may also use every earlier returned bit when choosing the
next address.

To the best of our knowledge, this work \textcolor{black}{along with our more practically-oriented companion paper
\cite{bhandari2026readmattersqueryadaptivequantization}} provide the first
minimax formulation of inner-product
estimation with separate storage and read budgets.

\begin{quote}
\centering
\emph{Can a reader use the query to choose which stored bits to inspect, and
does this reduce the smallest achievable worst-case MSE compared with
query-oblivious reading?}
\end{quote}

We show that query-dependent access can yield an unbounded improvement.
We begin by characterizing query-independent access.
Theorem~\ref{thm:oblivious-collapse}
shows that, for every finite \(d,w,r\), even a query-independent reader that
branches on returned bits is exactly equivalent to fully reading an
\(r\)-bit message.  \textcolor{black}{Query-independent reading can be
simulated before the query arrives, so storing the resulting \(r\)
returned bits suffices.}  The remaining \(w-r\) stored bits cannot improve its
risk.  Theorem~\ref{thm:full-read} then gives the exact asymptotic normalized
risk of the resulting full-read problem.  At every fixed read rate \(R>0\),
it converges to \(1/(4^R-1)\), independently of \(W\).
\textcolor{black}{We use the full-read upper bound of Ordentlich and
Polyanskiy \cite[Theorem~3(2)]{ordentlich-polyanskiy} and prove a matching
converse for arbitrary scalar decoders, including biased estimates and
nonlinear functions of the query.  Safaryan, Shulgin, and Richt\'arik
previously proved a matching lower bound for unbiased vector compressors
\cite{safaryan-compression}.}
\textcolor{black}{For decoders that are linear in the query for each fixed
choice of random seeds, an elementary reconstruction-and-rescaling
argument already gives the sharp asymptotic lower bound, even when the
estimates are biased (Proposition~\ref{prop:basis-reconstruction}).
Our general converse extends the conclusion to arbitrary nonlinear scalar
answers.}

\textcolor{black}{When the read budget covers the entire record,
query-independent reading can simulate any query-dependent protocol.
A strict advantage therefore requires a positive write surplus \(W-R>0\)
(Corollary~\ref{cor:write-surplus-necessary}).}
We next construct a query-dependent protocol that uses this surplus.
Protocol~\ref{prot:window-reader} stores one-bit thresholded Gaussian
projections.  The independent random threshold is called a
\emph{dither}.  The reader first fetches a fixed pilot batch.
Using the query and the pilot estimate, it then selects a second and
final batch with large query projections and nearby thresholds.
The decoder refines the estimate by inverting the selected bits'
average response.  Thus, the protocol uses only one adaptive transition
between returned bits and later addresses.
For sufficiently large \(W/R\), Theorem~\ref{thm:dither-reader} gives
asymptotic normalized MSE at most \((16+o(1))/[R\log(W/R)]\).

\textcolor{black}{Comparing these two error bounds gives our main result}
(see Theorem~\ref{thm:separation} for a formal statement):

\begin{informaltheorem}[Informal Main Separation]
Fix a read rate \(R>0\).  An encoder stores \(\lfloor Wd\rfloor\) bits
about \(x\in\Bdd\) before seeing the query \(y\in\Bdd\), and each query
reads at most \(\lfloor Rd\rfloor\) bits to estimate \(x^\top y\).
Measure error by \(d\) times the worst-case MSE, averaged over protocol
randomness.  Take \(d\to\infty\) at fixed rates before letting \(W/R\to\infty\).
The normalized risks satisfy
\[
  \begin{aligned}
    \text{query-independent optimum}
      &=\frac{1}{4^R-1},\\
    \text{two-batch query-dependent achievable risk}
      &\le\frac{16+o(1)}{R\log(W/R)}.
  \end{aligned}
\]
The first identity allows addresses to depend on previously returned
bits, provided they do not depend on the query.  An explicit protocol
with one adaptive transition satisfies the second bound; here \(o(1)\)
tends to zero as \(W/R\to\infty\), after the dimension limit.
Thus, at the same fixed read rate, additional storage makes the
query-dependent error arbitrarily smaller than the query-independent
optimum.
\end{informaltheorem}

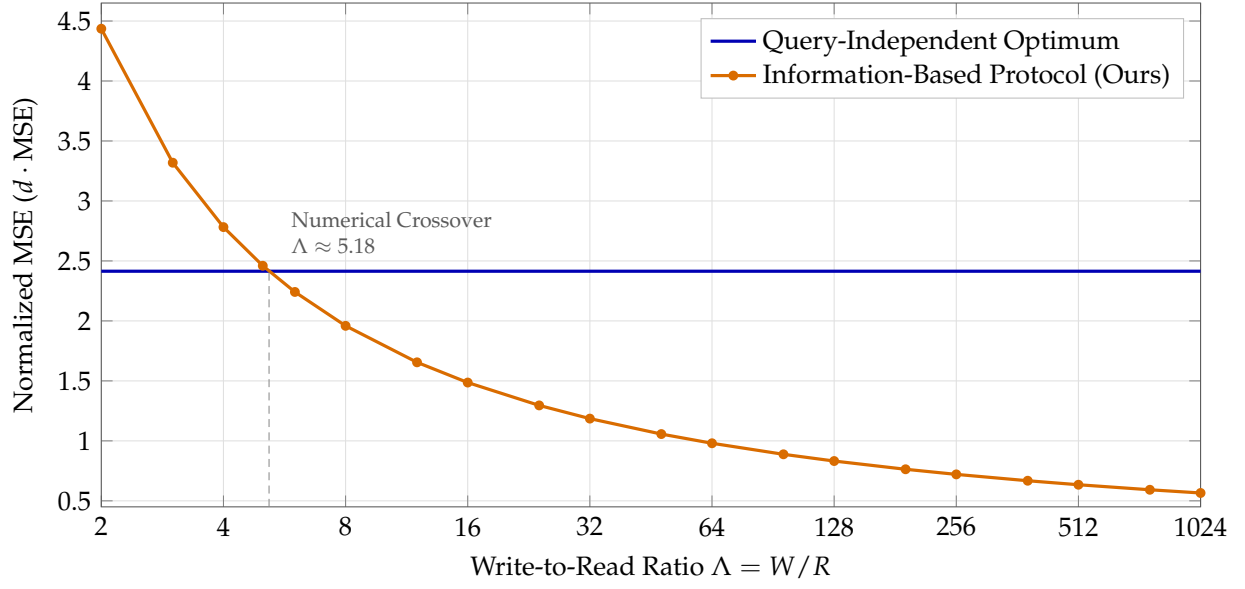
\begin{figure}[!htbp]
\centering
\begin{tikzpicture}
\begin{axis}[
  width=0.92\linewidth,
  height=0.47\linewidth,
  xmode=log,
  log basis x=2,
  xmin=2,
  xmax=1024,
  ymin=0.45,
  ymax=4.65,
  xtick={2,4,8,16,32,64,128,256,512,1024},
  xticklabels={2,4,8,16,32,64,128,256,512,1024},
  ytick={0.5,1,1.5,2,2.5,3,3.5,4,4.5},
  xlabel={Write-to-Read Ratio \(\Lambda=W/R\)},
  ylabel={Normalized MSE (\(d\cdot\mathrm{MSE}\))},
  grid=major,
  grid style={black!12},
  axis line style={black!65},
  tick label style={font=\small},
  label style={font=\small},
  legend style={
    at={(0.985,0.97)},
    anchor=north east,
    draw=black!25,
    fill=white,
    font=\small,
    cells={anchor=west}
  },
]

\addplot[
  blue!70!black,
  very thick
] coordinates {
  (2,2.4142135624)
  (1024,2.4142135624)
};
\addlegendentry{Query-Independent Optimum}

\addplot[
  orange!85!black,
  very thick,
  mark=*,
  mark size=1.25pt
] coordinates {
  (2,4.4357372014)
  (3,3.3188647164)
  (4,2.7820511041)
  (5,2.4597629351)
  (6,2.2414027434)
  (8,1.9590541370)
  (12,1.6547816891)
  (16,1.4861918008)
  (24,1.2956354074)
  (32,1.1855512798)
  (48,1.0567720102)
  (64,0.9800757022)
  (96,0.8880242376)
  (128,0.8319261520)
  (192,0.7632549237)
  (256,0.7206495305)
  (384,0.6676750723)
  (512,0.6343356962)
  (768,0.5923562822)
  (1024,0.5656274805)
};
\addlegendentry{Information-Based Protocol (\textcolor{black}{Ours})}

\addplot[
  black!45,
  densely dashed,
  thin,
  forget plot
] coordinates {
  (5.1815054358,0.45)
  (5.1815054358,2.4142135624)
};

\node[
  anchor=south west,
  align=left,
  font=\scriptsize,
  text=black!65
] at (axis cs:5.55,2.48) {Numerical Crossover\\\(\Lambda\approx5.18\)};

\end{axis}
\end{tikzpicture}
 \caption{Limiting normalized MSE (\(d\) times mean-squared error) at
read rate \(R=1/4\), as the write-to-read ratio \(\Lambda=W/R\) grows.
The blue line is the exact query-independent optimum
\(1/(4^{1/4}-1)=1+\sqrt2\).
The orange curve is a numerical estimate of the limiting worst-case risk
of our information-based protocol (Section~\ref{sec:adaptive-separation}).
\textcolor{black}{The crossover near \(\Lambda=5.18\) is estimated from
the sampled correlation grid.}}
\label{fig:mse-separation}
\end{figure}

\begin{newrevision}
The separation theorem keeps the read rate fixed while the write rate
grows.  Throughout our asymptotic comparisons, the dimension limit is
taken first at each fixed positive pair \((W,R)\).
Thus, even a small positive \(R\) corresponds to a number of probes
proportional to dimension, and the normalized risks describe the
coefficients of the \(1/d\) MSE scale.

The coefficient \(16\) suffices to establish the separation, but a
refined selection rule and decoder improve it to \(\pi/4\), using the
same stored bits and two read batches
(Section~\ref{sec:adaptive-separation}).  The refined protocol's
asymptotic normalized MSE is at most \((\pi+o(1))/[4R\log(W/R)]\).
Figure~\ref{fig:mse-separation} compares this refined protocol
numerically with the query-independent optimum at \(R=1/4\).
\end{newrevision}

\begin{newrevision}
The lower bound shows how close this construction comes to the best
possible query-dependent protocol.
Theorem~\ref{thm:finite-adaptive-converse} applies to arbitrary encoders
and decoders, with each read address allowed to depend on the query and
all bits returned so far.  In the doubly sparse regime
\(W/R\to\infty\) and \(R\log(W/R)\to0\), it gives normalized MSE at least
\((1-o(1))/[2R\log(W/R)]\).
Together, the bounds determine the optimal dependence on the budgets
in this regime: the refined two-batch construction is within a factor
\(\pi/2\) of the optimum in the leading constant.

In this regime, the read rate tends to zero.  For example, fix \(W>0\) and let
\(R\downarrow0\).  The query-independent normalized risk \(1/(4^R-1)\)
is asymptotic to \(1/(R\log4)\), while the optimal query-dependent normalized risk grows
like \(1/[R\log(W/R)]\), up to constant factors.
Both increase as fewer bits are read, but query-dependent schemes can improve the error by a factor of order \(\log(W/R)\).
\end{newrevision}

\subsection{AI Disclosure}

GPT-5.6 Sol Pro was used during manuscript preparation for brainstorming,
drafting, editing, and formatting assistance, including preliminary drafts
of proof arguments.  The tool materially affected
Sections~\ref{sec:query-independent-results}, \ref{sec:adaptive-results},
and~\ref{sec:adaptive-lower-bounds}.  All proof
arguments were independently checked, corrected, and finalized by the
authors, who take full responsibility for the correctness, originality, and
presentation of the paper.

\section{Formal Model}

\begin{revision}
\textcolor{black}{In this section, we define the protocol model, access
classes, and minimax risks.}
Let \(d\ge1\) be an integer,
let \(w\ge r\ge0\) be integer storage and probe budgets,
let \(\Bdd=\{x\in\mathbb R^d:\|x\|_2\le1\}\), and let
\([w]=\{1,\ldots,w\}\) be the cell addresses.  The encoder observes
\(x\in\Bdd\) and stores \(w\) bits before the query \(y\in\Bdd\) is known.
The query algorithm reads at most \(r\) cells to estimate \(x^\top y\).
All protocol maps are jointly measurable in their inputs and random
seeds.\footnote{\textcolor{black}{Throughout the paper, random seeds and
observations take values in standard Borel spaces.  These include discrete
spaces, Euclidean spaces, and their countable products.  This convention
ensures that the conditional distributions used in the proofs exist.}}
Computation is unrestricted, and the maps may be discontinuous.

A protocol may use a public seed shared by all components, encoder-private
coins, a query-time seed shared by the reader and decoder, and
decoder-private coins.  Their joint law is independent of \((x,y)\), and
each component uses only its designated seeds.
Fixing all seeds at \(\omega\) gives a deterministic protocol
\(\Pi_\omega=(\operatorname{Enc},\operatorname{Read},\operatorname{Dec})\).
We suppress \(\omega\) in the following maps.
The encoder stores
\[
  \operatorname{Enc}:\Bdd\to\{0,1\}^w,
  \qquad
  Z=\operatorname{Enc}(x)=(Z_1,\ldots,Z_w).
\]
The reader chooses addresses using
\[
  A_t:\Bdd\times\{0,1\}^{t-1}\to[w],
  \qquad t=1,\ldots,r.
\]
At probe \(t\), it chooses \(a_t=A_t(y,z_{<t})\) and receives
\(z_t=Z_{a_t}\).  The transcript is \(T=(z_1,\ldots,z_r)\).  Repeated
probes are allowed, and a reader using fewer than \(r\) probes may be padded
with ignored probes.

The decoder makes no additional probes and returns
\[
  \operatorname{Dec}:\Bdd\times\{0,1\}^r\to\mathbb R,
  \qquad
  \widehat s_{\Pi_\omega}(x,y)=\operatorname{Dec}(y,T).
\]
Because the decoder knows the reader's seeds, it can reconstruct the
address sequence recursively from \(y\) and \(T\).  These inputs therefore
contain the complete read transcript.

The minimax risk over randomized protocols \(\Pi\) with unrestricted,
fully branching query-adaptive access is
\begin{equation}
  \D_d^{\mathrm{ad}}(w,r)
  =
  \inf_{\Pi}
  \sup_{x,y\in\Bdd}
  \E\!\left[
    \bigl(x^\top y-\widehat s_\Pi(x,y)\bigr)^2
  \right].
  \label{eq:minimax-risk}
\end{equation}
The protocol is chosen first, subject to the storage and read budgets,
and then evaluated on its hardest source--query pair.  The expectation
averages over protocol randomness.
Codebooks and their distributions are fixed independently of the inputs.
An encoder may select a codeword using \(x\).
Every source-dependent choice made available to the reader or decoder
must be communicated through the stored record.
\end{revision}

\begin{revision}
\begin{definition}[Access Classes]
\label{def:access-classes}
With the public and query-time seeds fixed, we classify readers by the
information their address rules use.  The decoder has the same inputs
in every class.
\begin{itemize}[leftmargin=*]
\item \textbf{Query-Oblivious}: The entire address list is fixed before the
query and before any bit is returned.  The minimax risk is
\(\D_d^{\mathrm{obl}}(w,r)\).
\item \textbf{Query-Independent Branching}: Each address is a function
only of previously returned bits.  The first address is fixed, and the
later addresses may adapt to the record through those bits.
The minimax risk is \(\D_d^{\mathrm{ind}}(w,r)\).
\item \textbf{Query-Adaptive}: Addresses may depend on the query \(y\).
A \emph{fully branching} reader may also use every previously returned
bit; its minimax risk is \(\D_d^{\mathrm{ad}}(w,r)\) in
Equation~\eqref{eq:minimax-risk}.  We also call this access
\emph{query-dependent}.

For \(1\le J\le r\), a reader is \emph{\(\mathrm{staged}(J)\)} if it
uses \(J\) read batches with fixed boundaries
\[
  0=t_0<t_1<\cdots<t_J=r
\]
and chooses all addresses in batch \(j\) using \(y\) and
\(z_{\le t_{j-1}}\), the bits returned by earlier batches.
Thus, a \(\mathrm{staged}(1)\) reader chooses all addresses before the
first read, while a fully branching reader is \(\mathrm{staged}(r)\)
when \(r\ge1\).
\end{itemize}
When \(r=0\), every reader returns an empty transcript.
\textcolor{black}{In standard cell-probe terminology, a \emph{nonadaptive}
reader may choose addresses using the query, but not previously returned
bits.  For \(r\ge1\), this corresponds to our \(\mathrm{staged}(1)\) class.}
The classes are nested, so
\[
  \D_d^{\mathrm{ad}}(w,r)
  \leq
  \D_d^{\mathrm{ind}}(w,r)
  \leq
  \D_d^{\mathrm{obl}}(w,r).
\]
Theorem~\ref{thm:oblivious-collapse} will show that the final inequality is
always an equality for the inner-product problem studied here.
\end{definition}

For comparison, we write
\(\D_d^{\mathrm{full}}(m)\) for the minimax risk when the encoder stores an
arbitrary \(m\)-bit message and the reader receives it in full.
\end{revision}

\subsection{Notation}
\label{sec:notation}

The following table collects our main notation.
\begin{center}
\small
\begin{tabular}{ll}
\(d;\ x,y\in\Bdd\) & dimension; source and query \\
\(w,r;\ W,R\) & bit budgets; write and read rates per dimension \\
\(\D_d^{\mathrm{obl}}(w,r)\) & query-oblivious minimax MSE \\
\(\D_d^{\mathrm{ind}}(w,r)\) & query-independent branching minimax MSE \\
\(\D_d^{\mathrm{ad}}(w,r)\) & fully branching query-adaptive minimax MSE \\
\(\D_d^{\mathrm{full}}(m)\) & full-read minimax MSE for an \(m\)-bit message
\end{tabular}
\end{center}

\begin{revision}
Normalized risk is MSE multiplied by \(d\).  At fixed rates \(0<R\le W\),
we use \(w=\lfloor Wd\rfloor\) and \(r=\lfloor Rd\rfloor\).  These are
budgets for the entire vector: each stored bit may depend on all
coordinates.  We take \(d\to\infty\) at fixed \(W,R\) before studying
limits of the rates.  In dimension limits, constants in \(O(\cdot)\)
bounds may depend on the fixed rates and on a stated fixed parameter
interval.
\end{revision}

The norm \(\|\cdot\|_2\) is Euclidean, \(S^{d-1}\) denotes the unit sphere,
and \([n]=\{1,\ldots,n\}\), with \([0]=\varnothing\).
We write \(\E\), \(\Prb\), \(\Var\), and \(\Cov\) for expectation,
probability, variance, and covariance.  Unless stated otherwise,
expectations in a protocol's risk are over its randomness.
Logarithms \(\log\) are natural and \(\log_2\) is base two.

\section{Related Work}

We work in the static bit-probe model, the one-bit-cell specialization of
Yao's static cell-probe model \cite{yao-cell-probe}. This model charges memory probes while
leaving computation unrestricted.  Classical
bit-probe work includes static set membership with randomized or adaptive
few-probe schemes
\cite{buhrman-bitvectors,garg-bitprobe}.  Exact linear-query data structures
over \(\mathbb F_2\) and locally decodable source coding provide related
models \cite{ramamoorthy-rashtchian,ko-linear-operators,
makhdoumi-local,pananjady-local}.  In standard cell-probe terminology, a
nonadaptive reader may choose addresses from the query but not from returned
\textcolor{black}{bits}, corresponding to our \(\mathrm{staged}(1)\) class.  Our problem
allows queries throughout the Euclidean unit ball and measures the
worst-case MSE of scalar estimates under separate storage and probe budgets.

\begin{revision}
For full reading, we use the one-sided matrix-multiplication quantizer of
Ordentlich and Polyanskiy \cite[Theorem~3(2)]{ordentlich-polyanskiy}.
Applied to single-column matrices, it gives asymptotic worst-case error
\(1/[d(4^R-1)]\) on the Euclidean unit ball.
Their one-sided lower bound, Theorem~2(2), concerns independent Gaussian
matrices and gives the rate-distortion scale \(4^{-R}\).
Theorem~2 of Safaryan, Shulgin, and Richt\'arik \cite{safaryan-compression} gives the matching
coefficient \(1/(4^R-1)\) for unbiased finite-bit vector compressors.  Our full-read contribution is the matching
worst-case converse for arbitrary scalar decoders, which may be biased
and nonlinear in the query.
\end{revision}
For every finite \(d,w,r\)
with \(w\ge r\), Theorem~\ref{thm:oblivious-collapse} equates
query-independent \(r\)-probe access with full reading of an \(r\)-bit
message.  Together, these results identify the exact asymptotic risk for
query-independent local access.  Work on additive inner-product
approximation, compressed similarity signatures, distortion side
information, and functional quantization uses different error criteria or
side-information models
\cite{alon-klartag,ingber-similarity,martinian-side-information,
misra-functional}.  \textcolor{black}{We focus on how the query guides access
to a limited number of stored bits.}

Explicit full-read quantizers often combine random rotation, scalar
quantization, and ratio correction.  Examples include DRIVE and EDEN
\cite{drive,eden}, TurboQuant \cite{turboquant}, Block-Sphere Vector
Quantization \cite{blocksphere}, and RaBitQ \cite{rabitq}.
\textcolor{black}{Metric-universal vector quantization studies codebooks
that work across quadratic query metrics
\cite{metric-universality}.}  White and Singal optimize adaptive quantizers
for worst- and average-case inner-product variance \cite{white-singal}, but
their adaptation occurs when the record is written.  These works reconstruct
a vector, analyze a specified architecture, or adapt the write-time
quantizer rather than query-time addresses.

KV-cache systems such as Quest use the current query to select cached
data \cite{quest}.  Our companion paper
\cite{bhandari2026readmattersqueryadaptivequantization} studies KV-cache
quantization under separate storage and read budgets from a systems
perspective.  Here, we study how query-dependent access to bits within a
single record affects minimax inner-product error.

Product quantization represents a vector by indices into shared
subcodebooks \cite{jegou-pq}; QUIP learns such codebooks for inner-product
error \cite{guo-quip}.  In the usual scoring rule, every subcode index of a
scanned vector contributes to its score.  Our packet extension of
Theorem~\ref{thm:oblivious-collapse} shows that a query-independent reader
of selected subcodes can be simulated by fully reading a message
containing only the retrieved indices.

\textcolor{black}{Our adaptive converse follows the correlation-estimation
approach of Hadar, Liu, Polyanskiy, and Shayevitz
\cite{hadar-correlation}.  For the labels produced by local reads, we
establish the required information bounds using the pseudotranscript
framework of Prabhakaran and Prabhakaran
\cite{prabhakaran-pseudotranscript}.}  A \emph{pseudotranscript}
is a randomized summary whose conditional output probability factors into
a function of the source and a function of the query.  With the protocol's
random seeds fixed, our labels recording the cells read and their contents have this
property.  We prove an information bound for these labels under independent
copies of a doubly symmetric binary source (DSBS), a pair of correlated
fair bits.  We compare different correlations by conditioning on a randomly
selected subset of coordinate pairs and applying this bound to the
remaining pairs.  This connects to strong data-processing inequalities
(SDPIs), which quantify how much information is lost
under an observation channel
\textcolor{black}{\cite{polyanskiy-wu-sdpi,liu-cuff-verdu}}, and to
Fisher-information bounds under communication constraints
\cite{barnes-fisher}.

For the construction, dithered quantization, random-hyperplane sketches, and
order-statistic selection of random projections provide the closest
technical precedents
\cite{zamir-feder-dither,charikar-hyperplane,valsesia-magli}.  Our record is
persistent, each query selects a strict subset of its cells, and the
analysis is uniform over the unknown source-query correlation.

\section{Query-Independent Access}
\label{sec:query-independent-results}

In this section, we characterize query-independent readers in two steps.
We first show that a branching \(r\)-probe transcript can be stored as an
\(r\)-bit message without changing the output distribution.
\textcolor{black}{We then combine the known full-read upper bound with
our converse for arbitrary scalar decoders to obtain normalized minimax
risk \(1/(4^R-1)\) at rate \(R\).  We prove the finite identity below;
Appendix~\ref{app:full-read} gives the cited upper bound's specialization
to our model and the converse proof.}  Together they show that using the
write surplus \(W-R\) requires query-dependent addresses.

\subsection{\textcolor{black}{Query-Independent Collapse}}

The finite identity follows by simulating the reader before the query
arrives.  Once the random seeds are fixed, a query-independent reader is
a decision tree that the encoder can run before \(y\) is known.
The returned bits determine every later address.  Storing those bits
therefore reproduces the answer for any query.
Conversely, we can place a full \(r\)-bit message in fixed cells.
These two simulations give the following identity.
For the packet version, \(b\) is the number of bits per packet, \(N\) is
the number of stored packets, and \(k\le N\) is the packet-probe budget.
Packets are disjoint blocks with fixed boundaries;
each probe returns one entire block.

\begin{theorem}[Exact Query-Independent Collapse]
\label{thm:oblivious-collapse}
For every finite \(d\) and integers \(w\ge r\ge0\),
\[
  \D_d^{\mathrm{obl}}(w,r)
  =
  \D_d^{\mathrm{ind}}(w,r)
  =
  \D_d^{\mathrm{full}}(r).
\]
For aligned \(b\)-bit packets, \(N\ge k\), and at most \(k\) packet probes,
both corresponding query-oblivious and query-independent branching risks
equal \(\D_d^{\mathrm{full}}(kb)\).
\end{theorem}

\begin{proof}
\begin{revision}
Sample the query-time seed before encoding and reveal it to the encoder.
The seed is independent of the source and query, and the original encoder
ignores it.  Thus, for every fixed pair \((x,y)\), this rescheduling
preserves the joint law of the original record, seed, and answer.
Matching these pairwise distributions is enough because the minimax
criterion evaluates one source--query pair at a time.
\end{revision}
After conditioning
on the public and query-time seeds, a query-independent branching reader is
a deterministic decision tree.  The encoder can run its padded depth-\(r\)
tree on the stored record before the query arrives.  Let
\(T_{\mathrm{ind}}(x)\in\{0,1\}^r\) be \textcolor{black}{the sequence of returned bits}.  An \(r\)-bit
full-read encoder stores \(T_{\mathrm{ind}}(x)\).  Given the later query, the decoder
reconstructs every address from the public tree and preceding transcript
values and applies the original decoder.  The output distribution is
identical, proving
\(\D_d^{\mathrm{full}}(r)\le\D_d^{\mathrm{ind}}(w,r)\).

Every query-oblivious reader is query-independent, so
\(\D_d^{\mathrm{ind}}(w,r)\le\D_d^{\mathrm{obl}}(w,r)\).
Conversely, place any \(r\)-bit full message in the first \(r\) cells and
read those cells in a fixed order.  This is query-oblivious and proves
\(\D_d^{\mathrm{obl}}(w,r)\le\D_d^{\mathrm{full}}(r)\).
The three inequalities give the claimed identity.  The packet argument is
identical with alphabet cardinality \(2^b\).  A length-\(k\) packet
transcript contains exactly \(kb\) bits.
\end{proof}

{\color{black}
\begin{corollary}[Adaptive Comparison and Write Surplus]
\label{cor:write-surplus-necessary}
For every finite \(d\) and integers \(w\ge r\ge0\),
\[
  \D_d^{\mathrm{full}}(w)
  \le \D_d^{\mathrm{ad}}(w,r)
  \le \D_d^{\mathrm{full}}(r).
\]
When \(w=r\), every access class has risk
\(\D_d^{\mathrm{full}}(w)\).  Consequently, a strict minimax advantage
from query-dependent addressing requires \(w>r\).  In the fixed-rate
limit, it requires \(W>R\).
\end{corollary}

\begin{proof}
A decoder receiving the entire \(w\)-bit record can simulate any adaptive
reader, using the same query and protocol seeds.  The simulation preserves
the answer distribution, which proves the lower bound.
For the upper bound, store an \(r\)-bit full message in the first \(r\)
cells and read them in a fixed order; this is an admissible adaptive
protocol with the same risk.
When \(w=r\), the two bounds coincide.  Theorem~\ref{thm:oblivious-collapse}
gives the same risk for both query-independent classes.
If \(W=R\), the floored budgets agree for every \(d\).
\end{proof}
}

\subsection{\textcolor{black}{Full-Read Minimax Risk}}

\begin{revision}
The full-read upper bound is due to Ordentlich and Polyanskiy
\cite[Theorem~3(2)]{ordentlich-polyanskiy}.  Our contribution in
Theorem~\ref{thm:full-read} is the matching converse for arbitrary scalar
functions of the message and query.  This extends the lower bound for
unbiased vector compressors \cite{safaryan-compression}.
\end{revision}
Recent randomized-Hadamard quantizers,
including a pairwise-independent dithered construction, also give worst-case
expected inner-product guarantees for explicit full-read schemes
\cite{feng-hadamard,lin-pairwise-dither}.

\begin{newrevision}
The converse starts from the Shannon lower bound on reconstruction
error.  We apply it to an auxiliary Gaussian-query experiment to bound
the average Fisher information about the query amplitude.  A Gaussian
conditioning argument then turns this information bound into the sharp
scalar error bound for arbitrary decoders.  The outline after the
theorem explains the two estimates and how they fit together.
\end{newrevision}

\begin{theorem}[Exact Full-Read Constant]
\label{thm:full-read}
For every sequence of nonnegative integer budgets \(L_d\) satisfying
\(L_d/d\to R>0\),
\[
  \lim_{d\to\infty}
  d\,\D_d^{\mathrm{full}}(L_d)
  =
  \frac{4^{-R}}{1-4^{-R}}
  =
  \frac1{4^R-1}.
\]
\end{theorem}

{\color{black}
Together with Corollary~\ref{cor:write-surplus-necessary}, this gives a
storage-dependent lower bound for adaptive reading: for every fixed
\(0<R\le W\),
\[
  \liminf_{d\to\infty}
  d\,\D_d^{\mathrm{ad}}(\lfloor Wd\rfloor,\lfloor Rd\rfloor)
  \ge \frac1{4^W-1}.
\]
}

\textcolor{black}{Section~\ref{sec:cited-full-read-upper} specializes the
cited upper bound to the unit ball and the integer budget \(L_d\).
The remaining argument proves the converse.}

\begin{newrevision}
\textcolor{black}{For any scalar decoder with worst-case MSE at most \(D\), the answers to
the coordinate queries reconstruct a vector \(\widehat X\) with
mean-squared Euclidean error at most \(V=dD\), without requiring
linearity or unbiasedness.  The reconstruction lower bound therefore
gives \(\liminf_d dD_d\ge4^{-R}\) when \(L_d/d\to R>0\), weaker than
\(1/(4^R-1)\).  For linear decoders,
Proposition~\ref{prop:basis-reconstruction} sharpens this argument by
rescaling the reconstructed vector.}
The answer at \(y=x\) \textcolor{black}{then} also bounds
\(\bigl|\E\langle\widehat X-x,x\rangle\bigr|\) by \(\sqrt D\).
Multiplying the reconstructed vector by \(1/(1+V)\)
then gives a reconstruction error bound of
\[
  \frac{V}{1+V}+\frac{2V}{(1+V)^2}\sqrt D.
\]
For a source uniform on \(\Bdd\), every reconstruction from an \(L\)-bit
message has mean-squared error at least \(c_d2^{-2L/d}\), where \(c_d\to1\).
Along a subsequence with \(dD_d\to v<\infty\), these bounds give
\(4^{-R}\le v/(1+v)\), and hence \(v\ge1/(4^R-1)\).
We now discuss the converse for general scalar decoders, whose answers
need not come from a single reconstructed vector.
\end{newrevision}

\begin{revision}
The converse uses information to bound estimation accuracy.
We write \(H\) for entropy, \(I\) for mutual information, and
\(\KL(P\|Q)\) for Kullback--Leibler (KL) divergence.
These quantities use natural logarithms and are measured in nats;
one bit equals \(\log2\) nats.
For an observation \(O\) with density \(p_\theta\) relative to a fixed
measure, the \emph{score}
\(\partial_\theta\log p_\theta(O)\) describes how its log-likelihood
changes with the parameter \(\theta\), when the derivative exists.
Its expected square is the scalar \emph{Fisher information}:
\[
  J_\theta(O)
  =
  \E_\theta\!\left[
    \bigl(\partial_\theta\log p_\theta(O)\bigr)^2
  \right].
\]
This measures how sensitive the observation's distribution is to
changes in \(\theta\).
\end{revision}

\begin{newrevision}
We use an auxiliary experiment to connect the decoder's error to its
information about the source.  Take \(X\) uniform on \(S^{d-1}\), and
put the stored message and all protocol seeds in \(S\), so that
\(I(X;S)\le L_d\log2\).  Independently, draw
\(\Theta\sim N(0,A^2)\) and \(G\sim N(0,I_d)\), and set
\[
  Y=\Theta X+G,
  \qquad Z=X^\top Y,
  \qquad \lambda=A^{-2}.
\]
For a scalar decoder with worst-case MSE \(D_d\), define
\[
  g(S,Y)=\|Y\|_2\,
    \widehat s\!\left(S,Y/\|Y\|_2\right),
\]
with value zero at \(Y=0\).  Conditional on \(X,Y,\Theta\), the
protocol seeds retain their original law.  Applying the risk guarantee
to the fixed source and normalized query, then averaging, gives
\[
  \delta_d:=\E(g-Z)^2\le(d+A^2)D_d.
\]
The estimator \(g\) observes \(S,Y\), while \(\Theta\) is hidden.

For the first estimate, fix an amplitude \(\theta\) and write
\[
  Y_\theta=\theta X+G,
  \qquad
  \mu_\theta=\E[X\mid S,Y_\theta],
  \qquad
  C_\theta=\operatorname{Cov}(X\mid S,Y_\theta).
\]
The Gaussian observation adds at most \(\theta^2/2\) nats about \(X\),
as follows by comparing its conditional law with \(N(0,I_d)\).
Multiplying \(X\) by an independent radius produces a uniform ball
source, to which Lemma~\ref{lem:ball-shannon} applies.  With
\(a_d=d/(d+2)\), \(\beta_d=d/(d+1)\), and \(c_d\to1\) from that
lemma, this gives
\[
  \E_\theta\operatorname{tr}C_\theta
  \ge1-\frac{a_d-c_d2^{-2L_d/d}e^{-\theta^2/d}}{\beta_d^2}.
\]
The score in \((S,Y_\theta)\) is
\(s_\theta=Y_\theta^\top\mu_\theta-\theta\).
Let \(J_d(\theta)=\E_\theta s_\theta^2\) and
\(\overline J_{d,A}=\E J_d(\Theta)\).
Viewing the conditional moments as functions of \((S,y)\), the
posterior mean satisfies
\(\nabla_y\mu_\theta=\theta C_\theta\) and
\(\partial_\theta\mu_\theta=C_\theta y\).
Gaussian integration by parts in the observation and in the amplitude
then yields
\[
  \overline J_{d,A}
  \le1-\frac12\E\!\left[
      (1+\Theta^2/A^2)\operatorname{tr}C_\Theta\right].
\]
The weight \((1+\Theta^2/A^2)/2\) has mean one.  Under this weighted
Gaussian law, the second moment of \(\Theta\) is \(2A^2\), so Jensen's
inequality averages the exponential term above and gives
\[
  \overline J_{d,A}
  \le\min\left\{1,\,
    \frac{a_d-c_d2^{-2L_d/d}e^{-2A^2/d}}{\beta_d^2}\right\}.
\]
Proposition~\ref{prop:gaussian-averaged-information} derives this bound
using only first derivatives of the posterior mean.

For the second estimate, the Gaussian prior and likelihood give
\[
  \Theta\mid X,S,Y
  \sim N\!\left(\frac{Z}{1+\lambda},\frac1{1+\lambda}\right).
\]
Thus, the residual \(\eta=Z-(1+\lambda)\Theta\) has conditional mean
zero and conditional variance \(1+\lambda\).
Its conditional expectation given \(\Theta,S,Y\) is
\(B=s_\Theta-\lambda\Theta\), with
\(\E B^2=\overline J_{d,A}+\lambda\).
These conditional moment identities imply
\[
  \E[(g-(1+\lambda)\Theta)B]=1+\lambda,
  \qquad
  \E(g-(1+\lambda)\Theta)^2=\delta_d+1+\lambda.
\]
Cauchy--Schwarz therefore gives
\[
  \delta_d\ge
  \frac{(1+\lambda)(1-\overline J_{d,A})}
       {\overline J_{d,A}+\lambda}.
\]
Lemma~\ref{lem:gaussian-amplitude-risk} proves these identities for
arbitrary measurable \(g(S,Y)\) with finite error.

Finally, choose \(A^2=\sqrt d\), so \(\lambda=d^{-1/2}\).
The risk transfer becomes \(\delta_d\le d(1+\lambda)D_d\).
Canceling the common factor \(1+\lambda\) in the two error bounds gives
\[
  dD_d\ge
  \frac{1-\overline J_{d,A}}{\overline J_{d,A}+d^{-1/2}}.
\]
Since \(L_d/d\to R\), the information bound tends to \(1-4^{-R}\).
The right-hand side is decreasing in \(\overline J_{d,A}\), so
\[
  \liminf_{d\to\infty}dD_d
  \ge\frac{4^{-R}}{1-4^{-R}}=\frac1{4^R-1}.
\]
The finite-dimensional bounds hold for every protocol and hence for
the minimax risk.  Section~\ref{sec:full-read-proof} gives this final
step in full.
\end{newrevision}

\begin{corollary}[Query-Independent Fixed-Rate Constant]
\label{cor:oblivious-rate}
For every fixed \(0<R\le W\),
\[
  \lim_{d\to\infty}
  d\,\D_d^{\mathrm{ind}}
  (\lfloor Wd\rfloor,\lfloor Rd\rfloor)
  =
  \lim_{d\to\infty}
  d\,\D_d^{\mathrm{obl}}
  (\lfloor Wd\rfloor,\lfloor Rd\rfloor)
  =
  \frac1{4^R-1}.
\]
In particular, neither query-independent branching nor its fixed-layout
query-oblivious subclass can exploit \(W-R\).
\end{corollary}

\begin{proof}
Combine Theorems~\ref{thm:oblivious-collapse}
and~\ref{thm:full-read}.
\end{proof}

The small-read-rate expansion is
\[
  \frac1{4^R-1}
  =
  \frac1{R\log4}
  -
  \frac12
  +
  O(R),
  \qquad R\downarrow0.
\]
Hence, the query-independent normalized risk scales as
\(1/R\).  In the large-write-surplus regime, the refined construction
discussed in Section~\ref{sec:adaptive-separation} has achievable
normalized risk asymptotic to
\[
  \frac{\pi}{4R\log(W/R)}.
\]
\textcolor{black}{After the dimension limit, let \(R\downarrow0\) and
\(\Lambda=W/R\to\infty\).  The ratio of the query-independent optimum
to this achievable bound is then asymptotic to}
\[
  \frac{4}{\pi\log4}\log(W/R)
  =
  \frac{2}{\pi\log2}\log(W/R).
\]
The separation therefore comes from the additional logarithmic gain in the
write-to-read ratio.

\begin{remark}[Product Quantization]
Our results also have consequences for partial-read product quantization,
where a vector is stored as a tuple of indices into shared subcodebooks.
Suppose that only some of these indices may be fetched.  The packet extension
of Theorem~\ref{thm:oblivious-collapse} shows that a query-independent
packet reader is equivalent to fully reading only the fetched indices.
Consequently, unread indices cannot reduce worst-case MSE unless the query
influences which indices are selected.  Specifically, fetching at most
\(k\) aligned indices of \(b\) bits each gives risk at least
\(\D_d^{\mathrm{full}}(kb)\).  If \(k_db_d/d\to R>0\),
Theorem~\ref{thm:full-read} makes its normalized limit inferior at least
\(1/(4^R-1)\).  This bound applies regardless of the particular shared
subcodebooks.  We treat shared codebooks as fixed independently of the
evaluated source.  A source-dependent codebook would carry additional
information, whose storage and access must also be counted.
\end{remark}

\section{Query-Dependent Access and Separation}
\label{sec:adaptive-results}

\textcolor{black}{In this section, we prove the separation using an explicit
two-stage construction.}
Each stored bit records whether a Gaussian source projection exceeds
an independent public Gaussian threshold.
\textcolor{black}{The reader selects cells with large query projections and
thresholds near the source projections predicted by the pilot.
The decoder then inverts the selected bits' average response.
The sharper protocol uses the same stored record.}
Section~\ref{sec:explicit-construction} specifies the protocol,
Section~\ref{sec:adaptive-achievability} bounds its risk, and
Section~\ref{sec:adaptive-separation} compares it with query-independent
reading.  We then state the sharper refinement used in
Section~\ref{sec:adaptive-lower-bounds}.

\subsection{Explicit Two-Stage Protocol}
\label{sec:explicit-construction}

Protocol~\ref{prot:window-reader} starts with a small fixed pilot batch
to estimate the inner product.  It then looks for unused cells with large
query projections and thresholds close to the source projections predicted
by the pilot.  Large query projections make these bits sensitive to
changes in the inner product.  A randomized acceptance rule makes their
threshold offsets uniform in a fixed window, so their average response
remains well behaved even near inner products \(+1\) and \(-1\).
For each candidate inner product, the Gaussian model gives the expected
fraction of positive accepted bits.  The decoder inverts this monotone
relation to refine the pilot estimate.

\begin{revision}
For unit directions \(u,v\), write \(\rho=u^\top v\) for their true
correlation.  We use \(\theta\) for a candidate correlation and \(s\)
for a pilot estimate.  Let \(\phi,\Phi\) be the standard normal density
and distribution function, and put \(\Lambda=W/R\).
\end{revision}
The construction applies when \(\Lambda\)
is large enough that
\begin{equation}
  t=t_\Lambda=\sqrt{\log(\Lambda/\pi)-1}-2\ge2.
  \label{eq:window-threshold}
\end{equation}
\begin{revision}
Equivalently, this construction requires
\(\Lambda\ge\pi e^{17}\), and \(t_\Lambda^2\sim\log\Lambda\).
This explicit sufficient threshold proves the asymptotic separation.
The information-based refinement below
applies to every fixed \(0<R<W\).
\end{revision}
Appendix~\ref{app:dither} gives the shared preliminaries and complete
proofs.  The norm representation and finite-dimensional conventions are
in Section~\ref{sec:unit-ball-protocol}.

\begin{mdframed}[style=protocolshade]
\begin{protocol}[Gaussian-Dither Window Protocol]
\label{prot:window-reader}
Fix rates satisfying Equation~\eqref{eq:window-threshold} and integer budgets
\(0\le k_d\le N_d\), \(N_d/d\to W\), and \(k_d/d\to R\).
Here, \(N_d,k_d\) count directional cells and probes.  In the unit-ball
implementation, they are the budgets remaining after reserving
\(b_d=\lceil\log_2(d+1)\rceil\) stored and read bits for the source norm,
as in Appendix~\ref{sec:unit-ball-protocol}.
For unit directions, take \(x=u\), \(y=v\), and \(b_d=0\).
Set \(n_d=\lceil d^{3/4}\rceil\) and \(m_d=k_d-n_d\).
If \(d<2\) or \(m_d\le0\), the encoder stores an all-zero record,
the reader makes no probes, and the decoder returns \(\widehat s=0\).
Otherwise, use the following steps.

\begin{enumerate}[leftmargin=*]
\item \textbf{Encoder: Source Record.}
Before the query arrives, set \(u=x/\|x\|_2\) when \(x\ne0\), using a
fixed unit vector when \(x=0\).  If \(b_d>0\), round \(\|x\|_2\) to the
nearest point \(\widehat a_d\) in \(\{0,1/d,\ldots,1\}\), breaking ties
toward the smaller value, and store its grid index in the first \(b_d\)
cells.  If \(b_d=0\), take \(\widehat a_d=1\).
Draw independent public rows \(G_i\sim N(0,I_d)\) and thresholds
\(\Theta_i\sim N(0,1)\), and store
\[
  Z_i=\sgn(G_i^\top u-\Theta_i),\qquad i=1,\ldots,N_d.
\]
Measurement \(i\) occupies cell \(b_d+i\).
Use \(\sgn(a)=1\) for \(a\ge0\) and \(-1\) otherwise, representing the
two signs by binary cell values.

\item \textbf{Reader: Pilot Batch.}
Set \(v=y/\|y\|_2\) when \(y\ne0\), using a fixed unit vector otherwise,
and compute \(B_i=G_i^\top v\).
\textcolor{black}{The public projection--threshold pairs
\((B_i,\Theta_i)\) form the \emph{scalar design} used by the protocol.}
The first batch reads the reserved norm cells, if any, and the first
\(n_d\) directional measurement cells.  From the measurement bits, set
\begin{equation}
  \begin{aligned}
    \widehat q_d
    &=\frac1{n_d}\sum_{i\le n_d}
       \mathbf1\{Z_i\ne\sgn(B_i-\Theta_i)\},\\
    \widetilde\rho_d
    &=\proj_{[-1,1]}\{2\cos(\pi\widehat q_d)-1\}.
  \end{aligned}
  \label{eq:window-pilot}
\end{equation}
Here, \(\proj\) denotes projection onto the indicated interval:
values outside it are clipped to the nearest endpoint.

\item \textbf{Reader: Window Selection Batch.}
For every nonpilot cell, independently accept it with probability
\begin{equation}
  a_s(b,c)=
  \mathbf1_{\{t\le b\le t+1,\ |c-sb|\le2\}}\,
  \frac{\phi(t+1)\phi(t+3)}{\phi(b)\phi(c)},
  \qquad s=\widetilde\rho_d.
  \label{eq:window-acceptance}
\end{equation}
The acceptance coins are independent query-time randomness.
This selection uses only public quantities and the pilot.
If at least \(m_d\) cells are accepted, read the first \(m_d\) in address
order as the second batch, and call their set \(\mathcal I_d\).
Otherwise, make no informative second-batch probes.
Pad any unused second-batch budget with ignored probes to unused cells.

\item \textbf{Decoder: Mean-Response Inversion.}
Using the public scalar pairs \((B_i,\Theta_i)\), query-time seed, and
returned transcript, reconstruct the pilot and selection outcome.
The decoder makes no probes.
If selection fails, set \(\widehat\rho_d=\widetilde\rho_d\).
Otherwise, put
\[
  \overline Y_d=\frac1{m_d}\sum_{i\in\mathcal I_d}\frac{Z_i+1}{2}.
\]
For \(s\in[-1,1]\), define
\begin{equation}
  \begin{aligned}
    K_s&=[-1,1]\cap
        \left[s-\frac1{t+1},s+\frac1{t+1}\right],\\
    H_s(\theta)&=\frac14\int_t^{t+1}\int_{-2}^{2}
         \Phi\left(\frac{(\theta-s)b-u}{\sqrt{1-\theta^2}}\right)\,du\,db.
  \end{aligned}
  \label{eq:window-response}
\end{equation}
Here, \(\theta\in(-1,1)\) is a candidate correlation.
Use the continuous extension of \(H_s\) at \(\theta=\pm1\).
It is strictly increasing on \(K_s\).  On successful selection, project
\(\overline Y_d\) onto \(H_{\widetilde\rho_d}(K_{\widetilde\rho_d})\) and set
\begin{equation*}
  \widehat\rho_d
  =H_{\widetilde\rho_d}^{-1}
     \left(\proj_{H_{\widetilde\rho_d}(K_{\widetilde\rho_d})}
                  (\overline Y_d)\right).
\end{equation*}
The decoder averages over the accepted design distribution in
Equation~\eqref{eq:window-response};
\textcolor{black}{its estimate is a function of the pilot and read-bit mean.}
\end{enumerate}

\textbf{Decoder: Final Output.}
Reconstruct \(\widehat a_d\) from the norm cells read in the first batch,
or take \(\widehat a_d=1\) for unit directions.  After either branch of
Step 4, return
\[
  \widehat s=\widehat a_d\,\|y\|_2\,\widehat\rho_d.
\]
Here, \(|\widehat a_d-\|x\|_2|\le1/(2d)\).
For unit directions, \(\widehat s=\widehat\rho_d\).
The formula returns zero when either input is zero.
\end{protocol}
\end{mdframed}

All second-batch addresses are fixed before their bits are observed.
For all sufficiently large \(d\), the reader therefore has fixed
boundaries \(b_d+n_d,b_d+k_d\) and is \(\mathrm{staged}(2)\).
For unit directions, \(b_d=0\), so these boundaries are \(n_d,k_d\).

\subsection{Direct Query-Adaptive Upper Bound}
\label{sec:adaptive-achievability}

\begin{revision}
\textcolor{black}{We now analyze the risk of
Protocol~\ref{prot:window-reader}.  The argument rests on three facts:}
the pilot places the true
correlation in a short interval, enough cells pass the window test, and
the accepted bits' mean response has slope of order \(t\).
The first two facts control the probability of failure.  The third
turns error in the bits' sample mean into error in the estimated
correlation.  Both constructions use the same conditional Gaussian
model and pilot identities.
\end{revision}

We first bound the pilot error.
The mismatch probability is
\(q(\rho)=\pi^{-1}\arccos((1+\rho)/2)\).
Its inverse is \(2\pi\)-Lipschitz, so the shared pilot bound in
Lemma~\ref{lem:shared-pilot} gives
\[
  \Prb_\rho\left\{
    |\widetilde\rho_d-\rho|>\frac1{t+1}
  \right\}
  \le2e^{-n_d/[2\pi^2(t+1)^2]}
  =o(d^{-1}).
\]
Thus, the true correlation lies in \(K_{\widetilde\rho_d}\) with
sufficiently high probability, uniformly on \([-1,1]\).

We next show that enough cells pass the window test.
The density ratio in Equation~\eqref{eq:window-acceptance} cancels the
Gaussian design density on the acceptance window.  Conditional on the
pilot \(s\), the accepted design therefore has the simple law
\[
  B\sim\operatorname{Unif}[t,t+1],\qquad
  \Theta-sB\sim\operatorname{Unif}[-2,2],
  \quad\text{independently}.
\]
Integrating the constant density over this rectangle gives the acceptance
probability in Lemma~\ref{lem:window-selection}:
\begin{equation*}
  \alpha(t)=4\phi(t+1)\phi(t+3)
  =\frac2\pi e^{-(t+2)^2-1}
  =\frac2\Lambda.
\end{equation*}
The expected number of accepted cells is therefore asymptotic to \(2Rd\),
while \(m_d\sim Rd\).  Independence of the nonpilot designs makes their
acceptance count binomial.  A lower-tail bound therefore makes selection
failure exponentially unlikely.

Finally, we bound the slope of the mean response to control the estimation
error.
For a fixed accepted projection \(b\), average over its uniform threshold
offset:
\[
  F(z,\sigma)=\frac14\int_{-2}^{2}\Phi((z-u)/\sigma)\,du,
  \qquad z=(\theta-s)b,\quad \sigma=\sqrt{1-\theta^2}.
\]
When \(\theta\in K_s\), we have \(|z|\le1\).
Differentiating this integral gives
\(F_z\ge(2\Phi(1)-1)/4\) and
\(|\theta F_\sigma/\sigma|\le\phi(1)/4\).
Thus, for interior \(\theta\) and \(b\ge2\),
\[
  \begin{aligned}
    \frac{d}{d\theta}F((\theta-s)b,\sqrt{1-\theta^2})
    &=bF_z-\frac{\theta}{\sigma}F_\sigma\\
    &\ge b\left\{\frac{2\Phi(1)-1}{4}-\frac{\phi(1)}8\right\}
      \ge\frac b8.
  \end{aligned}
\]
Averaging over \(b\in[t,t+1]\) proves \(H_s'(\theta)\ge t/8\).
Lemma~\ref{lem:window-response} justifies these bounds and extends the
inverse Lipschitz conclusion to the endpoints.
Consequently, on a successful selection and a good pilot,
\[
  |\widehat\rho_d-\rho|
  \le\frac8t\,|\overline Y_d-H_{\widetilde\rho_d}(\rho)|.
\]
Conditional on the pilot and selection success, the selected bits are
iid Bernoulli after averaging over their accepted design.
Their mean has variance at most \(1/(4m_d)\).
The clipped squared error is at most four on the exceptional events, so
\[
  \sup_{\rho\in[-1,1]}
  \E_\rho(\widehat\rho_d-\rho)^2
  \le\frac{16}{m_dt^2}+o(d^{-1}).
\]
The common norm-rescaling argument in
Appendix~\ref{sec:unit-ball-protocol} transfers this bound to the unit ball.
Section~\ref{sec:window-risk-proof} gives the complete proof of the
following theorem.

\begin{theorem}[Direct Two-Stage Query-Dependent Bound]
\label{thm:dither-reader}
Fix \(0<R<W\) such that Equation~\eqref{eq:window-threshold} holds.
With the unit-ball implementation,
Protocol~\ref{prot:window-reader} is admissible in every dimension and,
for all sufficiently large \(d\), uses two read batches.
Writing
\[
  \mathcal E_d^{\mathrm{win}}(W,R)
  =\sup_{x,y\in\Bdd}
      \E\bigl[(x^\top y-\widehat s(x,y))^2\bigr],
\]
one has
\begin{equation}
  \begin{aligned}
    \limsup_{d\to\infty}
    d\,\D_d^{\mathrm{ad}}(\lfloor Wd\rfloor,\lfloor Rd\rfloor)
    &\le\limsup_{d\to\infty}d\,\mathcal E_d^{\mathrm{win}}(W,R)\\
    &\le\frac{16}{Rt_\Lambda^2}.
  \end{aligned}
  \label{eq:window-upper-bound}
\end{equation}
As \(\Lambda=W/R\to\infty\),
\begin{equation}
  \frac{16}{Rt_\Lambda^2}
  =\frac{16+o(1)}{R\log\Lambda}.
  \label{eq:window-sparse-risk}
\end{equation}
The dimension limit is taken at fixed \(W,R\), before the rate-ratio limit.
\end{theorem}

By linearity of expectation, the pairwise bound also bounds average MSE
over source and query collections fixed independently of the shared
Gaussian design.

\subsection{Unbounded Separation from Query-Independent Access}
\label{sec:adaptive-separation}

Theorem~\ref{thm:oblivious-collapse} identifies query-independent reading
with an \(r\)-bit full message.  Theorem~\ref{thm:full-read} gives that
message normalized risk \(1/(4^R-1)\), independently of the write surplus.
Theorem~\ref{thm:dither-reader} supplies a query-dependent upper bound
that tends to zero as \(W/R\to\infty\) at fixed \(R>0\).
These two facts give the separation.

\begin{theorem}[\textcolor{black}{Unbounded Minimax Separation}]
\label{thm:separation}
Fix \(R>0\) and set \(W=\Lambda R\).  As \(\Lambda\to\infty\),
\begin{equation}
  \lim_{d\to\infty}
  d\,\D_d^{\mathrm{ind}}
  (\lfloor \Lambda Rd\rfloor,\lfloor Rd\rfloor)
  =
  \lim_{d\to\infty}
  d\,\D_d^{\mathrm{obl}}
  (\lfloor \Lambda Rd\rfloor,\lfloor Rd\rfloor)
  =
  \frac1{4^R-1},
  \label{eq:independent-rate}
\end{equation}
whereas
\begin{equation}
  \limsup_{d\to\infty}
  d\,\D_d^{\mathrm{ad}}
  (\lfloor \Lambda Rd\rfloor,\lfloor Rd\rfloor)
  \le
  \frac{16+o(1)}{R\log\Lambda}.
  \label{eq:separation-upper-bound}
\end{equation}
Consequently, the ratio between the query-independent optimum and the
guaranteed query-dependent risk is at least
\begin{equation}
  \frac{R\log\Lambda}{16(4^R-1)}(1-o(1)),
  \label{eq:separation-factor}
\end{equation}
which diverges.  Equivalently,
\[
  \lim_{\Lambda\to\infty}\limsup_{d\to\infty}
  \frac{\D_d^{\mathrm{ad}}(\lfloor\Lambda Rd\rfloor,\lfloor Rd\rfloor)}
       {\D_d^{\mathrm{ind}}(\lfloor\Lambda Rd\rfloor,\lfloor Rd\rfloor)}
  =0.
\]
The preceding limit also holds if
\(\D_d^{\mathrm{ind}}(\lfloor\Lambda Rd\rfloor,\lfloor Rd\rfloor)\)
is replaced by
\(\D_d^{\mathrm{obl}}(\lfloor\Lambda Rd\rfloor,\lfloor Rd\rfloor)\).
\end{theorem}

\begin{proof}
Fix \(\Lambda\) satisfying Equation~\eqref{eq:window-threshold} before taking the
dimension limit.  Corollary~\ref{cor:oblivious-rate} gives
Equation~\eqref{eq:independent-rate}.  Theorem~\ref{thm:dither-reader} and
\(t_\Lambda^2\sim\log\Lambda\) give
Equation~\eqref{eq:separation-upper-bound}.  Since the query-independent normalized
risk converges to the positive constant \(1/(4^R-1)\),
\[
  \limsup_{d\to\infty}
  \frac{\D_d^{\mathrm{ad}}(\lfloor\Lambda Rd\rfloor,\lfloor Rd\rfloor)}
       {\D_d^{\mathrm{ind}}(\lfloor\Lambda Rd\rfloor,\lfloor Rd\rfloor)}
  \le
  \frac{16(4^R-1)}{Rt_\Lambda^2}
  \longrightarrow0.
\]
Taking the reciprocal of this guaranteed comparison gives
Equation~\eqref{eq:separation-factor}.
Theorem~\ref{thm:oblivious-collapse} equates the query-independent and
query-oblivious risks.
\end{proof}

\paragraph{Improved Bound.} We now turn to the sharper information-based construction.
The same window argument also lower-bounds the information available to
an \emph{oracle selector}, which knows the true correlation when choosing
addresses, as Proposition~\ref{prop:window-information} shows.
Appendix~\ref{sec:sharp-protocol} specifies
Protocol~\ref{prot:dithered-reader}, which uses the same stored bits but
ranks cells by Fisher information and applies a score correction.
\textcolor{black}{The window decoder uses the selected bits only through
their mean.  The refined decoder pairs each bit from the second batch
with its public projection and threshold when computing the score
correction, so the estimate accounts for their different sensitivities
to the correlation.}

{\color{black}
For every fixed \(0<R<W\), let \(\J_{W,R}^{\min}\) be the minimum,
over correlations, of the information per dimension obtained by selecting
the most informative \(R/W\) fraction of cells
(Equation~\eqref{eq:oracle-information}).
Theorem~\ref{thm:sharp-dither-reader}\textcolor{black}{, proved in
Appendix~\ref{sec:unit-ball-extension},} shows that the refined protocol's
normalized worst-case risk converges to \(1/\J_{W,R}^{\min}\).
We may choose between this protocol and a fully read \(r\)-bit encoding
before storing the source.  Together with
Corollary~\ref{cor:oblivious-rate}, this gives
\[
  \limsup_{d\to\infty}
  d\,\D_d^{\mathrm{ad}}(\lfloor Wd\rfloor,\lfloor Rd\rfloor)
  \le
  \min\left\{\frac1{4^R-1},\frac1{\J_{W,R}^{\min}}\right\}.
\]
}
\textcolor{black}{As \(W/R\to\infty\), the refined bound improves
the coefficient \(16\) to \(\pi/4\):}
\[
  \limsup_{d\to\infty}
  d\,\D_d^{\mathrm{ad}}(\lfloor Wd\rfloor,\lfloor Rd\rfloor)
  \le\frac{\pi+o(1)}{4R\log(W/R)}.
\]
Here, \(o(1)\to0\) as \(W/R\to\infty\), after the dimension limit
at each fixed rate pair.
\textcolor{black}{This refinement supplies the upper bound used for comparison in
Section~\ref{sec:adaptive-lower-bounds}.}
Figure~\ref{fig:mse-separation} numerically illustrates the limiting risk
of Protocol~\ref{prot:dithered-reader} at \(R=1/4\);
Appendix~\ref{sec:refinement-numerics} gives the numerical details and
limitations.

\section{Query-Adaptive Lower Bounds}
\label{sec:adaptive-lower-bounds}

\begin{revision}
We compare the refined construction with the unrestricted query-adaptive
optimum.  Our lower bound applies to arbitrary encoders and fully
branching readers.  In the doubly sparse regime, it matches the order
\(1/[dR\log(W/R)]\) achieved by the construction.
\end{revision}
We allow each stored cell to contain one of \(q\) symbols.  Let
\(\D_d^{(q)}(N,k)\) be the corresponding minimax risk with \(N\) stored
cells and at most \(k\) probes; the binary model is \(q=2\).  Define
\[
  \mathsf C_{N,k}(q)=\binom Nkq^k.
\]

\begin{revision}
The proof in Appendix~\ref{app:adaptive-converse} has four steps.
First, record which cells were read and their values.  We call this record
a \emph{canonical label}.  Given the query, it determines the ordered
transcript, and it has at most \(\mathsf C_{N,k}(q)\) possible values.
Second, for each label, any compatible source can be paired with any
compatible query.  The resulting set of pairs is a Cartesian product,
or \emph{rectangle}.  This structure lets us bound how much the label
and query reveal about the source--query correlation.
Third, a cosine-squared prior converts the information bound into an
estimation lower bound; optimizing its width gives
\(\mathfrak A(N,k,q)\).
Fourth, the inner product fluctuates around the correlation by at most
\(d^{-1/2}\) in root mean square.  Accounting for this fluctuation gives
the subtraction below.  We write \(t_+=\max\{t,0\}\) for the positive part.
\end{revision}

\begin{theorem}[Finite Fully Branching Converse]
\label{thm:finite-adaptive-converse}
Let \(d\ge1\), \(q\ge2\), \(N\ge0\), and \(0\le k\le N\) be integers.  Put
\[
  \mathfrak A(N,k,q)
  =
  \left\{
    \bigl(2\log\mathsf C_{N,k}(q)\bigr)^{1/3}
    +\pi^{2/3}
  \right\}^{3}.
\]
The minimax risk over randomized \(q\)-ary protocols with \(N\) stored
cells and at most \(k\) fully branching probes satisfies
\begin{equation}
  \D_d^{(q)}(N,k)
  \ge
  \left[
    \frac1{\sqrt{\mathfrak A(N,k,q)}}
    -
    \frac1{\sqrt d}
  \right]_+^2.
  \label{eq:finite-branching-converse}
\end{equation}
\end{theorem}

We first show why the reader enters the bound only through
\(\mathsf C_{N,k}(q)\).  Repeated probes can be answered from cached
values, and ignored probes to unused addresses can pad a reader to exactly
\(k\) distinct cells.  The resulting partial address-value assignments
number at most \(\binom Nkq^k\).  The argument applies to every measurable
encoder and reader in our model.  We fix the random seeds.
\textcolor{black}{Given a query and a partial assignment, we simulate the
reader using the recorded values.  This checks whether the query is
compatible with the label and recovers the ordered transcript.}
For each label, the source--query pairs that
produce it form a Cartesian product of a source set and a query set.
This holds even when the address order depends on both the query and
previously read values.

Next, use \(X,Y\in\{\pm d^{-1/2}\}^d\) with independent, doubly symmetric
binary coordinate pairs satisfying \(\E[X_jY_j]=\rho/d\).  The marginal
query law is independent of \(\rho\).
\begin{revision}
Write \(\Gamma\) for the canonical label and \(\Omega\) for the seeds.
With the seeds fixed, the rectangle property bounds the information
carried by the label.  For one coordinate pair,
Lemma~\ref{lem:scalar-dsbs} compares the source distributions before
and after observing a label \(T\) with this product structure.
\end{revision}
An extra term in that bound has mean zero over labels, leaving
\[
  I(T;X)+I(T;Y)\le(1+\rho^2)I(T;X,Y).
\]
The two block identities in Lemma~\ref{lem:two-block} extend this
inequality to independent coordinate pairs: the difference from summing
the two block bounds is a nonpositive combination of conditional mutual
informations.

To compare nearby correlations \(\rho_0,\rho_1\), freeze each coordinate
pair with probability \(|\rho_0|\).
The frozen pairs have matching signs for \(\rho_0\ge0\) and opposite
signs for \(\rho_0<0\).  Give the remaining pairs correlation
\[
  \xi=\frac{\rho_1-\rho_0}{1-|\rho_0|}.
\]
Conditioning on the frozen coordinates preserves the label's rectangle
factorization.  The information inequality and a
KL chain rule then
bound the observed divergence by
\(\xi^2\log\mathsf C_{N,k}(q)\).
The quadratic expansion of KL divergence in
Lemma~\ref{lem:finite-latent} gives
\[
  J_\rho(\Omega,\Gamma,Y)
  \le
  \frac{2\log\mathsf C_{N,k}(q)}{(1-|\rho|)^2}.
\]
\textcolor{black}{Thus, the number of possible labels bounds the Fisher
information available for estimating the correlation.}

\begin{revision}
Finally, draw \(\rho\) from the cosine-squared prior
\[
  \pi_\varepsilon(\rho)
  =\varepsilon^{-1}
     \cos^2\!\left(\frac{\pi\rho}{2\varepsilon}\right)
     \mathbf1_{\{|\rho|<\varepsilon\}}.
\]
Its Fisher information is \(\pi^2/\varepsilon^2\).
\end{revision}
\textcolor{black}{Van Trees' inequality combines information from the
observations and the prior: the Bayes MSE for estimating \(\rho\) is at
least the reciprocal of}
\[
  \frac{2\log\mathsf C_{N,k}(q)}{(1-\varepsilon)^2}
  +\frac{\pi^2}{\varepsilon^2}.
\]
The infimum of this denominator over \(0<\varepsilon<1\) is exactly
\(\mathfrak A(N,k,q)\); when \(\mathsf C_{N,k}(q)=1\), the infimum is
understood as \(\varepsilon\uparrow1\).
\begin{revision}
The protocol estimates \(X^\top Y\), whose conditional mean is \(\rho\)
and whose conditional variance is \((1-\rho^2)/d\).
Under the joint prior and data law, the \(L_2\) triangle inequality gives
\[
  \|\widehat s-X^\top Y\|_{L_2}
  \ge
  \|\widehat s-\rho\|_{L_2}
  -
  \|X^\top Y-\rho\|_{L_2}
  \ge
  \mathfrak A(N,k,q)^{-1/2}-d^{-1/2}.
\]
The square of the left side is the average MSE, which is at most the
worst-pair risk.
Taking the positive part and squaring gives
Equation~\eqref{eq:finite-branching-converse}.
\end{revision}
Appendix~\ref{sec:branching-risk-proof} completes the proof of
Theorem~\ref{thm:finite-adaptive-converse}.

The canonical assignments are analogous to transcripts in adaptive
bit-probe models \cite{garg-bitprobe}.
To bound their information, we prove inequalities for pseudotranscripts
\cite{prabhakaran-pseudotranscript} using strong data-processing methods
\cite{polyanskiy-wu-sdpi,hadar-correlation} for a shared stored record.
We then use the Fisher-information method
for communication-constrained estimation
\cite{barnes-fisher,van-trees}.

We now express the finite converse in terms of the write and read rates
for binary cells, then compare it with the construction.  Let
\begin{equation}
  h(p)=-p\log p-(1-p)\log(1-p),\qquad 0\log0:=0,
  \label{eq:binary-entropy}
\end{equation}
be binary entropy in nats.
At fixed rates, the logarithm of the label count per dimension has two
limiting contributions:
\(Wh(R/W)\) for the address sets and
\(R\log2\) for the returned bits.  Their sum is the exponent
\(\ell(W,R)\) below.

\begin{corollary}[Unrestricted Query-Adaptive Rate Bounds]
\label{cor:adaptive-rate-bounds}
For every fixed \(0<R<W\), put
\[
  \ell(W,R)=Wh(R/W)+R\log2.
\]
Then
\begin{equation}
  \liminf_{d\to\infty}
  d\,\D_d^{\mathrm{ad}}(\lfloor Wd\rfloor,\lfloor Rd\rfloor)
  \ge
  \left[(2\ell(W,R))^{-1/2}-1\right]_+^2.
  \label{eq:binary-rate-converse}
\end{equation}
For any sequence \(0<R_n<W_n\), write \(\Lambda_n=W_n/R_n\) and assume
\begin{equation}
  \Lambda_n\to\infty,
  \qquad R_n\log\Lambda_n\to0.
  \label{eq:doubly-sparse-regime}
\end{equation}
The normalized minimax risks then satisfy
\begin{equation}
  \liminf_{n\to\infty}
  \left[
    R_n\log\Lambda_n
    \liminf_{d\to\infty}
    d\,\D_d^{\mathrm{ad}}(\lfloor W_nd\rfloor,\lfloor R_nd\rfloor)
  \right]
  \ge\frac12,
  \label{eq:sparse-branching-converse}
\end{equation}
\begin{equation}
  \limsup_{n\to\infty}
  \left[
    R_n\log\Lambda_n
    \limsup_{d\to\infty}
    d\,\D_d^{\mathrm{ad}}(\lfloor W_nd\rfloor,\lfloor R_nd\rfloor)
  \right]
  \le\frac{\pi}{4}.
  \label{eq:sparse-branching-upper}
\end{equation}
In both bounds, the dimension limit precedes the rate limit.
\end{corollary}

\begin{proof}
Set \(N_d=\lfloor Wd\rfloor\) and \(k_d=\lfloor Rd\rfloor\).
For all sufficiently large \(d\), \(N_d\ge1\).
For \(N\ge1\) and \(0\le k\le N\), a binomial variable with success
probability \(p=k/N\) has a mode at \(k\), whose mass lies between
\(1/(N+1)\) and one.  Writing that mass
as \(\binom Nk e^{-Nh(p)}\) gives
\[
  \frac{e^{N_d h(k_d/N_d)}}{N_d+1}
  \le\binom{N_d}{k_d}
  \le e^{N_d h(k_d/N_d)}.
\]
Since \(k_d/N_d\to R/W\), these bounds imply
\[
  \frac1d\log\mathsf C_{N_d,k_d}(2)
  =\frac1d\log\binom{N_d}{k_d}+\frac{k_d}{d}\log2
  \longrightarrow\ell(W,R)>0.
\]
The optimized denominator therefore satisfies
\[
  \frac{\mathfrak A(N_d,k_d,2)}d
  =\left\{
    \left(\frac{2\log\mathsf C_{N_d,k_d}(2)}d\right)^{1/3}
    +\pi^{2/3}d^{-1/3}
  \right\}^{3}
  \longrightarrow2\ell(W,R).
\]
Multiplying Equation~\eqref{eq:finite-branching-converse} by \(d\) and taking the
lower limit proves Equation~\eqref{eq:binary-rate-converse}.

Along a rate sequence satisfying Equation~\eqref{eq:doubly-sparse-regime}, the
entropy expansion at zero gives
\begin{revision}
\[
  \ell(W_n,R_n)
  =R_n\log\Lambda_n+O(R_n)
  =R_n\log\Lambda_n(1+o(1)).
\]
\end{revision}
In particular, \(\ell(W_n,R_n)\to0\), and
\[
  \left[(2\ell)^{-1/2}-1\right]_+^2
  =\frac{1+o(1)}{2\ell},\qquad \ell\downarrow0.
\]
Substitution in the fixed-rate lower bound proves
Equation~\eqref{eq:sparse-branching-converse}.
For the upper bound, Theorem~\ref{thm:sharp-dither-reader} gives
\[
  \limsup_{d\to\infty}
  d\,\D_d^{\mathrm{ad}}(\lfloor W_nd\rfloor,\lfloor R_nd\rfloor)
  \le\frac{\pi+o(1)}{4R_n\log\Lambda_n}.
\]
Multiplication by \(R_n\log\Lambda_n\) and the outer rate limit prove
Equation~\eqref{eq:sparse-branching-upper}.
\end{proof}

The \textcolor{black}{correlated-input} fixed-rate lower bound is positive
exactly when \(\ell(W,R)<1/2\).
\textcolor{black}{In the doubly sparse regime, this exponent tends to zero
and the lower bound is asymptotic to \(1/(2\ell)\).}  Thus, the lower and upper bounds
agree on the scale \(1/[dR\log(W/R)]\), with a factor of \(\pi/2\)
between their leading constants.  Appendix~\ref{sec:finite-alphabet-rates}
gives the corresponding rate calculation for general finite alphabets,
including the zero-exponent case and fixed-width packets.

{\color{black}
We now compare this converse with a geometric bound from the same
rectangle structure.  Draw the source and query independently and
uniformly from \(\Bdd\).  With the seeds fixed, conditioning on a
canonical label leaves independent uniform vectors in the corresponding
source and query sets.  The decoder knows the query, so its conditional
MSE is at least the source's conditional variance in that direction.
The volumes of the two sets bound their covariance
determinants and hence this average error.
Proposition~\ref{prop:geometric-rectangle-bound} gives the
finite-dimensional bound.  Using the label-count limit from the
preceding proof, we obtain, for every fixed \(0<R<W\),
\begin{equation}
  \liminf_{d\to\infty}
  d\,\D_d^{\mathrm{ad}}(\lfloor Wd\rfloor,\lfloor Rd\rfloor)
  \ge e^{-2\ell(W,R)}.
  \label{eq:geometric-rate-converse}
\end{equation}

This comparison explains the role of correlated inputs in our converse.
Under the independent uniform-ball input law,
\(\E[XX^\top]=\E[YY^\top]=I_d/(d+2)\), so even the zero estimator
has average MSE \(d/(d+2)^2\).  This input law therefore cannot establish
a normalized lower bound that grows without bound.
As \(\ell\downarrow0\), the geometric bound tends to one, whereas the
correlated-input bound is asymptotic to \(1/(2\ell)\).
At general fixed rates, each bound remains available, together with
the storage lower bound from Section~\ref{sec:query-independent-results}.
Taking their maximum and combining it with the upper bounds in
Section~\ref{sec:adaptive-results} gives the following statement.

\begin{corollary}[Combined Fixed-Rate Bounds]
\label{cor:combined-rate-bounds}
For every fixed \(0<R<W\), write \(\ell=\ell(W,R)\), and let
\(\J_{W,R}^{\min}\) be as in Equation~\eqref{eq:oracle-information}.
Then
\[
\begin{aligned}
  \max\left\{
    \frac1{4^W-1},\ e^{-2\ell},\
    \left[(2\ell)^{-1/2}-1\right]_+^2
  \right\}
  &\le \liminf_{d\to\infty}
    d\,\D_d^{\mathrm{ad}}(\lfloor Wd\rfloor,\lfloor Rd\rfloor)\\
  &\le \limsup_{d\to\infty}
    d\,\D_d^{\mathrm{ad}}(\lfloor Wd\rfloor,\lfloor Rd\rfloor)\\
  &\le \min\left\{
    \frac1{4^R-1},\ \frac1{\J_{W,R}^{\min}}
  \right\}.
\end{aligned}
\]
\end{corollary}

\begin{proof}
Corollary~\ref{cor:write-surplus-necessary} places the adaptive risk
between the full-read risks with budgets \(\lfloor Wd\rfloor\) and
\(\lfloor Rd\rfloor\).  Theorem~\ref{thm:full-read} therefore gives
the storage lower bound and the full-read upper bound.
Equation~\eqref{eq:geometric-rate-converse} and
Corollary~\ref{cor:adaptive-rate-bounds} give the other two lower bounds.
Theorem~\ref{thm:sharp-dither-reader} supplies the refined protocol's
upper bound.  All these statements hold at the same fixed rate pair.
Taking the maximum of the lower bounds and the minimum of the upper
bounds proves the claim.
\end{proof}
\par}

\section{Discussion and Open Problems}

{\color{black}
Every query-independent branching reader is equivalent to a fully read
\(r\)-bit message, so the limiting normalized minimax risk is
\(1/(4^R-1)\), independent of the storage rate.
Our refined two-batch reader uses the query to select informative bits
and achieves normalized risk \((\pi+o(1))/[4R\log(W/R)]\) as
\(W/R\to\infty\).  At fixed \(R>0\), its ratio to the query-independent
benchmark tends to zero as storage grows.
In the doubly sparse regime \(W/R\to\infty\) and
\(R\log(W/R)\to0\), the same reader is within a factor
\(\pi/2+o(1)\) of the optimum over fully branching readers.
\textcolor{black}{In both comparisons, we first let \(d\to\infty\) at fixed
\(0<R\le W\), with both minimax MSEs vanishing on the order of \(1/d\).
The divergence in the subsequent doubly sparse limit concerns the
normalized MSE, obtained by multiplying the actual MSE by \(d\).}
\par}

Our main theoretical open problem is to determine the sharp leading
constant for unrestricted query-adaptive access.  In the doubly sparse
regime in Equation~\eqref{eq:doubly-sparse-regime}, our lower and upper bounds on
minimax MSE have coefficients \(1/2\) and \(\pi/4\) multiplying
\(1/[dR\log(W/R)]\).
\textcolor{black}{After fixing the random seeds, our converse uses the rectangle
structure of the canonical labels and the bound \(M=\binom wr2^r\) on
their number.  A relaxation retaining only these properties permits a
fully read \(\lfloor\log_2M\rfloor\)-bit message, whose label sets are
source sets paired with all queries.  By Theorem~\ref{thm:full-read},
such a message can attain the sparse leading coefficient \(1/2\),
but need not satisfy the original \(r\)-probe budget.
A converse improving this sparse leading coefficient must therefore
use further restrictions of actual readers, such as at most \(2^r\)
possible bit histories for each fixed query and seed realization,
and consistency between overlapping reads of the same stored record.}
\textcolor{black}{We conjecture that our refined two-batch reader is minimax
optimal in the doubly sparse regime, with leading coefficient \(\pi/4\).}

\begin{conjecture}[Sharp Unrestricted Converse]
\label{conj:sharp-adaptive-converse}
For every sequence \(0<R_n<W_n\) with \(W_n/R_n\to\infty\) and
\(R_n\log(W_n/R_n)\to0\),
\[
  \liminf_{d\to\infty}
  d\,\D_d^{\mathrm{ad}}(\lfloor W_nd\rfloor,\lfloor R_nd\rfloor)
  \ge
  \frac{\pi-o(1)}{4R_n\log(W_n/R_n)}.
\]
Here, \(o(1)\to0\) as \(n\to\infty\), with the dimension limit taken first
at each fixed rate pair \((W_n,R_n)\).
\end{conjecture}

Together with Equation~\eqref{eq:sparse-branching-upper},
Conjecture~\ref{conj:sharp-adaptive-converse} would establish the
asymptotic minimax optimality of Protocol~\ref{prot:dithered-reader}
among all query-adaptive protocols in this regime.

{\color{black}
At each fixed rate pair \(0<R<W\), another open problem is to establish
the existence and determine the value of
\[
  \lim_{d\to\infty}
  d\,\D_d^{\mathrm{ad}}(\lfloor Wd\rfloor,\lfloor Rd\rfloor).
\]
Corollary~\ref{cor:combined-rate-bounds} bounds the corresponding lower
and upper limits.  An exact characterization would determine which
rate pairs allow a strict improvement over \(1/(4^R-1)\) and whether
choosing between a fully read \(r\)-bit encoding and our refined
two-batch protocol attains the optimum.
\par}

\clearpage
\appendix

\section{Full-Read Results}
\label{app:full-read}

\begin{newrevision}
\subsection{Dependency Structure of the Full-Read Converse}

Section~\ref{sec:cited-full-read-upper} obtains achievability from
Ordentlich and Polyanskiy \cite[Theorem~3(2)]{ordentlich-polyanskiy},
with a reduction to the unit ball and the integer message budget.
The converse starts from Lemma~\ref{lem:ball-shannon}, which bounds
reconstruction error for a uniform ball source in terms of mutual
information.  Proposition~\ref{prop:basis-reconstruction} uses this
lemma to give a short proof for linear decoders, including biased ones.
The proof for arbitrary scalar decoders uses the same lemma through
two further estimates:
\begin{enumerate}[leftmargin=*]
\item Proposition~\ref{prop:gaussian-averaged-information} bounds the
average Fisher information about a Gaussian amplitude.  The Shannon
bound supplies conditional uncertainty about the source; first
derivatives of the posterior mean and Gaussian integration by parts
turn that uncertainty into an information bound.
\item Lemma~\ref{lem:gaussian-amplitude-risk} converts the information
bound into a scalar error bound.  Conditioning the amplitude on the
source and observation gives explicit Gaussian moments, which enter
a Cauchy--Schwarz inequality valid for any measurable decoder with
finite error.
\end{enumerate}
Section~\ref{sec:full-read-proof} applies these estimates to a protocol,
chooses amplitude variance \(\sqrt d\), and obtains a finite-dimensional
lower bound whose limit is \(1/(4^R-1)\).
\end{newrevision}
\subsection{\textcolor{black}{Full-Read Achievability from Prior Work}}
\label{sec:cited-full-read-upper}

\begin{revision}
We apply the one-sided quantization bound of Ordentlich and Polyanskiy
\cite[Theorem~3(2)]{ordentlich-polyanskiy} to two vectors, viewed as
single-column matrices.  Fix a coding rate \(0<R_-<R\) and
\(\varepsilon>0\).  The cited theorem allows zero entries and nonzero
entries with magnitudes in \([M_d^{-1},M_d]\), where
\(M_d=d^{10}2^{2000}\).  To meet this condition uniformly over
\(x,y\in\Bdd\), set coordinates of magnitude below \(d^{-2}\) to zero,
obtaining \(x',y'\).  Then
\[
  \|x'\|_2,\|y'\|_2\le1,
  \qquad
  \|x-x'\|_2,\|y-y'\|_2\le d^{-3/2},
  \qquad
  |x^\top y-x'^\top y'|\le2d^{-3/2}.
\]
Every nonzero coordinate of \(x',y'\) has magnitude in \([d^{-2},1]\),
so the cited theorem applies for all sufficiently large \(d\).
Its encoder stores a message about \(x'\), and its decoder uses that
message and \(y'\) to produce \(\widehat s\).  The shared randomness is
permitted by our model.  Since centering a vector can only decrease its
Euclidean norm, the cited error bound gives, uniformly over \(x,y\in\Bdd\),
\[
  \E(\widehat s-x'^\top y')^2
  \le
  \frac1d\left(\frac1{4^{R_-}-1}+\varepsilon\right)+d^{-8}.
\]
The message uses at most \(\lceil dR_-\rceil\) bits, including the scalar
metadata in the cited construction.  Since \(L_d/d\to R>R_-\), it fits
within \(L_d\) bits for all sufficiently large \(d\).
The triangle inequality in \(L_2\) now yields
\[
  \sqrt{d\,\sup_{x,y\in\Bdd}\E(\widehat s-x^\top y)^2}
  \le
  \sqrt{\frac1{4^{R_-}-1}+\varepsilon+d^{-7}}+\frac2d.
\]
First taking \(d\to\infty\), then \(\varepsilon\downarrow0\) and
\(R_-\uparrow R\), proves
\[
  \limsup_{d\to\infty}d\,\D_d^{\mathrm{full}}(L_d)
  \le\frac1{4^R-1}.
\]
\end{revision}

\subsection{Entropy-to-MMSE Bound for the Euclidean Ball}

\begin{newrevision}
The following form of the Shannon lower bound \cite{shannon1959}
relates the information available about a uniform ball source to its
reconstruction error.  It will supply the information bound in both the
linear-decoder argument and the general scalar converse.
\end{newrevision}

Let
\[
  v_d=\operatorname{vol}(\Bdd)
  =
  \frac{\pi^{d/2}}{\Gamma(d/2+1)},
  \qquad
  c_d=\frac{d\,v_d^{2/d}}{2\pi e}.
\]

\begin{lemma}[Ball Shannon Lower Bound]
\label{lem:ball-shannon}
\textcolor{black}{Let \(U\) be uniform on \(\Bdd\), and let \(O\) be any
observation.}  Then
\begin{equation}
  \E\left[
    \|U-\E[U\mid O]\|_2^2
  \right]
  \ge
  c_d
  \exp\left\{
    -\frac{2}{d}I(U;O)
  \right\}.
  \label{eq:ball-shannon-bound}
\end{equation}
Moreover, \(c_d\to1\) as \(d\to\infty\).
\end{lemma}

\begin{newrevision}
\begin{proof}
If \(I(U;O)=+\infty\), the right-hand side is zero.  Suppose therefore
that \(I(U;O)<\infty\), and write
\[
  \mu(O)=\E[U\mid O],
  \qquad
  \Delta=\E\|U-\mu(O)\|_2^2.
\]
For \(t>0\), let \(Q_t\) be the joint law that keeps the marginal law of
\(O\) and assigns \(U\mid O=o\) the Gaussian distribution
\(N(\mu(o),tI_d)\).  Finite mutual information ensures that the
conditional law of \(U\) has a density for \(P_O\)-almost every \(o\).
Since the density of \(U\) is \(1/v_d\) on the ball, expanding the
relative entropy gives
\[
  0\le \KL(P_{U,O}\|Q_t)
  =
  I(U;O)-\log v_d+\frac d2\log(2\pi t)+\frac{\Delta}{2t}.
\]
Here the last term is the expected squared distance from the Gaussian
mean, divided by \(2t\).  Choose
\[
  t=\frac{v_d^{2/d}}{2\pi e}
     \exp\!\left\{-\frac{2}{d}I(U;O)\right\}.
\]
The three terms preceding \(\Delta/(2t)\) then sum to \(-d/2\), so
\(\Delta\ge dt\).  This is Equation~\eqref{eq:ball-shannon-bound}.
Finally, the ball-volume formula and Stirling's formula give
\[
  c_d=\frac{d}{2e}\Gamma(d/2+1)^{-2/d}\longrightarrow1
\]
as \(d\to\infty\).
\end{proof}
\end{newrevision}

\begin{newrevision}
\subsection{Basis Reconstruction and Rescaling}
\label{sec:basis-reconstruction}

A full-read decoder answers every query from the same stored message.
Its answers to the coordinate queries therefore define a vector
reconstruction.  The scalar error bound controls the total squared error
of this vector in expectation.  If the decoder is linear in the query,
its answer at \(y=x\) also controls the expected inner product of the
reconstruction error with the source.  Rescaling the vector uses this
information to obtain the sharp asymptotic lower bound.

\begin{proposition}[Basis Reconstruction and Rescaling]
\label{prop:basis-reconstruction}
Let \(d\ge1\) and \(L\ge0\) be integers.  Consider an \(L\)-bit full-read
protocol with worst-case scalar MSE \(D<\infty\), and set \(V=dD\).
With \(c_d\) as in Lemma~\ref{lem:ball-shannon}, every such protocol
satisfies
\[
  c_d\,2^{-2L/d}\le V.
\]
If the decoder is linear in the query for each fixed message and
realization of the protocol seeds, then
\[
  c_d\,2^{-2L/d}
  \le
  \frac{V}{1+V}+\frac{2V}{(1+V)^2}\sqrt D.
\]
Consequently, any sequence of protocols with linear decoders and
\(L_d/d\to R>0\) satisfies
\[
  \liminf_{d\to\infty}dD_d\ge\frac{1}{4^R-1}.
\]
\end{proposition}

\begin{proof}
We first form a vector reconstruction from the scalar answers.
Let \(\Omega\) denote the tuple of all protocol seeds, \(M\) the stored
message, and \(\widehat s(M,y;\Omega)\) the scalar answer.  Using the same
message and seeds for all coordinate queries, define
\[
  \widehat X(M,\Omega)
  =
  \sum_{i=1}^d\widehat s(M,e_i;\Omega)e_i,
\]
where \(e_1,\ldots,e_d\) are the standard basis vectors.
For each fixed source \(x\in\Bdd\), the scalar risk guarantee gives
\[
  \E\|\widehat X-x\|_2^2
  =
  \sum_{i=1}^d
  \E\bigl(\widehat s(M,e_i;\Omega)-x_i\bigr)^2
  \le dD=V.
\]
Now take \(U\) uniform on \(\Bdd\), independently of \(\Omega\), and let
\(M\) be the message produced from \(U\).  The seed independence and
the \(L\)-bit message budget imply
\[
  I(U;M,\Omega)
  =
  I(U;M\mid\Omega)
  \le H(M\mid\Omega)
  \le L\log2.
\]
The conditional mean minimizes squared error among reconstructions
from \((M,\Omega)\).  In particular, writing
\(\mu=\E[U\mid M,\Omega]\), every square-integrable reconstruction
\(a=a(M,\Omega)\) satisfies
\[
  \E\|U-a\|_2^2
  =
  \E\|U-\mu\|_2^2+\E\|\mu-a\|_2^2
  \ge c_d\,2^{-2L/d}.
\]
The equality follows because the cross term has conditional mean zero;
the inequality follows from Lemma~\ref{lem:ball-shannon} and the
information bound above.  Applying this to \(a=\widehat X\) proves the
first claim.

We now assume that the decoder is linear in the query and use this to
sharpen the reconstruction bound.  In this case, every answer is an
inner product against the reconstructed vector:
\[
  \widehat s(M,y;\Omega)=\widehat X(M,\Omega)^\top y.
\]
For each fixed \(x\in\Bdd\), the query \(y=x\) is admissible.
Its scalar error is \(\langle\widehat X-x,x\rangle\), so the risk
guarantee and Cauchy--Schwarz give
\[
  \E\langle\widehat X-x,x\rangle^2\le D,
  \qquad
  \bigl|\E\langle\widehat X-x,x\rangle\bigr|\le\sqrt D.
\]
This is the cross term that appears when rescaling the reconstruction.
For \(0\le\alpha\le1\), expanding the error of the rescaled vector yields
\begin{align*}
  \E\|\alpha\widehat X-x\|_2^2
  &=
  \alpha^2\E\|\widehat X-x\|_2^2
  +(1-\alpha)^2\|x\|_2^2\\
  &\quad
  -2\alpha(1-\alpha)\E\langle\widehat X-x,x\rangle\\
  &\le
  \alpha^2V+(1-\alpha)^2
  +2\alpha(1-\alpha)\sqrt D.
\end{align*}
The first two terms are minimized at \(\alpha=(1+V)^{-1}\).
With this choice, the bound becomes
\[
  \E\|\alpha\widehat X-x\|_2^2
  \le
  \frac{V}{1+V}+\frac{2V}{(1+V)^2}\sqrt D.
\]
This holds for every \(x\in\Bdd\), hence also after averaging over \(U\).
Applying the reconstruction lower bound to \(a=\alpha\widehat X\)
proves the second claim.

We now derive the asymptotic bound.
Write \(v=\liminf_{d\to\infty}dD_d\).  If \(v=+\infty\), the result
follows directly.  Otherwise, take a subsequence along which
\(V_d=dD_d\to v<\infty\).  Then \(D_d\to0\), \(c_d\to1\), and
\(2^{-2L_d/d}\to4^{-R}\).  The second bound therefore gives
\[
  4^{-R}\le\frac{v}{1+v},
  \qquad\text{hence}\qquad
  v\ge\frac{4^{-R}}{1-4^{-R}}=\frac{1}{4^R-1}.
\]
\end{proof}

For a nonlinear decoder, the basis answers still define a vector, but
its inner product with \(x\) need not equal the decoder's answer at
\(y=x\).  The rescaling step uses this equality to control the bias.
The general scalar converse below obtains the same lower bound through
an auxiliary Gaussian observation.
\end{newrevision}

\begin{newrevision}
\subsection{Gaussian-Amplitude Information}

The general converse uses a Gaussian query with a random amplitude.
We first bound the average Fisher information about this amplitude in
terms of the information available about the source.  We then relate
this Fisher information to the error of an arbitrary scalar decoder.
Both estimates hold in finite dimension, so we can choose the amplitude
variance as a function of dimension when proving the asymptotic bound.

\begin{proposition}[Gaussian-Averaged Information Bound]
\label{prop:gaussian-averaged-information}
Let \(d\ge1\), \(Q\ge1\), and \(A>0\).  Let \(X\) be uniform on
\(S^{d-1}\), and let \(S\) be side information satisfying
\(I(X;S)\le\log Q\).  Independently of \((X,S)\), draw
\(\Theta\sim N(0,A^2)\) and \(G\sim N(0,I_d)\), independently of each
other.  Set \(Y=\Theta X+G\).  For fixed \(\theta\), let
\(p_\theta(y\mid S)\) be the conditional density of
\(Y_\theta=\theta X+G\), and define
\[
  s_\theta(S,Y_\theta)
  =\left.\partial_\theta\log p_\theta(y\mid S)\right|_{y=Y_\theta},
  \qquad
  J_d(\theta)=\E_\theta s_\theta^2,
  \qquad
  \overline J_{d,A}=\E J_d(\Theta).
\]
Here \(\E_\theta\) denotes expectation at fixed amplitude \(\theta\).
With \(c_d\) from Lemma~\ref{lem:ball-shannon}, write
\[
  a_d=\frac d{d+2},
  \qquad
  \beta_d=\frac d{d+1}.
\]
Then
\begin{equation*}
  0\le\overline J_{d,A}
  \le
  \min\left\{
    1,\,
    \frac{a_d-c_dQ^{-2/d}e^{-2A^2/d}}{\beta_d^2}
  \right\}.
\end{equation*}
\end{proposition}

\begin{proof}
In this proof, we first relate the amplitude score to the conditional
uncertainty about \(X\), then apply the Shannon bound to that uncertainty.
We begin with the conditional mean and covariance.  At fixed \(S=s\)
and \(Y_\theta=y\), put
\[
  \mu=\E[X\mid S=s,Y_\theta=y],
  \qquad
  C=\operatorname{Cov}(X\mid S=s,Y_\theta=y).
\]
The conditional law of \(X\) is proportional to
\(e^{\theta x^\top y}P_{X\mid S}(dx\mid s)\).  Indeed, the factor
\(e^{-\theta^2\|x\|_2^2/2}=e^{-\theta^2/2}\) in the Gaussian
likelihood cancels in the normalization because \(\|X\|_2=1\).
Differentiating the normalized mean with respect to \(y\) and
\(\theta\), respectively, gives
\begin{equation*}
  \nabla_y\mu=\theta C,
  \qquad
  \partial_\theta\mu=Cy,
  \qquad
  \operatorname{tr}C=1-\|\mu\|_2^2.
\end{equation*}
In the amplitude derivative, \(s,y\) are held fixed.  Differentiating
the mixture density also gives
\[
  s_\theta=y^\top\mu-\theta
  =\E[X^\top G\mid S,Y_\theta=y].
\]
Thus, \(\E_\theta s_\theta=0\), and conditional Jensen's inequality
yields \(J_d(\theta)\le\E(X^\top G)^2=1\).

We next relate this information to the conditional covariance.
Write \(Z_\theta=X^\top Y_\theta\) and
\(\widehat Z_\theta=Y_\theta^\top\mu
=\E[Z_\theta\mid S,Y_\theta]\).
Gaussian integration by parts, conditional on \(X,S\), gives
\begin{align*}
  \E_\theta\widehat Z_\theta^2
  &=\E_\theta[Z_\theta\widehat Z_\theta]\\
  &=\theta\E_\theta\widehat Z_\theta
    +\E_\theta[(X^\top G)\widehat Z_\theta]\\
  &=\theta^2
    +\E_\theta[X^\top\nabla_y\widehat Z_\theta]\\
  &=\theta^2+\E_\theta\|\mu\|_2^2
    +\theta\E_\theta[\mu^\top CY_\theta].
\end{align*}
The last line uses
\(\nabla_y\widehat Z_\theta=\mu+\theta CY_\theta\) and conditions
\(X\) on \(S,Y_\theta\).  Since
\(\E_\theta\widehat Z_\theta=\E_\theta Z_\theta=\theta\), it follows
that
\begin{equation}
  J_d(\theta)
  =\E_\theta\|\mu\|_2^2
   +\theta\E_\theta[\mu^\top CY_\theta].
  \label{eq:gaussian-score-first-derivative}
\end{equation}
For \(F=\|\mu\|_2^2\), a second application of the same integration by
parts identity gives
\begin{equation}
  \begin{aligned}
    \E_\theta[s_\theta F]
    &=\E_\theta[(Z_\theta-\theta)F]
      =\E_\theta[(X^\top G)F]\\
    &=\E_\theta[X^\top\nabla_yF]
      =2\theta\E_\theta[\mu^\top C\mu].
  \end{aligned}
  \label{eq:gaussian-score-posterior-norm}
\end{equation}

We now average over the Gaussian amplitude.  In the following
expectations, \(\mu,C,F\) are evaluated at \((\Theta,S,Y)\), and
\(\lambda=A^{-2}\).  The logarithmic derivative in \(\theta\) of the
joint density of \((\Theta,S,Y)\), relative to \(d\theta\,P_S(ds)\,dy\),
is \(s_\theta-\lambda\theta\).  Integrating the derivative of
\(\theta F\) against this density, and using
\(\partial_\theta F=2\mu^\top Cy\), yields
\[
  \E F+2\E[\Theta\mu^\top CY]
  =\lambda\E[\Theta^2F]-\E[\Theta s_\Theta F].
\]
Substituting Equations~\eqref{eq:gaussian-score-first-derivative}
and~\eqref{eq:gaussian-score-posterior-norm} gives
\begin{align}
  \overline J_{d,A}
  &=\frac12\E[(1+\lambda\Theta^2)\|\mu\|_2^2]
    -\E[\Theta^2\mu^\top C\mu]\notag\\
  &=1-\frac12\E[(1+\lambda\Theta^2)\operatorname{tr}C]
    -\E[\Theta^2\mu^\top C\mu]\notag\\
  &\le1-\frac12\E[(1+\lambda\Theta^2)\operatorname{tr}C].
  \label{eq:gaussian-information-uncertainty}
\end{align}
Here \(\lambda\E\Theta^2=1\), and the final inequality uses
\(C\succeq0\).

The differentiations and integrations above are justified by the bounded
support of \(X\).  The conditional mean is smooth, with
\(\|\mu\|_2\le1\) and \(\|C\|_{\mathrm{op}}\le1\); the displayed
derivatives and integrands have at most polynomial growth in
\(\theta,y\).  All required moments are finite under the Gaussian
experiment.  For the amplitude integration, after integrating over
\(S,Y_\theta\), the absolute boundary term is at most
\(|\theta|\pi_A(\theta)\), where \(\pi_A\) is the density of
\(N(0,A^2)\).  This tends to zero at both ends of the real line.

It remains to lower-bound the weighted conditional variance in
Equation~\eqref{eq:gaussian-information-uncertainty}.
For fixed \(\theta\), write
\(m_d(\theta)=\E_\theta\operatorname{tr}C\).
Let \(\rho\) be independent of \((X,S,\Theta,G)\), with density
\(d\rho^{d-1}\) on \([0,1]\).  Then \(U=\rho X\) is uniform on
\(\Bdd\), and
\(\E\rho^2=a_d\), \(\E\rho=\beta_d\).
Independence of \(\rho\) gives
\(\E[U\mid S,Y_\theta]=\beta_d\mu\), and hence
\begin{equation}
  \E\|U-\E[U\mid S,Y_\theta]\|_2^2
  =a_d-\beta_d^2\bigl(1-m_d(\theta)\bigr).
  \label{eq:gaussian-radius-reconstruction}
\end{equation}

The Gaussian observation adds at most \(\theta^2/2\) nats about
\(X\).  To see this, compare its conditional law with \(N(0,I_d)\):
\begin{align*}
  \frac{\theta^2}{2}
  &=\E\KL\!\left(N(\theta X,I_d)\,\middle\|\,N(0,I_d)\right)\\
  &=I(X;Y_\theta\mid S)
    +\E\KL\!\left(P_{Y_\theta\mid S}\,\middle\|\,N(0,I_d)\right)\\
  &\ge I(X;Y_\theta\mid S).
\end{align*}
The Markov chain \(U-X-(S,Y_\theta)\) therefore implies
\[
  I(U;S,Y_\theta)
  \le I(X;S,Y_\theta)
  \le\log Q+\frac{\theta^2}{2}.
\]
Applying Lemma~\ref{lem:ball-shannon} to
Equation~\eqref{eq:gaussian-radius-reconstruction} yields
\begin{equation}
  m_d(\theta)
  \ge
  1-\frac{a_d-c_dQ^{-2/d}e^{-\theta^2/d}}{\beta_d^2}.
  \label{eq:gaussian-spherical-uncertainty}
\end{equation}

To average this bound, consider the probability measure
\[
  \nu_A(d\theta)
  =\frac{1+\theta^2/A^2}{2}\,\pi_A(\theta)\,d\theta.
\]
It has total mass one and second moment
\[
  \int\theta^2\,\nu_A(d\theta)
  =\frac12\left(A^2+\frac{3A^4}{A^2}\right)=2A^2.
\]
Since \(u\mapsto e^{-u/d}\) is convex, Jensen's inequality gives
\(\int e^{-\theta^2/d}\,\nu_A(d\theta)\ge e^{-2A^2/d}\).
Averaging Equation~\eqref{eq:gaussian-spherical-uncertainty} under
\(\nu_A\) and substituting into
Equation~\eqref{eq:gaussian-information-uncertainty} now gives
\[
  \overline J_{d,A}
  \le1-\int m_d(\theta)\,\nu_A(d\theta)
  \le\frac{a_d-c_dQ^{-2/d}e^{-2A^2/d}}{\beta_d^2}.
\]
Together with \(0\le J_d(\theta)\le1\), this proves the proposition.
\end{proof}

\subsection{Scalar Risk under a Gaussian Amplitude}

We now turn the information bound into a scalar error bound.
The decoder observes \(S,Y\), while the amplitude \(\Theta\) remains
unobserved.  Conditioning on \(X,S,Y\) gives an explicit Gaussian law
for \(\Theta\), whose residual supplies the needed Cauchy--Schwarz
inequality.

\begin{lemma}[Gaussian-Amplitude Scalar Risk]
\label{lem:gaussian-amplitude-risk}
In the experiment of
Proposition~\ref{prop:gaussian-averaged-information}, let
\(Z=X^\top Y\), \(\lambda=A^{-2}\), and let \(g=g(S,Y)\) be any
measurable estimator with \(\delta=\E(g-Z)^2<\infty\).  Then
\begin{equation}
  \delta\ge
  \frac{(1+\lambda)(1-\overline J_{d,A})}
       {\overline J_{d,A}+\lambda}.
  \label{eq:gaussian-amplitude-scalar-risk}
\end{equation}
\end{lemma}

\begin{proof}
Write \(b=1+\lambda\).  Conditional on \(X,S,Y\), the density of
\(\Theta\) is proportional to
\[
  \exp\!\left\{-\frac{\lambda\theta^2}{2}
                  -\frac{\|Y-\theta X\|_2^2}{2}\right\}
  \propto
  \exp\!\left\{-\frac b2(\theta-Z/b)^2\right\}.
\]
Here the conditional law of \(S\) given \(X\) does not depend on
\(\theta\), and \(\|X\|_2=1\).  Thus,
\(\Theta\mid X,S,Y\sim N(Z/b,1/b)\).
For the residual \(\eta=Z-b\Theta\), this gives
\begin{equation}
  \E[\eta\mid X,S,Y]=0,
  \qquad
  \E[\eta^2\mid X,S,Y]=b,
  \qquad
  \E[\Theta\eta\mid X,S,Y]=-1.
  \label{eq:gaussian-amplitude-residual}
\end{equation}

The conditional mean of this residual given \(\Theta,S,Y\) is
\[
  B:=\E[\eta\mid\Theta,S,Y]
    =s_\Theta-\lambda\Theta.
\]
Since \(\E[s_\Theta\mid\Theta]=0\), its second moment is
\[
  \E B^2=\overline J_{d,A}+\lambda^2\E\Theta^2
        =\overline J_{d,A}+\lambda.
\]
Also, \(\E Z^2=1+A^2\), so finite \(\delta\) implies that \(g\) is
square-integrable.  We can therefore condition first on
\(\Theta,S,Y\), then on \(X,S,Y\), to obtain
\[
  \E[(g-b\Theta)B]
  =\E[(g-b\Theta)\eta]
  =b.
\]
The last equality follows from
\(\E[g\eta]=0\) and \(\E[\Theta\eta]=-1\), using
Equation~\eqref{eq:gaussian-amplitude-residual}.
Likewise, since both \(g\) and \(Z\) are determined by \(X,S,Y\),
the cross term below vanishes:
\[
  \E(g-b\Theta)^2
  =\E\bigl((g-Z)+\eta\bigr)^2
  =\delta+b.
\]
Cauchy--Schwarz now gives
\[
  b^2\le(\delta+b)(\overline J_{d,A}+\lambda).
\]
Dividing by the positive quantity
\(\overline J_{d,A}+\lambda\) and subtracting \(b\) proves
Equation~\eqref{eq:gaussian-amplitude-scalar-risk}.
\end{proof}

\subsection{Full-Read Minimax Limit}
\label{sec:full-read-proof}

\begin{proof}[Proof of Theorem~\ref{thm:full-read}]
In this proof, we establish the converse;
Section~\ref{sec:cited-full-read-upper} gives the upper bound.
Fix an \(L\)-bit full-read protocol in dimension \(d\)
with finite worst-case scalar MSE \(D\).  Let \(\Omega\) contain all
protocol seeds, draw \(X\) uniformly on \(S^{d-1}\) independently of
these seeds, and let \(M\) be the message produced from \(X\).
With \(S=(M,\Omega)\), seed independence and the message budget give
\[
  I(X;S)=I(X;M\mid\Omega)\le H(M\mid\Omega)\le L\log2.
\]
Thus, Proposition~\ref{prop:gaussian-averaged-information} applies with
\(Q=2^L\).

Draw \(\Theta\sim N(0,A^2)\) and \(G\sim N(0,I_d)\) independently
of each other and of \((X,\Omega)\), and set \(Y=\Theta X+G\).
From the protocol's scalar answer \(\widehat s(S,y)\), form
\[
  g(S,Y)=\|Y\|_2\,
     \widehat s\!\left(S,\frac{Y}{\|Y\|_2}\right),
\]
with value zero at \(Y=0\).  This estimator uses only \(S,Y\).
Conditional on \(X,Y,\Theta\), the protocol seeds retain their
original law.  The source and normalized query are then fixed admissible
inputs, so the worst-case risk guarantee implies
\[
  \E\!\left[(g-X^\top Y)^2\mid X,Y,\Theta\right]
  \le\|Y\|_2^2D.
\]
Averaging gives
\begin{equation}
  \delta:=\E(g-X^\top Y)^2\le(d+A^2)D.
  \label{eq:gaussian-query-risk-transfer}
\end{equation}

Choose \(A^2=\sqrt d\), so \(\lambda=d^{-1/2}\), and define
\[
  K_{d,L}
  =\min\left\{
    1,\,
    \frac{a_d-c_d2^{-2L/d}e^{-2/\sqrt d}}{\beta_d^2}
  \right\}.
\]
Proposition~\ref{prop:gaussian-averaged-information} gives
\(\overline J_{d,A}\le K_{d,L}\).
Combining Lemma~\ref{lem:gaussian-amplitude-risk} with
Equation~\eqref{eq:gaussian-query-risk-transfer} yields
\[
  d(1+\lambda)D
  \ge\delta
  \ge(1+\lambda)
      \frac{1-\overline J_{d,A}}{\overline J_{d,A}+\lambda}.
\]
The chosen variance makes the factor \(1+\lambda\) the same on both
sides.  Canceling it and using that
\(j\mapsto(1-j)/(j+\lambda)\) is decreasing for \(j\ge0\), we obtain
\begin{equation}
  dD\ge\frac{1-K_{d,L}}{K_{d,L}+d^{-1/2}}.
  \label{eq:finite-full-read-converse}
\end{equation}
The denominator is positive: Lemma~\ref{lem:ball-shannon} with a
constant observation gives \(c_d\le a_d\), so \(K_{d,L}\ge0\).

The right-hand side depends only on \(d,L\).  Taking the infimum over
protocols therefore proves the same lower bound for
\(d\D_d^{\mathrm{full}}(L)\).  If \(L_d/d\to R>0\), then
\(a_d,\beta_d,c_d\to1\) and \(K_{d,L_d}\to1-4^{-R}\).
Equation~\eqref{eq:finite-full-read-converse} consequently gives
\[
  \liminf_{d\to\infty}d\D_d^{\mathrm{full}}(L_d)
  \ge\frac{4^{-R}}{1-4^{-R}}=\frac1{4^R-1}.
\]
Together with the cited upper bound, this proves the theorem.
\end{proof}
\end{newrevision}
 
\section{Shared Analysis and Query-Dependent Constructions}
\label{app:dither}

\begin{revision}
The two constructions share the record, Gaussian conditional model,
pilot sign identities, and source-norm encoding.  The proofs use these
shared tools as follows.

For the direct construction, the pilot places \(\rho\) in an interval
of radius \(1/(t+1)\), except with probability exponentially small
in the pilot size \(n_d\).
Density cancellation makes the accepted design uniform on a rectangle,
and a binomial bound guarantees enough accepted cells.  On the pilot
interval the response derivative is at least \(t/8\).
Consequently, inversion multiplies the Bernoulli mean's MSE by at most
\((8/t)^2\), giving
\[
  \E_\rho(\widehat\rho_d-\rho)^2
  \le\frac{16}{m_dt^2}
       +8e^{-n_d/[2\pi^2(t+1)^2]}+4e^{-m_d/12}
\]
for all sufficiently large \(d\).
The same window supplies
a lower bound on the information of a selector using the true parameter.

For the sharp refinement, we first find the information available to a
selector that knows the true correlation, in the scalar experiment with
public state \((B,\Theta)\).
Conditional score increments are orthogonal, so the best budgeted sum
is the sum of the largest cell informations.  The trimmed-sum limit
defines \(\J_{W,R}(\rho)\), the information per dimension, and the
Gaussian tail calculation proves
\[
  \inf_{-1<\rho<1}\J_{W,R}(\rho)
  \sim\frac4\pi R\log(W/R),
  \qquad W/R\to\infty.
\]
This is an oracle comparison within the scalar experiment used by the
construction; the unrestricted converse is proved separately.

Next, we show that the protocol asymptotically achieves the oracle risk
without knowing \(\rho\).
On every fixed closed interval inside \((-1,1)\), and hence away from
both endpoints, the pilot has second and fourth error moments of orders
\(n_d^{-1}\) and \(n_d^{-2}\).  Constants in these bounds may depend on
the interval.
Bounds on the derivatives show that ranking cells at the pilot estimate
loses only \(o(d)\) information.  We identify a region where every cell
contributes a positive amount of information.  Counting the cells in
this region keeps the selected denominator \(D_d\) above
\(\gamma_{W,R}d\), except with probability \(o(d^{-1})\).
Holding the selected set fixed, expansion of the score sum \(U_d\)
gives
\[
  \widehat\rho_d^{\mathrm{raw}}-\rho
  =\frac{U_d(\rho)}{D_d}+o_{L_2}(d^{-1/2}),
  \qquad
  d\,\E_\rho\!\left[\frac{U_d(\rho)^2}{D_d^2}\right]
  \longrightarrow\frac1{\J_{W,R}(\rho)}
\]
on the good event, with ratios set to zero elsewhere.
The moment bounds use \(n_d/d\to0\) and \(d/n_d^2\to0\);
our choice \(n_d=\lceil d^{3/4}\rceil\) satisfies both.

Finally, we control the risk uniformly near the endpoints.
The mismatch probability is of order
\(\sqrt{1-|\rho|}\).  Small mismatch counts justify returning the nearby
endpoint; otherwise, rescaled derivative envelopes control the score
update.  These two cases give \(o(d^{-1})\) risk along every sequence
\(\rho_d\to\pm1\).  Combining the endpoint and interior results identifies
the worst-case limit at an interior minimizer of \(\J_{W,R}\).
The shared norm grid then changes root MSE by at most \(1/(2d)\),
which preserves both directional constants on the unit ball.
\end{revision}

\subsection{Unit-Ball Implementation and Finite-Dimensional Conventions}
\label{sec:unit-ball-protocol}

The following norm representation and budget conventions are shared by
Protocols~\ref{prot:window-reader} and~\ref{prot:dithered-reader}.
They extend either directional estimator to \(x,y\in\Bdd\).
Under the exact storage and read budgets
\(w_d=\lfloor Wd\rfloor\) and \(r_d=\lfloor Rd\rfloor\), set
\[
  b_d=\lceil\log_2(d+1)\rceil,
  \qquad
  N_d=w_d-b_d,
  \qquad
  k_d=r_d-b_d.
\]
The first \(b_d\) cells store the source norm, and measurement \(i\) is
stored at address \(b_d+i\).  For all sufficiently large \(d\), the
remaining budgets are positive and satisfy \(N_d/d\to W\) and
\(k_d/d\to R\).

Use a directional procedure only when \(d\ge2\) and \(n_d<k_d\).
Apply any additional fallback condition in its specification.
In every other dimension, store an all-zero record,
make no probes, and return zero.
\textcolor{black}{This convention also covers dimensions in which storing
and reading the source norm would exceed the budget.}  For directional inputs,
take \(b_d=0\), use the given nonnegative budgets \(N_d,k_d\), and apply
the same convention.  Thus, both protocols are defined at every finite
budget; the fallback is unused for all sufficiently large \(d\).

Given \(x\in\Bdd\), put \(a=\|x\|_2\).  Set \(u=x/a\) when \(a>0\), and
use a fixed unit vector when \(a=0\).  Round \(a\) to the nearest point in
\[
  \{0,1/d,\ldots,1\},
\]
breaking a tie toward the smaller grid point.  Store its index in the
reserved cells and denote the reconstructed value by \(\widehat a_d\).
The grid has \(d+1\) points, so its index fits in \(b_d\) bits, and
\[
  |\widehat a_d-a|\le\frac1{2d}.
\]
The remaining \(N_d\) cells contain the directional measurements
\(Z_i=\sgn(G_i^\top u-\Theta_i)\).

Given \(y\), set \(v=y/\|y\|_2\) when \(y\ne0\), and use a fixed unit
vector otherwise.  The first batch reads the norm cells and the first
\(n_d\) measurement cells.  Run the directional pilot and selection
rules on the measurement cells, then return
\[
  \widehat s=\widehat a_d\,\|y\|_2\,\widehat\rho_d.
\]
This formula returns zero when either input is zero.  In the directional
protocol, \(a=\widehat a_d=1\) and \(b_d=0\).

When the directional procedure is used, the unit-ball protocol stores
\(b_d+N_d=w_d\) bits and uses at most
\(b_d+k_d=r_d\) probes.  Its fixed batch boundaries are
\[
  t_1=b_d+n_d,\qquad t_2=b_d+k_d=r_d.
\]
The directional boundaries are \(t_1=n_d\), \(t_2=k_d\).
If a protocol returns without informative second-batch probes, fill
that batch with ignored probes to unused measurement addresses.  There are enough such addresses
because \(k_d\le N_d\).  All second-batch addresses are fixed before their
values are observed, as required by
Definition~\ref{def:access-classes}.

For either directional estimate clipped to \([-1,1]\), write
\(a=\|x\|_2\) and \(\rho=u^\top v\).  The error decomposes as
\[
  \widehat s-x^\top y
  =\|y\|_2\{a(\widehat\rho_d-\rho)
                   +(\widehat a_d-a)\widehat\rho_d\}.
\]
The triangle inequality in \(L_2\) therefore gives the uniform bound
\begin{equation}
  \{\E(\widehat s-x^\top y)^2\}^{1/2}
  \le \{\E(\widehat\rho_d-\rho)^2\}^{1/2}+\frac1{2d}.
  \label{eq:common-norm-transfer}
\end{equation}

\subsection{Shared Gaussian Model and Pilot Estimates}
\label{sec:shared-dither-preliminaries}

Both protocols use the same Gaussian-dither record.
For fixed unit directions \(u,v\), put \(\rho=u^\top v\).
In each row, write \(A=G^\top u\), \(B=G^\top v\), and
\(Z=\sgn(A-\Theta)\).  The joint Gaussian law gives
\begin{equation}
  A=\rho B+\sqrt{1-\rho^2}\,E,
  \qquad B,E,\Theta\ \text{independent standard normals}.
  \label{eq:shared-gaussian-model}
\end{equation}
\begin{revision}
At \(\rho=\pm1\), the residual term is zero.  Throughout this analysis,
conditioning on the scalar design \((B_i,\Theta_i)\) leaves the
independent Gaussian residuals in
Equation~\eqref{eq:shared-gaussian-model}.
Both readers use only the scalar design, their pilot, and independent
coins to select nonpilot cells.  The nonpilot residuals therefore retain
their independent standard normal laws after selection.
\end{revision}

\begin{lemma}[Shared Pilot Identities and Localization]
\label{lem:shared-pilot}
Let \(M_d^\pm\) count pilot mismatches between \(Z_i\) and
\(\sgn(\pm B_i-\Theta_i)\), respectively.  Define
\begin{equation}
  q_+(\rho)
  =\frac1\pi\arccos\left(\frac{1+\rho}{2}\right),
  \qquad
  q_-(\rho)
  =\frac1\pi\arccos\left(\frac{1-\rho}{2}\right).
  \label{eq:mismatch-probabilities}
\end{equation}
Then \(M_d^\pm\sim\operatorname{Bin}(n_d,q_\pm(\rho))\), and
\begin{equation}
  \rho
  =2\cos(\pi q_+(\rho))-1
  =1-2\cos(\pi q_-(\rho)).
  \label{eq:mismatch-inversion}
\end{equation}
The plus estimate in Equation~\eqref{eq:window-pilot} satisfies, for every
\(\eta>0\), uniformly over \(\rho\in[-1,1]\),
\begin{equation}
  \Prb_\rho\{|\widetilde\rho_d-\rho|>\eta\}
  \le2e^{-n_d\eta^2/(2\pi^2)}.
  \label{eq:shared-pilot-localization}
\end{equation}
The analogous bound holds for the minus inverse projected onto
\([-1,1]\).
\end{lemma}

\begin{proof}
The pairs
\[
  (A-\Theta,B-\Theta)
  \quad\text{and}\quad
  (A-\Theta,-B-\Theta)
\]
are centered Gaussian with correlations
\[
  \frac{1+\rho}{2},
  \qquad
  \frac{1-\rho}{2}.
\]
The Gaussian sign-mismatch identity, equivalently the random-hyperplane
angle formula \cite{charikar-hyperplane}, gives
Equation~\eqref{eq:mismatch-probabilities} through Equation~\eqref{eq:mismatch-inversion}.
For completeness, two standardized Gaussian linear forms with
correlation \(c\) can be represented by unit vectors at angle
\(\arccos c\) in a two-dimensional isotropic Gaussian plane.
Their signs differ on sectors of total angle \(2\arccos c\).
The polar angle is uniform, so the mismatch probability is
\(\arccos(c)/\pi\); the degenerate case follows by continuity.
Both inverse maps are \(2\pi\)-Lipschitz.  Projection onto \([-1,1]\)
cannot increase their distance from \(\rho\).  Hoeffding's inequality for
the empirical mismatch frequency therefore gives
Equation~\eqref{eq:shared-pilot-localization}.
\end{proof}

\subsection{Informative Windows and Their Mean Response}

\begin{revision}
The acceptance rule gives the selected projections and thresholds a
simple conditional distribution.  We also use this distribution to
lower-bound the oracle information available to the refined protocol.
\end{revision}

\begin{lemma}[Window Selection and Availability]
\label{lem:window-selection}
Fix \(t\ge2\) and \(s\in[-1,1]\).
For independent standard normals \(B,\Theta\), the rule
in Equation~\eqref{eq:window-acceptance} is a valid acceptance probability.
Its mean is
\[
  \alpha(t)=4\phi(t+1)\phi(t+3)
           =\frac2\pi e^{-(t+2)^2-1}.
\]
Conditional on acceptance,
\begin{equation}
  B\sim\operatorname{Unif}[t,t+1],\qquad
  U:=\Theta-sB\sim\operatorname{Unif}[-2,2],
  \qquad B\perp U.
  \label{eq:accepted-window-law}
\end{equation}
For the threshold in Equation~\eqref{eq:window-threshold}, conditional on any pilot,
the number of accepted nonpilot cells is
\(\operatorname{Bin}(N_d-n_d,2/\Lambda)\).
With \(m_d=k_d-n_d\), for all sufficiently large \(d\),
\begin{equation*}
  \Prb\{\text{fewer than }m_d\text{ accepted cells}\mid\text{pilot}\}
  \le e^{-m_d/12}.
\end{equation*}
Conditional on the pilot and selection success, the first \(m_d\)
accepted scalar pairs \((B_i,\Theta_i-sB_i)\) are iid with
the law in Equation~\eqref{eq:accepted-window-law}, independently of their residuals
in Equation~\eqref{eq:shared-gaussian-model}.
\end{lemma}

\begin{proof}
On the acceptance window, \(0<B\le t+1\) and
\(|\Theta|\le |s|B+2\le t+3\).  Thus, the density ratio in
Equation~\eqref{eq:window-acceptance} is at most one.
Multiplying it by the original density \(\phi(b)\phi(c)\) leaves the
constant \(\phi(t+1)\phi(t+3)\) on
\[
  \{t\le b\le t+1,\ |c-sb|\le2\}.
\]
The change of variables \((b,c)\mapsto(b,c-sb)\) has determinant one.
The transformed rectangle has area four, proving the acceptance
probability and Equation~\eqref{eq:accepted-window-law}.
Substitution of Equation~\eqref{eq:window-threshold} gives
\(\alpha(t)=2/\Lambda\).

Conditional on the pilot, the nonpilot rows and acceptance coins are
independent.  If \(\mu_d=(N_d-n_d)2/\Lambda\), then
\(\mu_d/m_d\to2\).  For all sufficiently large \(d\),
\(\mu_d\ge3m_d/2\), so a binomial lower-tail bound yields
\[
  \Prb\{\operatorname{Bin}(N_d-n_d,2/\Lambda)<m_d\}
  \le e^{-\mu_d/18}\le e^{-m_d/12}.
\]
\begin{revision}
To check that conditioning on success preserves this law, fix the pilot
and the entire acceptance-indicator vector.  The rows are independent,
so the accepted scalar pairs are independent, each with the law in
Equation~\eqref{eq:accepted-window-law}.  This product law is the same
for every indicator vector having at least \(m_d\) accepted addresses,
once its first \(m_d\) accepted addresses are listed in order.
Averaging over these indicator vectors leaves the product law unchanged.
Conditional on the pilot, acceptance depends on the scalar pair and an
independent coin, so it also preserves the independent residual laws.
\end{revision}
\end{proof}

\begin{lemma}[Uniform Window Response]
\label{lem:window-response}
For every \(t\ge2\) and \(s\in[-1,1]\), the response in
Equation~\eqref{eq:window-response} is continuous on \(K_s\), with
\begin{equation}
  H_s'(\theta)\ge t/8,\qquad \theta\in K_s\cap(-1,1).
  \label{eq:window-response-slope}
\end{equation}
Consequently, its inverse on \(H_s(K_s)\) is \(8/t\)-Lipschitz,
including at the endpoints.
\end{lemma}

\begin{proof}
For fixed \(b\in[t,t+1]\), put \(x=(\theta-s)b\) and
\(\sigma=\sqrt{1-\theta^2}\).
On \(K_s\), \(|x|\le1\).  Write the inner response as
\[
  F(x,\sigma)=\frac14\int_{-2}^{2}
       \Phi\left(\frac{x-u}{\sigma}\right)\,du.
\]
\begin{revision}
For \(\sigma>0\), the integrand is smooth on the compact integration
interval, so differentiation under this integral is valid.  Substituting
\(z=(x-u)/\sigma\) gives
\end{revision}
\[
  4F_x
  =\Phi\left(\frac{x+2}{\sigma}\right)
   -\Phi\left(\frac{x-2}{\sigma}\right),
  \qquad
  4F_\sigma
  =\phi\left(\frac{x+2}{\sigma}\right)
   -\phi\left(\frac{x-2}{\sigma}\right).
\]
Since \(x'=b\) and \(\sigma'=-\theta/\sigma\), the derivative with respect
to \(\theta\) is \(bF_x-\theta F_\sigma/\sigma\).
The inequalities \(x+2\ge1\), \(x-2\le-1\), and \(\sigma\le1\) give
\(4F_x\ge2\Phi(1)-1\).
Both densities in \(4F_\sigma\) lie between zero and
\(\phi(1/\sigma)\), so
\[
  \frac{|\theta F_\sigma|}{\sigma}
  \le\frac{\phi(1/\sigma)}{4\sigma}
  \le\frac{\phi(1)}4.
\]
The last step uses that \(z\phi(z)\) is decreasing for \(z\ge1\).
As \(b\ge2\), the derivative divided by \(b\) is at least
\[
  c_0=\frac{2\Phi(1)-1}{4}-\frac{\phi(1)}8>\frac18.
\]
For the last inequality, concavity of \(\phi\) on \([0,1]\) gives
\(2\Phi(1)-1\ge\phi(0)+\phi(1)\), and
\(2\phi(0)+\phi(1)>1\).
\begin{revision}
Averaging over \(b\) proves Equation~\eqref{eq:window-response-slope}.
For endpoint continuity, represent the response as
\[
  H_s(\theta)
  =\Prb\{U\le(\theta-s)B+\sqrt{1-\theta^2}\,E\},
\]
where \(B\) and \(U\) have the independent uniform laws in
Equation~\eqref{eq:accepted-window-law}, and \(E\) is an independent
standard normal.  Under convergent \(s,\theta\), the comparison value
converges almost surely and equals \(U\) with probability zero at its
limit.  Bounded convergence proves joint continuity, including
\(\theta=\pm1\).  Integrating the derivative bound between two interior
parameters and taking limits at any endpoint gives
\[
  H_s(\theta_2)-H_s(\theta_1)
  \ge(t/8)(\theta_2-\theta_1),
  \qquad \theta_1\le\theta_2,\quad \theta_1,\theta_2\in K_s.
\]
This proves strict monotonicity and the stated inverse Lipschitz bound
on the entire closed interval.
\end{revision}
\end{proof}

\subsection{Direct Uniform Risk Bound}
\label{sec:window-risk-proof}

\begin{proof}[Proof of Theorem~\ref{thm:dither-reader}]
\begin{revision}
All selection probabilities depend measurably on the pilot and public
design.  The response is jointly continuous in its center and parameter,
including at the endpoints, and the endpoints of \(K_s\) are continuous
in \(s\).  Its strictly monotone inverse and the projection in the
decoder are therefore measurable.  The address-order rule and independent
acceptance coins also define measurable choices, so the protocol is
measurable.
\end{revision}
The budget conventions in Appendix~\ref{sec:unit-ball-protocol}
give admissibility in every dimension and two fixed batches for all
sufficiently large \(d\).

First, consider unit directions.
Let \(\mathsf G_d=\{\rho\in K_{\widetilde\rho_d}\}\), and let
\(\mathsf S_d\) be the event that selection succeeds.
Lemma~\ref{lem:shared-pilot} gives
\[
  \Prb_\rho(\mathsf G_d^c)
  \le 2e^{-n_d/[2\pi^2(t+1)^2]}.
\]
By Lemma~\ref{lem:window-selection}, conditional on the pilot and
\(\mathsf S_d\), the selected bits \(Y_i=(Z_i+1)/2\) are iid Bernoulli
with mean \(H_{\widetilde\rho_d}(\rho)\), after averaging over their
accepted designs.  Their empirical mean has variance at most
\(1/(4m_d)\).
On \(\mathsf G_d\cap\mathsf S_d\), projection and
Lemma~\ref{lem:window-response} yield
\[
  |\widehat\rho_d-\rho|
  \le\frac8t\,|\overline Y_d-H_{\widetilde\rho_d}(\rho)|.
\]
Both the true and returned values lie in \([-1,1]\), so the squared
error on either exceptional event is at most four.
Thus, uniformly in \(\rho\),
\begin{equation*}
  \E_\rho(\widehat\rho_d-\rho)^2
  \le
  \frac{16}{m_dt^2}
  +8e^{-n_d/[2\pi^2(t+1)^2]}
  +4e^{-m_d/12}.
\end{equation*}
At fixed rates, \(t\) is fixed, \(m_d/d\to R\), and both exceptional
terms are \(o(d^{-1})\).
This proves the directional upper bound \(16/(Rt^2)\).
This argument only needs \(n_d/\log d\to\infty\) and \(n_d/d\to0\).
The common choice \(n_d=\lceil d^{3/4}\rceil\) also supports the
one-step analysis of the refined protocol.

For general \(x,y\in\Bdd\), the common norm transfer
in Equation~\eqref{eq:common-norm-transfer} gives the same normalized limit superior.
The minimax risk is at most this protocol's worst-case risk, proving
Equation~\eqref{eq:window-upper-bound}.
Finally, \(t_\Lambda^2\sim\log\Lambda\) proves
Equation~\eqref{eq:window-sparse-risk}.
\end{proof}

\subsection{Information Supplied by the Shared Window}
\label{sec:window-information}

\begin{revision}
The same window gives a lower bound on the Fisher information available
to the refined protocol.  The following quantities are needed only for
that refinement.
\end{revision}

\begin{revision}
Retain the scalar design and correlation notation from
Equation~\eqref{eq:shared-gaussian-model}.  Conditional on
\((B,\Theta)=(b,c)\),
\end{revision}
\[
  A-\Theta
  \sim
  N(\rho b-c,1-\rho^2).
\]
For \(\rho\in(-1,1)\), standardize the difference between the conditional
mean and the threshold:
\begin{equation}
  \alpha_\rho(b,c)=\frac{\rho b-c}{\sqrt{1-\rho^2}},
  \label{eq:standardized-probit-argument}
\end{equation}
Then
\[
  \Prb_\rho\{Z=z\mid B=b,\Theta=c\}
  =
  \Phi(z\alpha_\rho(b,c)).
\]
This is a \emph{probit model}: a binary observation whose success
probability is given by the standard normal distribution function.

Differentiation at fixed \((b,c)\) gives
\[
  \partial_\rho\alpha_\rho(b,c)
  =
  \frac{b-\rho c}{(1-\rho^2)^{3/2}},
\]
so the conditional score is
\begin{equation*}
  s_\rho(z;b,c)
  =
  \frac{
    z(b-\rho c)\phi(\alpha_\rho(b,c))
  }{
    (1-\rho^2)^{3/2}
    \Phi(z\alpha_\rho(b,c))
  }.
\end{equation*}
Put
\begin{equation}
  g(t)=\frac{\phi(t)^2}{\Phi(t)\Phi(-t)}.
  \label{eq:probit-envelope}
\end{equation}
Averaging the score's square over \(z\in\{-1,1\}\) gives the
conditional Fisher information in one stored bit:
\begin{equation}
  \iota_\rho(b,c)
  =
  \frac{(b-\rho c)^2}{(1-\rho^2)^3}\,
  g\!\left(\alpha_\rho(b,c)\right).
  \label{eq:probit-information}
\end{equation}
When the arguments are omitted, \(\iota_\rho\) means the random
variable \(\iota_\rho(B,\Theta)\).

For a nonnegative random variable \(V\) and \(p\in[0,1]\), define
\[
  \T_p(V)
  =
  \sup\{
    \E[aV]:0\le a\le1,\ \E a=p
  \}.
\]
Here, \(\T_p(V)\) is the total expected contribution of the largest \(p\)
fraction of \(V\).  An optimal rule selects values above a threshold,
using fractional selection at the threshold if it has positive
probability.
\begin{revision}
The optimal value depends only on the distribution of \(V\):
replacing any selector \(a\) by \(\E[a\mid V]\) preserves both its mass
and its expected contribution.  Put
\end{revision}
\begin{equation}
  \J_{W,R}(\rho)
  =
  W\,\T_{R/W}(\iota_\rho),
  \qquad
  \J_{W,R}^{\min}
  =
  \inf_{\rho\in(-1,1)}
  \J_{W,R}(\rho).
  \label{eq:oracle-information}
\end{equation}
Thus, \(\J_{W,R}(\rho)\) is the oracle information per dimension at
correlation \(\rho\), and \(\J_{W,R}^{\min}\) is its value at the
least informative correlation.

\begin{proposition}[Information Lower Bound from the Shared Window]
\label{prop:window-information}
Fix \(0<R<W\) with \(t=t_\Lambda\ge2\) as in
Equation~\eqref{eq:window-threshold}.  The information functional satisfies
\begin{equation}
  \J_{W,R}^{\min}\ge\frac4{25}Rt_\Lambda^2,
  \qquad
  \frac1{\J_{W,R}^{\min}}
  \le\frac{25}{4Rt_\Lambda^2}
  =\frac{25/4+o(1)}{R\log\Lambda}.
  \label{eq:window-information-bound}
\end{equation}
\textcolor{black}{The \(o(1)\) term tends to zero as \(\Lambda=W/R\to\infty\).}
\end{proposition}

\begin{proof}
Fix \(\rho\in(-1,1)\).  Center the window at \(s=\rho\), holding this
center fixed when differentiating the statistical parameter.
Lemma~\ref{lem:window-selection} gives acceptance probability \(2/\Lambda\).
Independently retain half the accepted cells, obtaining a selector of
mass \(p=R/W\) with the same accepted design law.

The averaged Bernoulli response at the reference parameter is
\[
  H_\rho(\rho)=\frac12,\qquad
  H_\rho'(\rho)
  =\frac{\E B}{4}
    \left\{2\Phi\left(\frac2{\sqrt{1-\rho^2}}\right)-1\right\}
  \ge\frac t5.
\]
This follows from the derivative calculation in
Lemma~\ref{lem:window-response}.  For the last inequality,
\(\E B\ge t\), and a standard normal scalar \(G\) satisfies
\(\Prb\{|G|\le2\}\ge13/16>4/5\), by
\(\E G^4=3\) and Markov's inequality.

\begin{revision}
Write \(p_\theta(b,c)=\Phi(\alpha_\theta(b,c))\).
At the fixed interior reference parameter, the accepted design has
compact support and the likelihood is differentiable on a neighborhood
of that parameter.  We may therefore differentiate its averaged
response under the integral.  Cauchy--Schwarz applied to
\(\partial_\theta p_\theta/
\sqrt{p_\rho(1-p_\rho)}\) and
\(\sqrt{p_\rho(1-p_\rho)}\), under this fixed accepted design law, gives
\end{revision}
\[
  \E[\iota_\rho\mid\mathrm{accepted}]
  \ge
  \frac{
    \{\E[\left.\partial_\theta p_\theta\right|_{\theta=\rho}
          \mid\mathrm{accepted}]\}^2
  }{
    \E[p_\rho(1-p_\rho)\mid\mathrm{accepted}]
  }
  \ge4H_\rho'(\rho)^2
  \ge\frac{4t^2}{25}.
\]
The top-\(p\) information functional is at least the information of this
particular randomized selector.  Hence,
\(\T_p(\iota_\rho)\ge4pt^2/25\), uniformly in \(\rho\).
Multiplying by \(W\) and taking the infimum proves
Equation~\eqref{eq:window-information-bound}.
\end{proof}

\begin{revision}
The window thus gives logarithmic growth of the available information.
The analysis below shows how the score decoder converts this information
into a risk bound.
\end{revision}
 
\subsection{Information-Based Protocol and Sharp Guarantee}
\label{sec:sharp-protocol}

Protocol~\ref{prot:dithered-reader} refines
Protocol~\ref{prot:window-reader} through three replacements:
an endpoint-aware pilot, information-based selection, and a score
correction.  The endpoint checks make the sharp risk bound uniform up to
correlations \(\pm1\).

\begin{mdframed}[style=protocolshade]
\begin{protocol}[Information-Based Refinement of Protocol~\ref{prot:window-reader}]
\label{prot:dithered-reader}
\textcolor{black}{Fix arbitrary \(0<R<W\).}
Retain the sign convention, budget definitions
\(b_d,N_d,k_d,n_d,m_d\), input normalization, query projections \(B_i\),
pilot addresses, \textcolor{black}{bits reserved for the source norm}, and
padding rule of Protocol~\ref{prot:window-reader}.
Set
\[
  h_d=\lceil96\log d\rceil,
  \qquad
  \epsilon_d=(h_d/n_d)^2.
\]
If \(d<2\), \(m_d\le0\), or \(h_d\ge n_d/4\), the encoder stores an
all-zero record, the reader makes no probes, and the decoder returns
\(\widehat s=0\).  Otherwise, retain Step 1 and replace Steps 2--4 of
Protocol~\ref{prot:window-reader} as follows.

\textbf{\textcolor{black}{Encoder: Source Record.}}
Use the source record from Protocol~\ref{prot:window-reader}, including
the same Gaussian measurements and source-norm encoding.

\begin{enumerate}[leftmargin=*,start=2]
\item \textbf{Reader: Endpoint-Aware Pilot.}
Compute the same query projections and read the same first batch as in
Protocol~\ref{prot:window-reader}, including any reserved norm cells.
From its \(n_d\) directional measurement bits, compute the mismatch frequency
\(\widehat q_d\) in Equation~\eqref{eq:window-pilot} and form
\begin{equation*}
  M_d^+
  =n_d\widehat q_d,
  \qquad
  M_d^-
  =
  \sum_{i=1}^{n_d}
  \mathbf1\{Z_i\ne\sgn(-B_i-\Theta_i)\}.
\end{equation*}
If the smaller count is below \(h_d\), make no informative second-stage
probes.  The unused budget may be padded with ignored probes.  Otherwise, form
\begin{equation}
\begin{aligned}
  \widetilde\rho_d^+
  &=
  \proj_{[-1+\epsilon_d,1-\epsilon_d]}
  \left\{
    2\cos\left(\pi M_d^+/n_d\right)-1
  \right\},\\
  \widetilde\rho_d^-
  &=
  \proj_{[-1+\epsilon_d,1-\epsilon_d]}
  \left\{
    1-2\cos\left(\pi M_d^-/n_d\right)
  \right\},
\end{aligned}
  \label{eq:pilot-inverse}
\end{equation}
and let \(\widetilde\rho_d\) be the candidate associated with the smaller
mismatch count, choosing \(\widetilde\rho_d^+\) when the counts tie.

\item \textbf{Reader: Fisher-Information Selection.}
If the endpoint test did not stop the reader, consider the nonpilot
addresses.  Replace window acceptance by probing the \(m_d\) cells with
the largest values of
\(\iota_{\widetilde\rho_d}(B_i,\Theta_i)\).  Break ties by increasing
address, and denote the selected set by \(\mathcal I_d\).

\item \textbf{Decoder: One-Step Score Correction.}
Replace mean-response inversion by the following score update.
The decoder reconstructs the reader's choices from the public scalar
pairs \((B_i,\Theta_i)\) and returned transcript; it makes no probes.
If the pilot endpoint test stopped the reader, it sets
\(\widehat\rho_d=+1\) when
\(M_d^+\le M_d^-\), and \(\widehat\rho_d=-1\) otherwise.  In the remaining
case, use the score function
\begin{equation*}
  s_\rho(z;b,c)
  =
  \frac{
    z(b-\rho c)\phi(\alpha_\rho(b,c))
  }{
    (1-\rho^2)^{3/2}
    \Phi(z\alpha_\rho(b,c))
  }.
\end{equation*}
The decoder computes
\[
  D_d
  =
  \sum_{i\in\mathcal I_d}
  \iota_{\widetilde\rho_d}(B_i,\Theta_i),
  \qquad
  U_d
  =
  \sum_{i\in\mathcal I_d}
  s_{\widetilde\rho_d}
  (Z_i;B_i,\Theta_i).
\]
Here, \(U_d\) is the log-likelihood slope of the selected bits at the
pilot estimate, and \(D_d\) is their total Fisher information.
The decoder makes one correction to the pilot:
\begin{equation}
  \widehat\rho_d
  =
  \proj_{[-1,1]}
  \left(
    \widetilde\rho_d+
    \frac{U_d}{D_d}
  \right)
  \label{eq:scoring-update}
\end{equation}
when \(D_d\ge\gamma_{W,R}d\), where \(\gamma_{W,R}>0\) is the explicit
constant in Equation~\eqref{eq:denominator-cutoff}.  Otherwise, it sets
\(\widehat\rho_d=\widetilde\rho_d\).
\end{enumerate}

\textbf{\textcolor{black}{Decoder: Final Output.}}
Reconstruct \(\widehat a_d\) from the norm cells read in the first batch,
or take \(\widehat a_d=1\) for unit directions, and return
\[
  \widehat s=\widehat a_d\,\|y\|_2\,\widehat\rho_d.
\]
Use this rule for every branch of Step 4.
For unit directions, \(\widehat s=\widehat\rho_d\); if either input is
zero, the formula returns zero.
\end{protocol}
\end{mdframed}

The cutoff keeps the score correction stable when the selected bits
carry little information.
Let \(g\) be as in Equation~\eqref{eq:probit-envelope}, and
define
\[
  q_0
  =
  \Prb\{
    1\le|B|\le2,\ |\Theta|\le1/4
  \},
\qquad
  c_{\mathrm{int}}
  =
  \frac14g(\sqrt7),
\]
\[
  a_0=2\phi(2)\Prb\{|T|\le1\},
\qquad
  c_{\mathrm{end}}
  =
  \frac18g(2\sqrt2),
\]
\begin{equation}
  \gamma_{W,R}
  =
  \min\left\{
    c_{\mathrm{int}}
    \min\left(\frac{Wq_0}{4},\frac R2\right),
    c_{\mathrm{end}}
    \min\left(\frac{Wa_0}{4},\frac R2\right)
  \right\}.
  \label{eq:denominator-cutoff}
\end{equation}
Here, \(B,\Theta,T\) are independent standard normals.  Numerically,
\[
  q_0=0.0536587809\ldots,
  \quad
  c_{\mathrm{int}}=0.0089390828\ldots,
\]
\[
  a_0=0.0737181310\ldots,
  \quad
  c_{\mathrm{end}}=0.0028601289\ldots.
\]
The decoder in Equation~\eqref{eq:scoring-update} uses the scoring update only when
\begin{equation}
  D_d
  \ge
  \gamma_{W,R}d.
  \label{eq:denominator-test}
\end{equation}

We count the arithmetic operations for the information-based protocol.
\textcolor{black}{Write \(N=N_d\sim Wd\) and \(k=k_d\sim Rd\) as
\(d\to\infty\) at fixed \(W,R\)}, and let \(G\in\mathbb R^{N\times d}\) have rows
\(G_i^\top\).  Count evaluations of \(\phi\) and \(\Phi\) as constant-time
scalar operations.  Once the public Gaussian design is available, encoding
one source computes \(Gu\) and thresholds its entries, which takes
\(O(Nd)\) arithmetic operations.  Processing one query first computes all
\(B_i=G_i^\top v\), also in \(O(Nd)\) time.
The pilot costs \(O(n_d)\), and evaluating all selection scores costs
\(O(N)\).  A linear-time order-statistic algorithm selects the
top-\((k-n_d)\) scores in \(O(N)\) operations.  The final score update
costs \(O(k)\).  Sorting the scores would make the selection step
\(O(N\log N)\).

With this implementation, encoding and query processing each take
\(O(Nd)=O(Wd^2)\) time, while the
reader probes only \(O(Rd)\) stored bits.  When the public design is shared
across many source records, \(B=Gv\) is computed once per query and reused.
After this shared matrix-vector product, the additional selection and
decoding work is \textcolor{black}{\(O(N)=O(Wd)\) per record, since
\(k\le N\)}.  Dense Gaussian
matrix-vector products dominate the arithmetic cost.

\begin{theorem}[Sharp Uniform Risk of the Information-Based Protocol]
\label{thm:sharp-dither-reader}
For every fixed \(0<R<W\), Protocol~\ref{prot:dithered-reader} is
admissible in every dimension and, for all sufficiently large \(d\), has
a two-batch \(\mathrm{staged}(2)\) reader.  Let
\[
  \mathcal E_d^{\mathrm{prot}}(W,R)
  =
  \sup_{x,y\in\Bdd}
  \E\!\left[
    \bigl(x^\top y-\widehat s(x,y)\bigr)^2
  \right]
\]
be its worst-case risk.  Then
\[
  \lim_{d\to\infty}
  d\,\mathcal E_d^{\mathrm{prot}}(W,R)
  =
  \frac1{\J_{W,R}^{\min}}.
\]
Consequently,
\begin{equation}
  \limsup_{d\to\infty}
  d\,\D_d^{\mathrm{ad}}(\lfloor Wd\rfloor,\lfloor Rd\rfloor)
  \le
  \frac1{\J_{W,R}^{\min}}.
  \label{eq:adaptive-upper-bound}
\end{equation}
Along arbitrary rate pairs with \(R/W\to0\),
\begin{equation}
  \J_{W,R}^{\min}
  \sim
  \frac4\pi R\log(W/R),
  \qquad
  \frac1{\J_{W,R}^{\min}}
  \sim
  \frac{\pi}{4R\log(W/R)}.
  \label{eq:sparse-oracle-information}
\end{equation}
\end{theorem}
\begin{revision}
The proof has three parts.  First,
Proposition~\ref{prop:dither-oracle} identifies the information in a
top-\(R/W\) fraction of cells and proves
\(\inf_\rho\T_p(\iota_\rho)\sim(4/\pi)p\log(1/p)\).
Second, Proposition~\ref{prop:dither-uniform} shows that selection at the
pilot estimate followed by one score correction has risk
\(1/[d\J_{W,R}(\rho)]+o(d^{-1})\) uniformly on each fixed closed interval
inside \((-1,1)\).
The same proposition proves \(o(d^{-1})\) risk along every
sequence \(|\rho_d|\to1\).  Together, the interior and endpoint estimates
establish the claimed worst-case limit.  Finally,
Section~\ref{sec:unit-ball-extension} extends the result to the unit ball
using the common norm bound.
Every dimension limit is taken with \(W,R\) fixed.  The sparse
equivalences then follow from the scalar information calculation as
\(R/W\to0\).
\end{revision}

\subsection{Oracle Information and Selection}

\begin{revision}
The randomized thresholds draw on dithered quantization
\cite{zamir-feder-dither}.
In the scalar experiment, the reader observes
\((B_i,\Theta_i)\) and selected signs.  At each correlation, we determine
the largest information available to a reader that chooses \(k\) signs
using this scalar state.  A general branching chain rule gives a
top-\(k\) information sum, and a trimmed-sum limit turns it into \(\T_{R/W}(\iota_\rho)\).
The window comparison already gives a lower bound on its growth when
the read fraction is small.
To identify the sharp constant, we calculate the information tail on
compact correlation intervals and then exclude minimizing sequences
that approach either endpoint.
\end{revision}

\begin{lemma}[Branching Information Chain Rule]
\label{lem:dither-chain-rule}
Let \(N,k\) be integers with \(0\le k\le N\), and fix
a public state \(\mathcal P\).  Conditional on \(\mathcal P\), let
\(Z_1,\ldots,Z_N\) be independent cells whose likelihoods
\(p_{\rho,i}(\cdot\mid\mathcal P)\) are differentiable and positive
on a common finite support in a neighborhood of the parameter.  Constant
cells are allowed and have zero information.  Write
\[
  s_{\rho,i}
  =
  \partial_\rho
  \log p_{\rho,i}(Z_i\mid\mathcal P),
  \qquad
  \iota_{\rho,i}
  =
  \E_\rho[s_{\rho,i}^2\mid\mathcal P].
\]
\begin{revision}
Every depth-\(k\) fully branching reader whose address rule is independent
of \(\rho\) and uses only \(\mathcal P\), earlier returned cells, and
an independent parameter-independent seed has conditional transcript
Fisher information at most the sum
\end{revision}
of the \(k\) largest values among
\(\iota_{\rho,1},\ldots,\iota_{\rho,N}\).  At a fixed reference
\(\rho_0\), a reader that freezes the top-\(k\) rule evaluated at
\(\rho_0\) attains equality at \(\rho=\rho_0\).
At each fixed correlation, the oracle value is the
largest information achieved by any of these parameter-independent rules.

\begin{revision}
For the dithered scalar experiment, take
\(\mathcal P=(B_i,\Theta_i)_{i\le N}\).  Its design law is independent
of \(\rho\), and
\end{revision}
\[
  \iota_{\rho,i}
  =
  \iota_\rho(B_i,\Theta_i),
  \qquad i=1,\ldots,N.
\]
\begin{revision}
This application concerns readers using the scalar state
\(\mathcal P\).  The general statement of the
lemma also applies to other independent finite-support cell kernels.
\end{revision}
\end{lemma}

\begin{proof}
First, fix any reader seed, and let \(I_t\) be the address chosen at
round \(t\).  The address \(I_t\) is measurable with respect to
\(\mathcal P\) and the earlier returned values.  It is therefore fixed when
the conditional likelihood at round \(t\) is differentiated.  If \(I_t\)
was probed earlier, its returned value is already determined by the
transcript.  The corresponding conditional likelihood is one and its score
increment is zero.

Let \(\mathsf T\) denote the returned-value transcript, and let
\(\mathcal N\) be the set of rounds that probe a new address.  For every
realized transcript with positive probability, conditional independence
gives the factorization
\[
  \log p_\rho(z_{1:k}\mid\mathcal P)
  =
  \sum_{t\in\mathcal N}
  \log p_{\rho,I_t}(z_t\mid\mathcal P).
\]
In this factorization, the addresses are fixed by the realized
transcript.  Differentiating the cell likelihoods gives the transcript score
\[
  S_\rho
  =
  \sum_{t\in\mathcal N}s_{\rho,I_t}.
\]
Given the preceding transcript, a newly selected cell is independent of
the previously returned cells.  Its score has conditional mean zero and
conditional variance \(\iota_{\rho,I_t}\).  The increments are thus
martingale differences, so their cross terms vanish and
\[
  J_\rho(\mathsf T\mid\mathcal P)
  =
  \E_\rho\left[
    \sum_{t\in\mathcal N}\iota_{\rho,I_t}
    \,\middle|\,\mathcal P
  \right].
\]
Every path uses at most \(k\) distinct addresses.  Its information sum is
therefore bounded by the sum of the \(k\) largest
\(\iota_{\rho,i}\).  Probing a fixed top-\(k\) set makes the score
increments independent and attains this sum.
\begin{revision}
If the reader seed is private, the transcript score is the conditional
expectation of the score obtained when the seed is also revealed, given
the transcript alone.  Jensen's inequality shows that revealing the
seed can only increase Fisher information.  Averaging the preceding
bound over the seed proves the same upper bound for randomized readers.
The maximizing top-\(k\) reader uses a fixed reference parameter to
choose its addresses; that reference is held fixed when differentiating
the likelihood.
\end{revision}
\end{proof}

The next lemma is a law of large numbers for the sum of the largest
\(k_d\) observations.  The term \emph{trimmed sum} refers to omitting
observations below the selection cutoff.

\begin{lemma}[Trimmed-Sum Limit]
\label{lem:trimmed-sum}
Let \(V_1,V_2,\ldots\) be iid nonnegative integrable random variables.
For integer budgets \(0\le k_d\le N_d\),
if \(N_d/d\to W\), \(k_d/d\to R\), and \(0<R<W\), then
\begin{equation}
  \frac1d
  \sum_{j=1}^{k_d}V_{(j;N_d)}
  \longrightarrow
  W\T_{R/W}(V)
  \label{eq:trimmed-sum-limit}
\end{equation}
in \(L_1\), where the order statistics are decreasing.
\end{lemma}

\begin{proof}
For every nonnegative vector \(v\in\mathbb R^N\)
and integer \(0\le k\le N\),
\begin{equation}
  \sum_{j=1}^k v_{(j)}
  =
  \inf_{\tau\ge0}
  \left\{
    k\tau+\sum_i(v_i-\tau)_+
  \right\}.
  \label{eq:top-sum-variational-formula}
\end{equation}
\begin{revision}
The population analogue is
\[
  \T_p(V)=\inf_{\tau\ge0}\{p\tau+\E(V-\tau)_+\},
  \qquad 0<p<1.
\]
Indeed, every selector of mass \(p\) satisfies
\(aV\le a\tau+(V-\tau)_+\), which proves one inequality.
Choose an upper \(p\)-quantile \(\tau\), so that
\(\Prb\{V>\tau\}\le p\le\Prb\{V\ge\tau\}\).
Accept all \(V>\tau\) and the fraction
\((p-\Prb\{V>\tau\})/\Prb\{V=\tau\}\) of the atom at \(\tau\),
omitting this fraction when the atom has zero mass.
The resulting selector has mass \(p\) and attains equality in the
pointwise bound.  The formula therefore includes distributions with
atoms, including an atom at zero.
\end{revision}
Write \(p_d=k_d/N_d\), so \(p_d\to p=R/W\).  First, suppose
\(0\le V\le M\).  A minimizer in Equation~\eqref{eq:top-sum-variational-formula} may then be restricted to
\([0,M]\).  The functions \(\tau\mapsto(V_i-\tau)_+\) are
one-Lipschitz.  Apply the law of large numbers on a finite mesh and then
refine the mesh.  This gives almost sure uniform convergence on \([0,M]\) of the
empirical objective divided by \(N_d\) to
\[
  p\tau+\E(V-\tau)_+,
  \qquad p=R/W.
\]
The infima are bounded by \(M\), so dominated convergence
also gives \(L_1\) convergence in the bounded case to
\[
  \inf_{\tau\ge0}
  \{p\tau+\E(V-\tau)_+\}
  =
  \T_p(V).
\]

For general integrable \(V\), put \(V^{(M)}=V\wedge M\).  The difference
between the top-\(k_d\) sums for \(V\) and \(V^{(M)}\) is at most
\[
  \sum_{i=1}^{N_d}(V_i-M)_+,
\]
whereas
\[
  0\le
  \T_p(V)-\T_p(V^{(M)})
  \le
  \E(V-M)_+.
\]
After division by \(N_d\), the expectations of both error bounds tend to
zero uniformly in \(d\) as \(M\to\infty\).  This transfers the bounded
result to integrable \(V\).  Multiplication by
\(N_d/d\to W\) proves Equation~\eqref{eq:trimmed-sum-limit} in \(L_1\).
\end{proof}

\begin{lemma}[Gaussian Probit Bounds]
\label{lem:gaussian-probit-envelope}
The function \(g\) defined in Equation~\eqref{eq:probit-envelope}
is even and strictly decreasing on \([0,\infty)\).  In particular,
\begin{equation*}
  0<g(x)\le g(0)=\frac2\pi.
\end{equation*}
Moreover,
\(\sup_x x^2g(x)\le8\) and \(\sup_x|g'(x)|\le80\).
\end{lemma}

\begin{proof}
The Gaussian log likelihood involves the logarithmic
derivative of \(\Phi\), called the inverse Mills ratio:
\[
  \lambda(x)=\frac{\phi(x)}{\Phi(x)}.
\]
The function \(g\) is even.  For \(x>0\),
\[
  (\log g)'(x)=-2x+\lambda(-x)-\lambda(x).
\]
The derivatives of \(\lambda(-x)\) and \(-\lambda(x)\) are respectively
\(1-\Var(G\mid G>x)\) and \(1-\Var(G\mid G\le x)\), for
\(G\sim N(0,1)\).  For example, integration by parts gives
\(\E[G\mid G>x]=\lambda(-x)\) and
\(\E[G^2\mid G>x]=1+x\lambda(-x)\).
Differentiating the first mean gives the stated variance identity.
Both conditional variances are positive.
Since \(\lambda(-x)-\lambda(x)=0\) at zero, integration shows that this
difference is strictly smaller than \(2x\) for \(x>0\).  Hence, \(g\)
is strictly decreasing.

We will also use a precise Gaussian-tail bound.  For \(x>0\),
integration by parts gives
\[
  \Phi(-x)=\frac{\phi(x)}x
              -\int_x^\infty\frac{\phi(t)}{t^2}\,dt.
\]
The integral is between zero and \(\Phi(-x)/x^2\), so
\begin{equation}
  \frac{x}{1+x^2}\phi(x)\le\Phi(-x)\le\frac{\phi(x)}x.
  \label{eq:gaussian-tail-bounds}
\end{equation}
\textcolor{black}{In particular,
\(0\le\phi(x)/\Phi(-x)-x\le1/x\) for \(x>0\).}

The following explicit envelopes will also be useful.  For \(x\ge0\),
\[
  \frac{\Phi(-x)}{\phi(x)}
  =\int_0^\infty e^{-xt-t^2/2}\,dt
  \ge \frac{e^{-3/2}}{1+x},
  \qquad \lambda(x)\le2\phi(x).
\]
For the integral lower bound, restrict to
\(0\le t\le(1+x)^{-1}\), where \(xt+t^2/2\le3/2\).
Thus, \(\lambda(-x)\le5(1+x)\).  The identities
\[
  \lambda'=-x\lambda-\lambda^2,
  \qquad
  \lambda''=(x^2-1)\lambda+3x\lambda^2+2\lambda^3
\]
give, for all real \(x\),
\begin{equation}
  \max\{|\lambda(x)|,|\lambda'(x)|,|\lambda''(x)|\}
  \le512(1+|x|)^3e^{-(x_+)^2/2}.
  \label{eq:mills-envelope}
\end{equation}
The polynomial bounds on the negative half-line have respective
constants \(5,30,330\).  On the positive half-line, we can use
\(2\phi(x),2(1+x)\phi(x),6(1+x)^2\phi(x)\).
Since \(g(x)=\lambda(x)\lambda(-x)\), the same estimates yield
\begin{equation}
  g(x)\le4(1+|x|)e^{-x^2/2},\qquad
  |g'(x)|\le32(1+|x|)^2e^{-x^2/2}.
  \label{eq:probit-derivative-envelope}
\end{equation}
Maximizing the elementary functions in these envelopes gives the stated
constants:
\[
  \sup_x x^2g(x)
  \le4\{2e^{-1}+3\sqrt3\,e^{-3/2}\}<8,\qquad
  \sup_x|g'(x)|\le128e^{-1/2}<80.
\]
\end{proof}

\begin{proposition}[Oracle Information and Sparse Asymptotics]
\label{prop:dither-oracle}
Let integer budgets \(0\le k_d\le N_d\) satisfy
\(N_d/d\to W\), \(k_d/d\to R\), and \(0<R<W\).
Conditional on the scalar public design, the pointwise
oracle of Lemma~\ref{lem:dither-chain-rule} has information equal to the
sum of the \(k_d\) largest cell informations.  Its normalized value
satisfies
\begin{equation}
  \frac1d\sum_{j=1}^{k_d}\iota_{\rho,(j;N_d)}
  \xrightarrow{L_1}
  \J_{W,R}(\rho)
  =
  W\T_{R/W}
  (\iota_\rho).
  \label{eq:oracle-limit}
\end{equation}
Here, the order statistics are decreasing and convergence
is over the Gaussian design.  In particular, the expected normalized
oracle information has the same limit.
This function is continuous on \((-1,1)\), diverges as
\(|\rho|\uparrow1\), and has a positive attained minimum.

As \(p\downarrow0\),
\begin{equation}
  \inf_{\rho\in(-1,1)}
  \frac{
    \T_p(\iota_\rho)
  }{
    p\log(1/p)
  }
  \longrightarrow
  \frac4\pi,
  \label{eq:sparse-oracle-minimum}
\end{equation}
and every minimizing correlation tends to zero.
\end{proposition}

\begin{proof}
\begin{revision}
For every fixed \(\rho\in(-1,1)\),
Lemma~\ref{lem:gaussian-probit-envelope} gives the integrability bound
\[
  \E\iota_\rho
  \le\frac{2\,\E(B-\rho\Theta)^2}{\pi(1-\rho^2)^3}
  =\frac{2(1+\rho^2)}{\pi(1-\rho^2)^3}<\infty.
\]
Equation~\eqref{eq:oracle-limit} now follows from
Lemmas~\ref{lem:dither-chain-rule} and~\ref{lem:trimmed-sum}.
Three claims remain.  We prove the tail asymptotics uniformly on each
compact interval inside the parameter domain, show that correlations
approaching the endpoints cannot minimize the information in the sparse limit, and establish
continuity with endpoint divergence at a fixed sampling fraction.
Together, these claims justify minimizing the asymptotic expression over
all correlations.
\end{revision}

\begin{revision}
A rotation of the Gaussian design gives bounds on the upper tail of the
single-cell information.
\end{revision}
Put \(\delta=1-\rho^2\) and make the orthogonal change
of variables
\begin{equation}
  T
  =
  \frac{B+\rho\Theta}{\sqrt{1+\rho^2}},
  \qquad
  U
  =
  \frac{\rho B-\Theta}{\sqrt{1+\rho^2}}.
  \label{eq:rotated-design}
\end{equation}
Then \(T,U\) are independent standard normals and
\begin{equation}
  \iota_\rho
  =
  \frac{
    \{\delta T+2\rho U\}^2
  }{
    (1+\rho^2)\delta^3
  }
  g\left(
    \sqrt{\frac{1+\rho^2}{\delta}}\,U
  \right).
  \label{eq:rotated-information}
\end{equation}
Write \(\iota_\rho=h_\rho(U)(T+2\rho U/\delta)^2\).  As a quadratic
in \(T\), its leading coefficient is at most
\[
  \kappa_\rho
  =
  \frac2{\pi(1-\rho^4)},
\]
with equality at \(U=0\).  The explicit envelope
\(\sup_xx^2g(x)\le8\) gives
\[
  h_\rho(U)(2\rho U/\delta)^2
  \le M_\rho:=\frac{32\rho^2}{(1+\rho^2)^2\delta^2}.
\]
Young's inequality bounds the cross term by
\(\varepsilon T^2+\kappa_\rho M_\rho/\varepsilon\).  Thus, for every
\(0<\varepsilon<\kappa_\rho\),
\begin{equation}
  \iota_\rho
  \le
  (\kappa_\rho+\varepsilon)T^2
  +
  C_{\rho,\varepsilon},
  \qquad
  C_{\rho,\varepsilon}
  =\left(1+\frac{\kappa_\rho}{\varepsilon}\right)M_\rho.
  \label{eq:information-tail-majorant}
\end{equation}
For the reverse bound, Equation~\eqref{eq:probit-derivative-envelope} gives
\(\sup|g'|\le80\).  The choice
\[
  \eta=\min\left\{1,\frac{\varepsilon\sqrt{1+\rho^2}\,
                              \delta^{3/2}}{160}\right\}
\]
ensures \(h_\rho(U)\ge\kappa_\rho-\varepsilon/2\) on \(|U|\le\eta\).
On \(|U|\le\eta\), bounding the absolute cross term by
\((\varepsilon/2)T^2+8\kappa_\rho^2\rho^2\eta^2/(\varepsilon\delta^2)\)
therefore gives
\begin{equation}
  \iota_\rho
  \ge
  (\kappa_\rho-\varepsilon)T^2
  -
  C'_{\rho,\varepsilon},
  \qquad
  C'_{\rho,\varepsilon}
  =\frac{8\kappa_\rho^2\rho^2\eta^2}{\varepsilon\delta^2}.
  \label{eq:information-tail-minorant}
\end{equation}
On a compact correlation interval, a common positive \(\eta\) and finite
upper bounds for both constants follow directly from these expressions.

For \(x>C_{\rho,\varepsilon}\), Equation~\eqref{eq:information-tail-majorant}
and Equation~\eqref{eq:information-tail-minorant} imply
\[
  \Prb\{|U|\le\eta\}
  \Prb\!\left\{T^2>\frac{x+C'_{\rho,\varepsilon}}
                         {\kappa_\rho-\varepsilon}\right\}
  \le\Prb\{\iota_\rho>x\}
  \le
  \Prb\!\left\{T^2>\frac{x-C_{\rho,\varepsilon}}
                         {\kappa_\rho+\varepsilon}\right\}.
\]
The lower bound uses the independence of \(T\) and \(U\).
Equation~\eqref{eq:gaussian-tail-bounds} gives
\(x^{-1}\log\Prb\{T^2>x\}\to-1/2\).
Letting \(\varepsilon\downarrow0\) in the two comparisons therefore gives
\begin{equation}
  \lim_{x\to\infty}
  \frac1x
  \log
  \Prb\{
    \iota_\rho>x
  \}
  =
  -\frac1{2\kappa_\rho}
  =
  -\frac{\pi(1-\rho^4)}4.
  \label{eq:information-tail-exponent}
\end{equation}
\begin{revision}
For uniformity on a compact interval \(K\subset(-1,1)\), first fix
\(0<\varepsilon<\min_{\rho\in K}\kappa_\rho\).
The displayed choice of \(\eta\) is bounded below by a positive number
on \(K\), and both additive constants are bounded above there.
The two Gaussian tail comparisons therefore hold with a common
probability prefactor and common additive bounds.  The coefficients
\(\kappa_\rho\pm\varepsilon\) also stay in a compact subset of
\((0,\infty)\), so the Gaussian logarithmic tail limit is uniform
over these coefficients.  Taking \(x\to\infty\), followed by
\(\varepsilon\downarrow0\), proves the asserted compact-uniform
version of Equation~\eqref{eq:information-tail-exponent}.
\end{revision}

\begin{revision}
These tail bounds determine the expected contribution of the largest
fraction of information values.
\end{revision}
If a nonnegative random variable \(V\) satisfies
\[
  x^{-1}\log\Prb\{V>x\}\to-\beta,
  \qquad 0<\beta<\infty,
\]
then
\begin{equation}
  \T_p(V)
  \sim
  \frac{p\log(1/p)}{\beta}.
  \label{eq:exponential-tail-trimming}
\end{equation}
Indeed, fix \(\varepsilon\in(0,1)\) and write
\[
  a_-=\frac{1-\varepsilon}{\beta}\log(1/p),
  \qquad
  a_+=\frac{1+\varepsilon}{\beta}\log(1/p).
\]
The tail assumption implies
\(\Prb\{V>a_-\}>p\) for all sufficiently small \(p\), so an upper-tail
selector of mass \(p\) gives
\(\T_p(V)\ge pa_-\).  For the reverse bound, every selector \(0\le a\le1\)
with \(\E a=p\) satisfies
\[
  \E[aV]
  \le
  pa_+
  +
  \E[V\mathbf1_{\{V>a_+\}}].
\]
Choose a tail exponent
\(\beta'<\beta\) sufficiently close to \(\beta\) that
\((1+\varepsilon)\beta'/\beta>1\).  For all large \(x\),
\(\Prb\{V>x\}\le e^{-\beta'x}\), and tail integration shows that the
last expectation is
\[
  O\!\left(
    \log(1/p)\,
    p^{(1+\varepsilon)\beta'/\beta}
  \right)
  =
  o(p\log(1/p)).
\]
Letting \(\varepsilon\downarrow0\) proves Equation~\eqref{eq:exponential-tail-trimming}.
Applying Equation~\eqref{eq:exponential-tail-trimming} to Equation~\eqref{eq:information-tail-exponent}, locally uniformly in \(\rho\),
\begin{equation}
  \frac{
    \T_p(\iota_\rho)
  }{
    p\log(1/p)
  }
  \longrightarrow
  \frac4{\pi(1-\rho^4)}.
  \label{eq:compact-sparse-information}
\end{equation}

\begin{revision}
The compact-interval bounds leave open the possibility that minimizing
correlations approach an endpoint.  To rule this out, let
\end{revision}
\(p_j\downarrow0\), \(|\rho_j|\uparrow1\), and
\(\delta_j=1-\rho_j^2\).  Using the coordinates in Equation~\eqref{eq:rotated-design}, put
\[
  H_j=\{|U_j|\le\delta_j/8\},
  \qquad
  \frac{\phi(1/8)}4\delta_j
  \le q_j:=\Prb(H_j)
  \le\frac{\phi(0)}4\delta_j.
\]
On \(H_j\cap\{|T_j|\ge1\}\), the numerator in
Equation~\eqref{eq:rotated-information} has absolute value at least
\(3\delta_j|T_j|/4\), and the argument of \(g\) has absolute value at
most \(1/8\).  Hence,
\[
  \iota_{\rho_j}
  \ge
  c_\star\frac{T_j^2}{\delta_j},
  \qquad c_\star=\frac9{32}g(1/8)>0.
\]
Let \(r_0=\Prb\{1\le|T|\le2\}>0\).  If \(q_jr_0\le p_j\), include
\(H_j\cap\{1\le|T_j|\le2\}\) in the selector and fill the remaining mass.
Then
\begin{equation}
  \frac{
    \T_{p_j}
    (\iota_{\rho_j})
  }{p_j}
  \ge
  \frac{c_\star r_0\phi(1/8)}{4p_j}.
  \label{eq:sparse-endpoint-small-band}
\end{equation}
If \(q_jr_0>p_j\), choose \(t_j>1\) so that
\[
  q_j\Prb\{|T|\ge t_j\}=p_j.
\]
Selecting \(H_j\cap\{|T_j|\ge t_j\}\) gives
\begin{equation}
  \frac{
    \T_{p_j}
    (\iota_{\rho_j})
  }{p_j}
  \ge
  c_\star\frac{t_j^2}{\delta_j}.
  \label{eq:sparse-endpoint-large-band}
\end{equation}
If \(p_j/q_j\) is bounded away from zero, the right side of
Equation~\eqref{eq:sparse-endpoint-large-band}, divided by \(\log(1/p_j)\), diverges because
\(\delta_j=O(p_j)\).  If \(p_j/q_j\to0\), then
\[
  t_j^2\sim2\log(q_j/p_j),
\]
and
\[
  \frac{
    \log(q_j/p_j)
  }{
    \delta_j\log(1/p_j)
  }
  \to\infty.
\]
Indeed, put \(a_j=\log(q_j/p_j)\to\infty\), so
\(\log(1/p_j)=\log(1/q_j)+a_j\).  If
\(a_j\ge\log(1/q_j)\), the displayed ratio is at least
\((2\delta_j)^{-1}\).  Otherwise, it is at least
\[
  \frac{a_j}{
    2\delta_j\log(1/q_j)
  },
\]
which also diverges because
\(\delta_j\asymp q_j\), \(q_j\log(1/q_j)\to0\), and
\(a_j\to\infty\).
Together with Equation~\eqref{eq:sparse-endpoint-small-band}, this proves the required
divergence near the endpoints.  These alternatives cover arbitrary sequences: every subsequence
of \(p_j/q_j\) has a further subsequence bounded away from zero or tending
to zero.

For fixed \(p\), use the transition band
\[
  \mathcal B_\rho
  =\{|T|\le1,\ \sqrt\delta\le|U|\le2\sqrt\delta\}.
\]
On this band, \(\alpha_\rho\) remains bounded, so both
signs have probabilities bounded away from zero.
The band narrows near an endpoint, while its per-cell information grows.
For \(|\rho|\ge3/4\), its probability and information satisfy
\[
  2\phi(2)\Prb\{|T|\le1\}\sqrt\delta
  \le\Prb(\mathcal B_\rho)
  \le2\phi(0)\sqrt\delta,
  \qquad
  \iota_\rho\mathbf1_{\mathcal B_\rho}
  \ge\frac{g(2\sqrt2)}{8\delta^2}
        \mathbf1_{\mathcal B_\rho}.
\]
Indeed, the Gaussian density on the band is at least \(\phi(2)\);
the argument of \(g\) is at most \(2\sqrt2\); and
\[
  |2\rho U+\delta T|
  \ge\sqrt\delta(2|\rho|-\sqrt\delta)
  \ge\sqrt\delta/2.
\]
The band has mass below \(p\) for all sufficiently small \(\delta\),
so an upper-\(p\) selector can include it.  Its contribution is at least
\begin{revision}
\[
  \frac{\phi(2)\Prb\{|T|\le1\}g(2\sqrt2)}4\,\delta^{-3/2}.
\]
This contribution diverges as \(\delta\downarrow0\), proving the
claim for fixed \(p\).
Lemma~\ref{lem:dither-denominator} uses the same band to bound the
selected information at finite sample sizes.
\end{revision}

\begin{revision}
The interior and endpoint bounds together determine the minimum
information.
On a compact \(\rho\)-interval, \(1-\rho^2\) is bounded away from zero,
and \(g\le2/\pi\) bounds every \(\iota_\rho\) by a fixed multiple of
\((|B|+|\Theta|)^2\).  Pointwise continuity and dominated convergence
therefore give
\(\iota_{\rho_n}\to\iota_\rho\) in \(L_1\) whenever
\(\rho_n\to\rho\).  Since
\end{revision}
\[
  |\T_p(V)-\T_p(W)|
  \le
  \E|V-W|,
\]
\begin{revision}
the oracle functional is continuous there.  It is positive at every
interior parameter because \(g>0\) and \(B-\rho\Theta\ne0\) almost
surely; for example, the constant selector \(a=p\) gives
\(\T_p(\iota_\rho)\ge p\E\iota_\rho>0\).
Endpoint divergence confines every minimizing sequence to a compact
interior interval.  Continuity on that interval then gives a positive
attained minimum.
\end{revision}

At \(\rho=0\), Equation~\eqref{eq:compact-sparse-information} gives the upper bound \(4/\pi\) in
Equation~\eqref{eq:sparse-oracle-minimum}.  For the reverse bound, take minimizing correlations along an
arbitrary sequence \(p_j\downarrow0\).  The endpoint bounds keep them in a
compact subset of \((-1,1)\).  Passing to a limit \(\rho\) and using the
locally uniform form of Equation~\eqref{eq:compact-sparse-information},
\[
  \liminf_j
  \frac{
    \T_{p_j}
    (\iota_{\rho_j})
  }{
    p_j\log(1/p_j)
  }
  \ge
  \frac4{\pi(1-\rho^4)}
  \ge
  \frac4\pi.
\]
Equality forces \(\rho=0\), proving both claims.
\end{proof}

\subsection{Pilot Estimation and Uniform Efficiency}

Here, statistical efficiency means that the leading normalized MSE
equals the reciprocal of the oracle information per dimension.
The oracle ranks cells at the true correlation; our reader ranks them
at the pilot \(\widetilde\rho_d\).
To prove the same limiting risk, we need to control the effect of this
replacement on both the selected information and the score correction.

\textcolor{black}{For correlations in a fixed closed interval inside
\((-1,1)\), put}
\(\Delta_d=\widetilde\rho_d-\rho\).
Pilot concentration gives
\(\E\Delta_d^2=O(n_d^{-1})\) and
\(\E\Delta_d^4=O(n_d^{-2})\).
\begin{revision}
On the event that the pilot lies in a fixed larger interval inside
\((-1,1)\), the difference between the best information sum at \(\rho\)
and the sum selected at the pilot is at most
\end{revision}
\[
  2|\Delta_d|\sum_{i>n_d}
      \sup_\theta|\partial_\theta\iota_\theta(B_i,\Theta_i)|.
\]
\textcolor{black}{The supremum is over that larger interval.}
A common integrable bound on the derivatives makes this loss, divided
by \(d\), tend to zero.
The trimmed-sum limit therefore gives
\(D_d/d\to\J_{W,R}(\rho)\).
Conditional on the pilot and scalar design, the selected scores at the
true parameter are independent and centered.  Taylor expansion of the
score correction gives a leading term \(U_d(\rho)/D_d\), with
conditional variance \(I_d/D_d^2\), where \(I_d\) is the information in
the selected cells at \(\rho\).  The three remainder terms have total
squared contribution
\begin{revision}
\[
  d\,\E[\text{remainder}^2]
  =O(d/n_d^2)=o(1),
\]
using \(n_d=o(d)\) and \(d/n_d^2\to0\).
\end{revision}
A lower cutoff for \(D_d\) and a fourth-moment bound justify taking
expectations and clipping.  These steps give the exact interior risk
\(1/[d\J_{W,R}(\rho)]+o(d^{-1})\).

\textcolor{black}{For interior correlations approaching an endpoint, let
\(\sigma=\sqrt{1-\rho^2}\to0\) and consider the branch that proceeds
to the second batch.}
\begin{revision}
Let \(E\) be the good pilot event used in the endpoint proof: the
near mismatch frequency differs from its true probability by at most
half that probability, the near count is at least \(h_d\), and the
far count is at least \(n_d/4\).
On \(E\), the near candidate is selected and its projection is inactive.
A transition band has probability of order \(\sigma\), and each cell
there has information of order \(\sigma^{-4}\).
The band-count event, intersected with \(E\), gives
\(D_d\ge c\,d\sigma^{-3}\).
The full-population moment bound gives
\(\E[U_d(\rho)^2]\le C\,d\sigma^{-3}\), with explicit constants below.
Thus, the leading term contributes \(O(\sigma^3)\) to scaled risk on
this intersection.
The inverse mismatch map improves the restricted pilot moments to
\[
  \E[\Delta_d^2\mathbf1_E]=O(\sigma^3/n_d),
  \qquad
  \E[\Delta_d^4\mathbf1_E]=O(\sigma^6/n_d^2).
\]
The endpoint derivative envelopes then bound the scaled remainder by
\(O(d\sigma^2/n_d^2)=o(1)\), again using \(n_d=o(d)\).
\end{revision}
When there are too few pilot mismatches for this calculation, the
protocol returns the corresponding endpoint; binomial concentration
controls both its error and the probability of returning the wrong one.
This covers every sequence approaching either endpoint.
The denominator cutoff is specified beside the complete protocol in
Appendix~\ref{sec:sharp-protocol}.

\begin{lemma}[Endpoint Pilot Moments]
\label{lem:dither-pilot}
\begin{revision}
Let \(d\ge2\), and use the mismatch probabilities and identities in
\end{revision}
Lemma~\ref{lem:shared-pilot}.
Near \(\rho=1\), assume \(q=q_+(\rho)\le1/32\).  On
\[
  E_q
  =
  \left\{
    \left|
      \frac{M_d^+}{n_d}-q
    \right|
    \le\frac q2
  \right\},
\]
if \(n_dq\ge h_d/4\), then
\begin{equation}
  \E[
    |\widetilde\rho_d^+-\rho|^2
    \mathbf1_{E_q\cap\{M_d^+\ge h_d\}}
  ]
  \le
  9\pi^4\frac{q^3}{n_d},
  \label{eq:endpoint-pilot-second-moment}
\end{equation}
\begin{equation}
  \E[
    |\widetilde\rho_d^+-\rho|^4
    \mathbf1_{E_q\cap\{M_d^+\ge h_d\}}
  ]
  \le
  324\pi^8\frac{q^6}{n_d^2},
  \label{eq:endpoint-pilot-fourth-moment}
\end{equation}
and
\begin{equation}
  \Prb(E_q^c)
  \le
  2e^{-n_dq/12}.
  \label{eq:relative-pilot-tail}
\end{equation}
If \(n_dq<h_d/4\), then
\begin{equation}
  \Prb\{M_d^+\ge h_d\}
  \le
  (e/4)^{h_d}.
  \label{eq:small-mean-pilot-tail}
\end{equation}
The analogous statements hold near \(\rho=-1\).
\end{lemma}

\begin{proof}
The binomial marginals and inverse maps follow from
Lemma~\ref{lem:shared-pilot}.
The inverse map \(f(t)=2\cos(\pi t)-1\) satisfies
\[
  |f(\widehat q)-f(q)|
  \le
  3\pi^2q|\widehat q-q|
\]
on \(|\widehat q-q|\le q/2\).  A binomial variable has variance at most
\(n_dq\) and fourth centered moment at most
\(3(n_dq)^2+n_dq\).  Consequently, the second moment is at most
\(9\pi^4q^3/n_d\).  The fourth moment is at most
\[
  (3\pi^2)^4\frac{q^6}{n_d^2}
  \left(3+\frac1{n_dq}\right)
  \le324\pi^8\frac{q^6}{n_d^2},
\]
because \(n_dq\ge h_d/4\ge1\).  This proves
Equation~\eqref{eq:endpoint-pilot-second-moment} through Equation~\eqref{eq:endpoint-pilot-fourth-moment}.  On
\(E_q\cap\{M_d^+\ge h_d\}\), the upper-end projection in Equation~\eqref{eq:pilot-inverse} is
inactive because
\[
  1-\{2\cos(\pi M_d^+/n_d)-1\}
  \ge
  4(h_d/n_d)^2.
\]
Since \(q\le1/32\), the event \(E_q\) also keeps
\(M_d^+/n_d\le3/64<1/2\), so the lower-end projection is inactive.
The multiplicative Chernoff bound gives Equation~\eqref{eq:relative-pilot-tail}.  If the mean is below
\(h_d/4\), the standard binomial upper-tail bound
\(\Prb\{X\ge h\}\le(e\mu/h)^h\) gives Equation~\eqref{eq:small-mean-pilot-tail}.
\end{proof}

\begin{lemma}[Uniform Denominator Event]
\label{lem:dither-denominator}
Condition on any pilot value
\(\theta=\widetilde\rho_d\in[-1+\epsilon_d,1-\epsilon_d]\).  For all
sufficiently large \(d\), define
\begin{equation*}
  p_{\mathrm{den},d}
  =e^{-Wq_0d/16}
   +e^{-Wa_0dh_d/(16n_d)}.
\end{equation*}
With conditional probability at least \(1-p_{\mathrm{den},d}\), uniformly
over all such pilot values,
\begin{equation}
  D_d
  \ge
  \gamma_{W,R}d.
  \label{eq:denominator-event}
\end{equation}
For every sequence of pilot values with \(|\theta| \to1\), if
\(a_\theta=\sqrt{1-\theta^2}\), then on the same event
\begin{equation}
  D_d
  \ge
  \frac{c_{\mathrm{end}}Wa_0}{4}
  d\,a_\theta^{-3}.
  \label{eq:endpoint-denominator-event}
\end{equation}
\end{lemma}

\begin{proof}
If \(|\theta|<3/4\), then on
\[
  \mathcal B_0
  =
  \{1\le|B|\le2,\ |\Theta|\le1/4\}
\]
one has
\[
  |B-\theta\Theta|\ge1/2,
  \qquad
  |\alpha_\theta(B,\Theta)|\le\sqrt7.
\]
Equation~\eqref{eq:probit-information} therefore gives
\begin{equation}
  \iota_\theta(B,\Theta)
  \ge
  c_{\mathrm{int}}.
  \label{eq:interior-band-information}
\end{equation}

If \(|\theta|\ge3/4\), use
\[
  T_\theta
  =
  \frac{B+\theta\Theta}{\sqrt{1+\theta^2}},
\qquad
  U_\theta
  =
  \frac{\theta B-\Theta}{\sqrt{1+\theta^2}}.
\]
These are independent standard normals.  On
\[
  \mathcal B_\theta
  =
  \{
    |T_\theta|\le1,\
    a_\theta\le|U_\theta|\le2a_\theta
  \},
\]
the transition-band calculation in the proof of
Proposition~\ref{prop:dither-oracle} gives
\begin{equation}
  \iota_\theta(B,\Theta)
  \ge
  c_{\mathrm{end}}a_\theta^{-4},
  \label{eq:endpoint-band-information}
\end{equation}
and
\begin{equation}
  \Prb(\mathcal B_\theta)
  \ge
  a_0a_\theta.
  \label{eq:endpoint-band-probability}
\end{equation}

For all large \(d\), the nonpilot population is at least \(Wd/2\) and the
remaining probe budget is at least \(Rd/2\).  Binomial lower-tail bounds
show that, except with probability at most
\[
  \exp(-Wq_0d/16)
\]
in the interior and
\[
  \exp(-Wa_0da_\theta/16)
\]
near the endpoints, the corresponding band contains at least
\[
  Wq_0d/4
  \quad\text{or}\quad
  Wa_0da_\theta/4
\]
addresses.
\begin{revision}
If a band contains \(Q_d\) addresses and each supplies information
at least \(c\), the top-\(m_d\) sum is at least
\(c\min(Q_d,m_d)\).
Apply Equation~\eqref{eq:interior-band-information}
and Equation~\eqref{eq:endpoint-band-information} to the count bounds above,
using the endpoint band probability in
Equation~\eqref{eq:endpoint-band-probability}.  This gives
\[
  D_d\ge
  \begin{cases}
    c_{\mathrm{int}}d\min(Wq_0/4,R/2),
        &|\theta|<3/4,\\[2mm]
    c_{\mathrm{end}}d\,a_\theta^{-4}
       \min(Wa_0a_\theta/4,R/2),
        &|\theta|\ge3/4.
  \end{cases}
\]
These inequalities apply directly to integer counts, without rounding
the deterministic lower bounds.  Since \(a_\theta\le1\), the endpoint
line is at least
\(c_{\mathrm{end}}d\min(Wa_0/4,R/2)\).
This proves Equation~\eqref{eq:denominator-event} with the specified
cutoff.  The pilot projection gives
\end{revision}
\[
  a_\theta\ge h_d/n_d,
\]
so \(p_{\mathrm{den},d}=o(d^{-1})\), uniformly over
the projected pilot interval.  When \(a_\theta\to0\), the lower bound
\(Wa_0da_\theta/4\) for the band count is eventually smaller than
the remaining budget \(Rd/2\).
Multiplying its size by the information lower bound in
Equation~\eqref{eq:endpoint-band-information} gives Equation~\eqref{eq:endpoint-denominator-event}.
\end{proof}

\begin{revision}
The interior and endpoint bounds use the same derivative identities.
\end{revision}
For fixed \(b,c\) and \(-1<\theta<1\), put
\[
  a_\theta=\sqrt{1-\theta^2},\qquad
  \alpha_\theta=\frac{\theta b-c}{a_\theta},\qquad
  n_\theta=b-\theta c.
\]
Direct differentiation gives
\begin{equation}
  \begin{aligned}
    \alpha_\theta'&=n_\theta a_\theta^{-3},\\
    \alpha_\theta''&=-ca_\theta^{-3}
                      +3\theta n_\theta a_\theta^{-5},\\
    \alpha_\theta'''&=3n_\theta a_\theta^{-5}
                      -6\theta ca_\theta^{-5}
                      +15\theta^2n_\theta a_\theta^{-7}.
  \end{aligned}
  \label{eq:probit-argument-derivatives}
\end{equation}
With \(\lambda(x)=\phi(x)/\Phi(x)\) and \(g\) from
Equation~\eqref{eq:probit-envelope},
\[
  s_\theta(z;b,c)=z\alpha_\theta'\lambda(z\alpha_\theta),\qquad
  \iota_\theta(b,c)=(\alpha_\theta')^2g(\alpha_\theta).
\]
The product and chain rules therefore give
\begin{equation}
  \begin{aligned}
    \partial_\theta s_\theta
      &=z\alpha_\theta''\lambda(z\alpha_\theta)
          +(\alpha_\theta')^2\lambda'(z\alpha_\theta),\\
    \partial_\theta^2s_\theta
      &=z\alpha_\theta'''\lambda(z\alpha_\theta)
          +3\alpha_\theta'\alpha_\theta''\lambda'(z\alpha_\theta)
          +z(\alpha_\theta')^3\lambda''(z\alpha_\theta),\\
    \partial_\theta\iota_\theta
      &=2\alpha_\theta'\alpha_\theta''g(\alpha_\theta)
          +(\alpha_\theta')^3g'(\alpha_\theta).
  \end{aligned}
  \label{eq:probit-score-derivatives}
\end{equation}

\begin{revision}
The interior Taylor expansion needs a common bound on the score and
information derivatives.  The next lemma gives one polynomial bound for
all of them.
\end{revision}

\begin{lemma}[Compact Derivative Envelope]
\label{lem:dither-compact-envelope}
For a nonempty compact \(K\subset(-1,1)\), put
\[
  \delta_K=\min_{\rho\in K}(1-\rho^2),\qquad
  A_K=24\delta_K^{-7/2},\qquad
  C_K=2^{13}(1+A_K)^6.
\]
For \(\rho\in K\), \(z\in\{-1,1\}\), and \(b,c\in\mathbb R\),
\begin{equation}
  |s_\rho|+|\partial_\rho s_\rho|
  +|\partial_\rho^2s_\rho|+\iota_\rho
  +|\partial_\rho\iota_\rho|
  \le C_K(1+|b|+|c|)^6.
  \label{eq:compact-derivative-envelope}
\end{equation}
The arguments of the score and information are \((z;b,c)\) and \((b,c)\),
respectively.
\end{lemma}

\begin{proof}
Writing \(m=1+|b|+|c|\), differentiation of
\(\alpha_\rho\) in
Equation~\eqref{eq:standardized-probit-argument} gives
\[
  |\alpha_\rho|\le\delta_K^{-1/2}m,\quad
  |\alpha_\rho'|\le\delta_K^{-3/2}m,\quad
  |\alpha_\rho''|\le4\delta_K^{-5/2}m,\quad
  |\alpha_\rho'''|\le24\delta_K^{-7/2}m.
\]
All four are at most \(A_Km\).
The differentiation formulas in Equation~\eqref{eq:probit-score-derivatives} apply to
every interior parameter.  They write the five terms in
Equation~\eqref{eq:compact-derivative-envelope} as sums of products.
Each product uses at most three of the derivatives above and one of
\(\lambda,\lambda',\lambda'',g,g'\).
By Equation~\eqref{eq:mills-envelope} through Equation~\eqref{eq:probit-derivative-envelope},
the sum is at most
\[
  512(1+A_K)^3(3A_K+7A_K^2+2A_K^3)m^6
  \le C_Km^6.
\]
This gives both the stated degree and an explicit constant.
\end{proof}

Near an endpoint, the conditional Gaussian variance vanishes and the
derivatives grow.  The next lemma tracks their powers of
\(\sigma=\sqrt{1-\rho^2}\) after averaging over the design.
\textcolor{black}{Its common constant bounds the remainder terms.}

\begin{lemma}[Endpoint Derivative Envelopes]
\label{lem:dither-endpoint-envelope}
Let
\[
  0<q\le\frac1{32},
  \qquad
  \rho=2\cos(\pi q)-1,
  \qquad
  \sigma=\sqrt{1-\rho^2}.
\]
For
\[
  \theta=2\cos(\pi\tau)-1,
  \qquad
  \tau\in[q/2,2q],
\]
the moment bounds below hold with the common
constant
\begin{equation}
  C_\star
  =2^{214}3^{11}
   \left\{\pi\,2^{22}
      +2\sqrt\pi\,\Gamma(13/2)\,2^{154}\right\}.
  \label{eq:endpoint-envelope-constant}
\end{equation}
Specifically,
\begin{equation}
  \E[
    \iota_\rho(B,\Theta)
  ]
  \le
  C_\star\sigma^{-3},
  \label{eq:endpoint-information-moments}
\end{equation}
\[
  \E\left[
    \sup_\theta
    |
      \partial_\theta
      \iota_\theta(B,\Theta)
    |
  \right]
  \le
  C_\star\sigma^{-5},
\]
\[
  \E\left[
    \sup_\theta
    |
      \partial_\theta
      \iota_\theta(B,\Theta)
    |^2
  \right]
  \le
  C_\star\sigma^{-11},
\]
\[
  \E_\rho\left[
    \{
      \partial_\rho
      s_\rho(Z;B,\Theta)
      +
      \iota_\rho(B,\Theta)
    \}^2
  \right]
  \le
  C_\star\sigma^{-7},
\]
and, for \(j=1,2\),
\begin{equation}
  \E_\rho\left[
    \sup_\theta
    |
      \partial_\theta^2
      s_\theta(Z;B,\Theta)
    |^j
  \right]
  \le
  C_\star\sigma^{1-6j}.
  \label{eq:endpoint-derivative-moments}
\end{equation}
The same bounds hold near \(-1\) after the corresponding sign reversal.
\end{lemma}

\begin{proof}
Use the true-parameter orthogonal coordinates
\[
  T=\frac{B+\rho\Theta}{\sqrt{1+\rho^2}},
  \qquad
  U=\frac{\rho B-\Theta}{\sqrt{1+\rho^2}}.
\]
They are independent standard normals.  In the integrals below, change
variables \(U=\sigma t\).  The elementary bounds on sine and cosine in
\(1-\rho=4\sin^2(\pi q/2)\) give, for
\(\tau\in[q/2,2q]\),
\[
  2q\le\sigma\le5q,\qquad
  \frac15\le\frac{a_\theta}{\sigma}\le5,\qquad
  |\theta-\rho|\le10\sigma^2.
\]
The same sine bound applies to \(a_\theta\), with \(q\) replaced by
\(\tau\le1/16\).  In the rotated coordinates,
\[
  |n_\theta|
  \le11\sigma^2|T|+2\sigma|t|,\qquad
  |\Theta|\le|T|+\sigma|t|.
\]
Substitution into Equation~\eqref{eq:probit-argument-derivatives} gives the explicit common bounds
\begin{equation}
  |\alpha_\theta|\le50(1+|t|+|T|),\qquad
  |\partial_\theta^j\alpha_\theta|
  \le2^{25}\sigma^{-2j}(1+|t|+|T|),\quad j=1,2,3.
  \label{eq:endpoint-rescaled-derivatives}
\end{equation}
At the true parameter,
\begin{equation}
  \alpha_\rho=\sqrt{1+\rho^2}\,t,\qquad
  \Prb_\rho\{Z=z\mid T,t\}
  =\Phi\!\left(z\sqrt{1+\rho^2}\,t\right).
  \label{eq:true-endpoint-sign-probability}
\end{equation}

To bound the decay uniformly over \(\theta\), we write
\[
  \alpha_\theta=c_{\rho,\theta}t+e_{\rho,\theta}\sigma T,
  \qquad
  \frac18\le c_{\rho,\theta}\le10,
  \quad |e_{\rho,\theta}|\le50,
\]
where
\[
  c_{\rho,\theta}
  =\frac{(1+\theta\rho)\sigma}{\sqrt{1+\rho^2}\,a_\theta},\qquad
  e_{\rho,\theta}
  =\frac{\theta-\rho}{\sqrt{1+\rho^2}\,a_\theta\sigma}.
\]
On the region \(|T|\le |t|/(800\sigma)\), every \(\alpha_\theta\)
has the sign of \(t\) and satisfies \(|\alpha_\theta|\ge|t|/16\).
For the matching sign of \(Z\), the inverse-Mills profiles in
Equation~\eqref{eq:mills-envelope} decay at least as \(e^{-t^2/512}\).
For the opposite sign they have polynomial growth, but the true sign
probability in Equation~\eqref{eq:true-endpoint-sign-probability} is at most \(e^{-t^2/2}/2\).
The functions \(g,g'\) have Gaussian decay for either sign.
On the complementary region, regardless of the sign of \(\alpha_\theta\),
\[
  e^{-T^2/2}
  \le e^{-T^2/4}e^{-t^2/(4\cdot800^2\sigma^2)}.
\]
Because \(\sigma\le1\), this decay is uniform as well.

\begin{revision}
It remains to bound the polynomial factors.  Put
\end{revision}
\(w_0=1+|T|+|t|\).  Each factor
\(\lambda,\lambda',\lambda'',g,g'\), after its indicated decay is
removed, is at most \(512\cdot51^3w_0^3\le2^{27}w_0^3\).
We use Equation~\eqref{eq:probit-score-derivatives} and Equation~\eqref{eq:endpoint-rescaled-derivatives}
to factor out the powers of \(\sigma\) listed in the table below.
\begin{revision}
Each remaining expression is bounded by \(2^{105}w_0^6\) times the
applicable decay.  For example, the three terms in
\(\partial_\theta^2s_\theta\) have powers
\(\sigma^{-6}\), \(\sigma^{-2}\sigma^{-4}\), and
\((\sigma^{-2})^3\), respectively.  Their common polynomial coefficient
is bounded by
\[
  2^{27}\bigl(2^{25}+3\cdot2^{50}+2^{75}\bigr)
  \le2^{105}.
\]
The information derivative has only the last two product types,
and \(\partial_\rho s_\rho+\iota_\rho\) has at most two derivative
factors.  They satisfy the same coefficient bound with the powers
listed below.

For either moment \(j\le2\), we square the coefficient when necessary.
For score expressions, we weight each sign's contribution by its
probability under the true parameter.  We also include the Gaussian
design densities.
In the first region, the resulting
decay is bounded by a constant times
\(e^{-T^2/2-t^2/512}\); the probability of the unlikely sign gives at
least this decay even when its score derivatives grow polynomially.
In the complementary region, it is bounded by a constant times
\(e^{-T^2/4-t^2/(4\cdot800^2)}\).
Since \(2^{22}>4\cdot800^2\), a common majorant after removing the
stated power of \(\sigma\) is
\end{revision}
\[
  H(T,t)=2^{214}w_0^{12}
          e^{-(T^2+t^2)/2^{22}}.
\]
This majorant also bounds the suprema over \(\theta\), since the region
split and all constants hold for every \(\theta\) simultaneously.
\begin{revision}
Since \(w_0^{12}\le3^{11}(1+|T|^{12}+|t|^{12})\), its integral can
be bounded explicitly using, for \(L>0\),
\[
  \int_{\mathbb R}e^{-x^2/L}\,dx=\sqrt{\pi L},
  \qquad
  \int_{\mathbb R}|x|^{12}e^{-x^2/L}\,dx
      =L^{13/2}\Gamma(13/2).
\]
Setting \(L=2^{22}\) gives
\(\int_{\mathbb R^2}H\le C_\star\) with exactly the value in
Equation~\eqref{eq:endpoint-envelope-constant}.
\end{revision}

\begin{revision}
The integrable envelope lets us track the powers of \(\sigma\) through
the integral.  The change of variables transforms the design measure as
follows:
\end{revision}
\begin{equation}
  \phi(T)\phi(U)\,dT\,dU
  =
  \sigma\phi(T)\phi(\sigma t)\,dT\,dt.
  \label{eq:transition-coordinate-measure}
\end{equation}
Equation~\eqref{eq:probit-score-derivatives} and Equation~\eqref{eq:endpoint-rescaled-derivatives} determine the explicit pointwise
powers of \(\sigma\).  Equation~\eqref{eq:transition-coordinate-measure} contributes one additional
factor \(\sigma\) after integration in the transition coordinate.  Once
the fixed envelope \(H\) is removed, the power count is
\[
\begin{array}{c|c|c|c}
\text{profile}
&
\text{pointwise scale}
&
\text{moment used}
&
\text{integrated scale}
\\ \hline
\iota_\rho
&
\sigma^{-4}
&
j=1
&
\sigma^{-3}
\\
\partial_\theta\iota_\theta
&
\sigma^{-6}
&
j=1,2
&
\sigma^{1-6j}
\\
\partial_\rho s_\rho+\iota_\rho
&
\sigma^{-4}
&
j=2
&
\sigma^{-7}
\\
\partial_\theta^2s_\theta
&
\sigma^{-6}
&
j=1,2
&
\sigma^{1-6j}.
\end{array}
\]
The first row gives Equation~\eqref{eq:endpoint-information-moments}.
The second gives the information-derivative scales \(\sigma^{-5}\)
and \(\sigma^{-11}\).  The third bounds the second moment of the
centered score derivative by \(C_\star\sigma^{-7}\), and the fourth
gives Equation~\eqref{eq:endpoint-derivative-moments}.
Together these prove all the stated bounds.
The negative endpoint follows by
\[
  (\rho,\theta,B)\mapsto(-\rho,-\theta,-B).
\]
\end{proof}

\begin{proposition}[Uniform Efficiency of the Explicit Protocol]
\label{prop:dither-uniform}
Use the protocol in
Appendix~\ref{sec:sharp-protocol}, with the cutoff in Equation~\eqref{eq:denominator-test}.
For every fixed \(0<R<W\),
\begin{equation}
  \lim_{d\to\infty}
  \sup_{\rho\in[-1,1]}
  d\,\E_\rho[
    (\widehat\rho_d-\rho)^2
  ]
  =
  \frac1{\J_{W,R}^{\min}}.
  \label{eq:uniform-protocol-risk}
\end{equation}
\end{proposition}

\begin{revision}
Uniformity over the full correlation interval \([-1,1]\) requires both
the interior and endpoint estimates in the proof.
\end{revision}

\begin{proof}
The deterministic tie rules and fixed pilot addresses make every address
map measurable.  The pilot and selected nonpilot sets are disjoint and use at
most \(k_d\) probes.

\begin{revision}
On compact interior intervals, we control three sources of error:
the pilot, selection at the pilot estimate, and the score remainder.
We then bound the risk for correlations approaching the endpoints and
combine the two results.
\end{revision}

Fix compact intervals \(K\subset\operatorname{int}(K')\subset(-1,1)\),
and put
\[
  q_K=\min_{\rho\in K,\,\nu\in\{+,-\}}q_\nu(\rho)>0,
  \qquad
  \eta_K=\operatorname{dist}(K,\mathbb R\setminus\operatorname{int}K')>0.
\]
\begin{revision}
For the proof, define the pilot estimate and its candidate selected set
on the whole stored record, even when the endpoint test stops the reader.
On those outcomes, these variables are used only in the proof; the
protocol returns the endpoint after the pilot batch.
\end{revision}
Let \(\mathsf G_d=\{M_d^+\ge h_d,M_d^-\ge h_d\}\).
Each inverse mismatch map is \(2\pi\)-Lipschitz.  The projection interval
contains \(K\) for all sufficiently large \(d\), so projection cannot
increase the error.  Hoeffding's bound, integrated against
\(2jt^{2j-1}\), gives
\(\E|M_d^\nu/n_d-q_\nu|^{2j}\le2j!/(2n_d)^j\).
The selected pilot's error is bounded by the larger of the two candidate
errors; summing their moments therefore gives, for \(j=1,2\),
\begin{equation}
  \sup_{\rho\in K}\E_\rho|\widetilde\rho_d-\rho|^{2j}
  \le\frac{P_j}{n_d^j},\qquad
  P_j=4j!(2\pi^2)^j.
  \label{eq:interior-pilot-moments}
\end{equation}
For all sufficiently large \(d\), \(h_d/n_d\le q_K/2\), and the same
concentration bound gives
\begin{equation}
  \sup_{\rho\in K}
  \Prb_\rho\{\mathsf G_d^c\text{ or }\widetilde\rho_d\notin K'\}
  \le
  2e^{-n_dq_K^2/2}+4e^{-n_d\eta_K^2/(2\pi^2)}
  =:p_{\mathrm{pil},d}.
  \label{eq:interior-pilot-failure}
\end{equation}

\begin{revision}
We compare the information selected at the pilot with the best sum at the
true correlation.
\end{revision}
Let \(N'_d=N_d-n_d\), and use the explicit envelope
\[
  L_K(b,c)=C_{K'}(1+|b|+|c|)^6
\]
from Lemma~\ref{lem:dither-compact-envelope}.
For \(j=1,2,4\), its moments satisfy
\[
  \E L_K(B,\Theta)^j\le M_{K,j}
  :=C_{K'}^j3^{6j-1}
    \left\{1+\frac{2^{3j+1}\Gamma(3j+1/2)}{\sqrt\pi}\right\}.
\]
This follows from \((a+b+c)^{6j}\le3^{6j-1}(a^{6j}+b^{6j}+c^{6j})\)
and the absolute moments of a standard normal.  Define
\[
  \mathcal T_d(\theta)
  =\frac1d\sum_{j=1}^{k_d-n_d}
     \bigl(\iota_\theta(B_i,\Theta_i)\bigr)_{(j;N'_d)},
  \qquad \theta\in K',
\]
where the order statistics use nonpilot indices only.  The top-sum
variational formula in Equation~\eqref{eq:top-sum-variational-formula} and the mean-value inequality give
\begin{equation}
  |\mathcal T_d(\theta)-\mathcal T_d(\eta)|
  \le |\theta-\eta|\frac1d\sum_{i>n_d}L_K(B_i,\Theta_i).
  \label{eq:empirical-information-modulus}
\end{equation}
Apply Lemma~\ref{lem:trimmed-sum} on a finite mesh of \(K'\).
The empirical modulus between mesh points is bounded by the mesh width
times the sum in Equation~\eqref{eq:empirical-information-modulus}; its expectation is at most
\((N'_d/d)M_{K,1}\) times that width.  The population modulus obeys the
same bound with \(W\) in place of \(N'_d/d\).  Taking the dimension
limit on the fixed mesh and then refining it proves
\begin{equation}
  \E\sup_{\theta\in K'}
  |\mathcal T_d(\theta)-\J_{W,R}(\theta)|\longrightarrow0.
  \label{eq:uniform-trimmed-sum}
\end{equation}

Let \(\mathcal I_d\) be the pilot-selected set, and put
\[
  I_d=\sum_{i\in\mathcal I_d}\iota_\rho(B_i,\Theta_i),\qquad
  \Delta_d=\widetilde\rho_d-\rho,\qquad
  \Lambda_{K,d}=\sum_{i>n_d}L_K(B_i,\Theta_i).
\]
If \(S_d(\rho)\) is the true-parameter top set, then on
\(\widetilde\rho_d\in K'\),
\begin{equation}
  0\le\sum_{i\in S_d(\rho)}\iota_\rho(B_i,\Theta_i)-I_d
     \le2|\Delta_d|\Lambda_{K,d},\qquad
  |D_d-I_d|\le|\Delta_d|\Lambda_{K,d}.
  \label{eq:selection-stability}
\end{equation}
For the first inequality, add and subtract the two sums evaluated at the
pilot.  Their difference is nonpositive because \(\mathcal I_d\)
maximizes the pilot sum; the two remaining differences are each bounded
by \(|\Delta_d|\Lambda_{K,d}\).
Since \(\E\Lambda_{K,d}^j\le(N'_d)^jM_{K,j}\),
the bounds in Equation~\eqref{eq:interior-pilot-moments}
through Equation~\eqref{eq:selection-stability} imply, uniformly on \(K\),
\begin{equation}
  I_d/d-\J_{W,R}(\rho)\longrightarrow0,\qquad
  D_d/d-\J_{W,R}(\rho)\longrightarrow0
  \quad\text{in probability}.
  \label{eq:information-consistency}
\end{equation}

\begin{revision}
We expand the selected score around the true correlation, keeping the
selected set fixed.
Let \(\mathcal F_d\) be generated by the pilot observations and the
scalar pairs \((B_i,\Theta_i)\).
Conditional on this state, the nonpilot residuals remain independent
standard normals.  The selected set is \(\mathcal F_d\)-measurable,
so its payloads have independent Bernoulli laws conditional on
\(\mathcal F_d\).  With this set held fixed during differentiation, put
\end{revision}
\[
  U_d(\theta)=\sum_{i\in\mathcal I_d}s_\theta(Z_i;B_i,\Theta_i),
  \qquad V_d=U_d'(\rho)+I_d.
\]
\begin{revision}
For a fixed scalar pair, write
\(p_z(\theta)=\Phi(z\alpha_\theta)\), so
\(s_\theta(z)=p_z'(\theta)/p_z(\theta)\).
Positivity and smoothness hold throughout the interior parameter
interval, and the support has only two points.  Differentiating
\(\sum_zp_z(\theta)=1\) therefore gives
\[
  \sum_zp_z(\rho)s_\rho(z)=\sum_zp_z'(\rho)=0,\qquad
  \sum_zp_z(\rho)\partial_\rho s_\rho(z)
     =\sum_zp_z''(\rho)
       -\sum_z\frac{p_z'(\rho)^2}{p_z(\rho)}
     =-\iota_\rho.
\]
Conditional independence of the selected cells now implies
\end{revision}
\[
  \E_\rho[U_d(\rho)\mid\mathcal F_d]=0,\quad
  \E_\rho[U_d(\rho)^2\mid\mathcal F_d]=I_d,\quad
  \E_\rho[V_d\mid\mathcal F_d]=0.
\]
Moreover, conditional independence and
\(\Var_\rho(\partial_\rho s_\rho\mid B_i,\Theta_i)
\le\E_\rho[(\partial_\rho s_\rho)^2\mid B_i,\Theta_i]\) give
\begin{equation}
  \E_\rho[V_d^2\mid\mathcal F_d]
    \le\sum_{i>n_d}L_K(B_i,\Theta_i)^2,\qquad
  \sup_{\theta\in K'}|U_d''(\theta)|\le\Lambda_{K,d}.
  \label{eq:conditional-score-variance}
\end{equation}
Bounding the selected sums by the nonpilot population makes these
inequalities uniform over the selection rule.

Let
\[
  \mathsf A_{K,d}
  =\mathsf G_d\cap\{\widetilde\rho_d\in K'\}
       \cap\{D_d\ge\gamma_{W,R}d\}.
\]
Its complement has probability at most
\(p_{\mathrm{pil},d}+p_{\mathrm{den},d}=o(d^{-1})\), uniformly on \(K\).
On \(\mathsf A_{K,d}\), Taylor expansion at the true parameter gives
\begin{equation}
  \widehat\rho_d^{\mathrm{raw}}-\rho
  =\frac{U_d(\rho)}{D_d}
   +\underbrace{\Delta_d(1-I_d/D_d)}_{E_{1,d}}
   +\underbrace{\Delta_dV_d/D_d}_{E_{2,d}}+R_d,
  \qquad
  |R_d|\le\frac{\Delta_d^2\Lambda_{K,d}}{2D_d}.
  \label{eq:score-taylor-expansion}
\end{equation}
\begin{revision}
All ratios multiplied by \(\mathbf1_{\mathsf A_{K,d}}\) below are defined
as zero off that event.  The pilot is independent of the nonpilot
population.  We combine this independence with the pilot moments in
Equation~\eqref{eq:interior-pilot-moments}, the selection bound in
Equation~\eqref{eq:selection-stability}, and the conditional variance bound
in Equation~\eqref{eq:conditional-score-variance} to obtain
\end{revision}
\[
\begin{aligned}
 d\E_\rho[E_{1,d}^2\mathbf1_{\mathsf A_{K,d}}]
 &\le\frac{P_2(N'_d)^2M_{K,2}}{\gamma_{W,R}^2d\,n_d^2},\\
 d\E_\rho[E_{2,d}^2\mathbf1_{\mathsf A_{K,d}}]
 &\le\frac{P_1N'_dM_{K,2}}{\gamma_{W,R}^2d\,n_d},\\
 d\E_\rho[R_d^2\mathbf1_{\mathsf A_{K,d}}]
 &\le\frac{P_2(N'_d)^2M_{K,2}}{4\gamma_{W,R}^2d\,n_d^2}.
\end{aligned}
\]
For the middle bound, we first condition on \(\mathcal F_d\), since
\(V_d\) depends on the pilot through the selected set.
We apply Equation~\eqref{eq:conditional-score-variance} before averaging over the
nonpilot design.  Each displayed bound tends to
zero because \(N'_d=O(d)\), \(n_d\to\infty\), and
\(d/n_d^2\to0\).  Thus, the sum of the three remainders is
\(o(d^{-1/2})\) in \(L_2\), uniformly on \(K\).

\begin{revision}
The leading score term determines the limiting risk.  We compute its
second moment and then verify that clipping preserves the limit.
Define
\end{revision}
\(\xi_d=\mathbf1_{\mathsf A_{K,d}}U_d(\rho)/D_d\).
Conditional score variance gives
\[
  d\E_\rho\xi_d^2
  =\E_\rho\!\left[
      \mathbf1_{\mathsf A_{K,d}}\frac{dI_d}{D_d^2}\right]
  \longrightarrow\frac1{\J_{W,R}(\rho)}
\]
\begin{revision}
uniformly on \(K\).
Equation~\eqref{eq:information-consistency} gives convergence in
probability.  The bound
\[
  \mathbf1_{\mathsf A_{K,d}}\frac{dI_d}{D_d^2}
  \le\frac{\Lambda_{K,d}}{\gamma_{W,R}^2d}
\]
gives a uniform second-moment bound, and hence uniform integrability.
\end{revision}
We use a fourth-moment bound to show that clipping preserves the limiting
risk.  Conditional independence and the fourth-moment formula for
centered sums give
\[
  \E_\rho\xi_d^4
  \le\frac{N'_dM_{K,4}+3(N'_d)^2M_{K,2}}
           {\gamma_{W,R}^4d^4}
  =O(d^{-2}).
\]
Put \(a_K=\min_{\rho\in K}(1-|\rho|)>0\).  Then
\[
  \E_\rho\!\left[
    \bigl\{\proj_{[-1,1]}(\rho+\xi_d)-\rho-\xi_d\bigr\}^2\right]
  \le\E_\rho[\xi_d^2\mathbf1_{\{|\xi_d|\ge a_K\}}]
  \le a_K^{-2}\E_\rho\xi_d^4.
\]
The projection is one-Lipschitz, so adding the \(L_2\)-negligible
remainders from Equation~\eqref{eq:score-taylor-expansion} preserves this approximation on
\(\mathsf A_{K,d}\).  Off that event the clipped error is at most two,
and \(d\Prb(\mathsf A_{K,d}^c)\to0\).  It follows that
\(d\E_\rho(\widehat\rho_d-\rho-\xi_d)^2\to0\), uniformly on \(K\).
Cauchy--Schwarz now gives the exact limit
\begin{equation}
  \sup_{\rho\in K}\left|
     d\E_\rho(\widehat\rho_d-\rho)^2
     -\frac1{\J_{W,R}(\rho)}\right|\longrightarrow0.
  \label{eq:interior-risk-limit}
\end{equation}

It remains to control every sequence \(\rho_d\to1\).  The same
good-event argument, with the two mismatch counts interchanged, will
handle the other endpoint.  Put
\[
  q_d=q_+(\rho_d),\qquad \sigma_d=\sqrt{1-\rho_d^2}.
\]
For all large \(d\), \(q_d\le1/32\),
\(2q_d\le\sigma_d\le5q_d\), and \(q_-(\rho_d)\ge3/8\).
Hoeffding's inequality therefore gives the explicit wrong-endpoint bound
\[
  \Prb_{\rho_d}\{M_d^-<n_d/4\}\le e^{-n_d/32}.
\]
If \(n_dq_d<h_d/4\), Lemma~\ref{lem:dither-pilot} implies that the
protocol returns \(1\) except on an event of probability at most
\((e/4)^{h_d}+e^{-n_d/32}\).  Since
\[
  (1-\rho_d)^2
  =4\{1-\cos(\pi q_d)\}^2\le\pi^4q_d^4,
\]
its scaled risk is at most
\[
  \pi^4d(h_d/(4n_d))^4
  +4d\{(e/4)^{h_d}+e^{-n_d/32}\}\longrightarrow0.
\]
This case includes \(\rho_d=1\).

Now, suppose \(n_dq_d\ge h_d/4\).  The relative-frequency event
\(E_{q_d}\) in Lemma~\ref{lem:dither-pilot} fails with probability at
most \(2e^{-h_d/48}\le2d^{-2}\).
On \(E_{q_d}\cap\{M_d^-\ge n_d/4\}\), the plus count is smaller.
If it is below \(h_d\), the protocol returns \(1\), and
\(q_d<2h_d/n_d\); this branch contributes at most
\(16\pi^4d(h_d/n_d)^4=o(1)\) to scaled risk.
Otherwise, define the good pilot event
\[
  \mathsf E_d
  =E_{q_d}\cap\{M_d^-\ge n_d/4\}\cap\{M_d^+\ge h_d\}.
\]
On this event the projection in the pilot formula is inactive,
\(\widetilde\rho_d=2\cos(\pi M_d^+/n_d)-1\), and its inverse-mismatch
coordinate lies in \([q_d/2,2q_d]\).
Let \(\mathsf A_{e,d}\) be \(\mathsf E_d\) intersected with the
band-count event in Lemma~\ref{lem:dither-denominator}.  Outside the
endpoint-return branch and \(\mathsf A_{e,d}\), the total exceptional
probability is at most
\(2d^{-2}+e^{-n_d/32}+p_{\mathrm{den},d}=o(d^{-1})\).

\begin{revision}
On the good event, the score correction satisfies the following bounds.
\end{revision}
On \(\mathsf A_{e,d}\), the denominator event gives both the protocol
cutoff and, for all sufficiently large \(d\),
\begin{equation}
  D_d\ge\kappa_e d\sigma_d^{-3},\qquad
  \kappa_e=\frac{c_{\mathrm{end}}Wa_0}{500}.
  \label{eq:scoring-endpoint-denominator}
\end{equation}
Indeed, \(a_{\widetilde\rho_d}\le5\sigma_d\), so Equation~\eqref{eq:endpoint-denominator-event} implies
this constant.  Also,
\begin{equation}
  d\sigma_d\ge\frac{dh_d}{2n_d}\longrightarrow\infty.
  \label{eq:endpoint-sample-size}
\end{equation}

Let \(I_d,U_d,V_d\) denote the same selected sums as in the interior,
now at the true parameter \(\rho_d\), and put
\(\Delta_d=\widetilde\rho_d-\rho_d\).
Define the deterministic parameter neighborhood
\[
  K_d^e
  =\{2\cos(\pi\tau)-1:q_d/2\le\tau\le2q_d\},
\]
and full-population envelopes
\[
  \Lambda_d=\sum_{i>n_d}\sup_{\theta\in K_d^e}
      |\partial_\theta\iota_\theta(B_i,\Theta_i)|,\qquad
  H_d=\sum_{i>n_d}\sup_{\theta\in K_d^e}
      |\partial_\theta^2s_\theta(Z_i;B_i,\Theta_i)|.
\]
The neighborhood \(K_d^e\) is deterministic, so these nonpilot envelopes
are independent of the pilot.  On
\(\mathsf E_d\), this neighborhood contains the entire Taylor segment.
Write
\[
  B_W=2WC_\star+4W^2C_\star^2.
\]
For all sufficiently large \(d\), \(N'_d\le2Wd\) and \(d\sigma_d\ge1\).
The first and second moments in Lemma~\ref{lem:dither-endpoint-envelope}
then give
\begin{equation}
  \E\Lambda_d^2\le B_Wd^2\sigma_d^{-10},\quad
  \E H_d^2\le B_Wd^2\sigma_d^{-10},\quad
  \E[V_d^2\mid\text{pilot}]\le2WC_\star d\sigma_d^{-7}.
  \label{eq:endpoint-population-moments}
\end{equation}
For example, the first bound before simplification is
\(N'_dC_\star\sigma_d^{-11}+(N'_d)^2C_\star^2\sigma_d^{-10}\).
For the last bound, we condition on \(\mathcal F_d\) and add the
conditional variances of the selected centered score derivatives.
We bound the resulting sum by the whole nonpilot population, then average
over its independent design.

All constants in the remaining endpoint estimates may be taken to be
\[
  C_e=\frac{324\pi^8(B_W+2WC_\star+1)}{\kappa_e^2}.
\]
The leading score term obeys
\begin{equation*}
  d\E_{\rho_d}\!\left[
     \frac{U_d(\rho_d)^2}{D_d^2}\mathbf1_{\mathsf A_{e,d}}\right]
  \le\frac{2WC_\star}{\kappa_e^2}\sigma_d^3
  \longrightarrow0.
\end{equation*}
Here, the conditional second moment is \(I_d\), which is at most the
full information sum; its expectation is at most
\(2WdC_\star\sigma_d^{-3}\).

Selection stability gives \(|D_d-I_d|\le|\Delta_d|\Lambda_d\) on
\(\mathsf E_d\).  The pilot bounds in Equation~\eqref{eq:endpoint-pilot-second-moment} through Equation~\eqref{eq:endpoint-pilot-fourth-moment} imply
\[
  \E[\Delta_d^2\mathbf1_{\mathsf E_d}]
     \le9\pi^4q_d^3/n_d,\qquad
  \E[\Delta_d^4\mathbf1_{\mathsf E_d}]
     \le324\pi^8q_d^6/n_d^2.
\]
Combining these with Equation~\eqref{eq:scoring-endpoint-denominator} through Equation~\eqref{eq:endpoint-population-moments} gives
\begin{equation*}
\begin{aligned}
 d\E\!\left[\Delta_d^2(1-I_d/D_d)^2
               \mathbf1_{\mathsf A_{e,d}}\right]
   &\le C_e\frac{dq_d^2}{n_d^2},\\
 d\E\!\left[\frac{\Delta_d^2V_d^2}{D_d^2}
               \mathbf1_{\mathsf A_{e,d}}\right]
   &\le C_e\frac{q_d^2}{n_d},\\
 d\E\!\left[\frac{\Delta_d^4H_d^2}{D_d^2}
               \mathbf1_{\mathsf A_{e,d}}\right]
   &\le C_e\frac{dq_d^2}{n_d^2}.
\end{aligned}
\end{equation*}
For the first and third inequalities, factor the pilot fourth moment
from the independent envelope second moment.  The powers are
\(q_d^6\sigma_d^{-4}\le q_d^2/16\).
For the second, use the conditional variance bound in Equation~\eqref{eq:endpoint-population-moments}
before the pilot second moment; the powers are
\(q_d^3/\sigma_d\le q_d^2/2\).
The last expression bounds four times the squared Taylor remainder.
\begin{revision}
Every right side tends to zero since \(d/n_d^2\to0\).
The chosen \(n_d=\lceil d^{3/4}\rceil\) satisfies all pilot
requirements.  More generally, the argument would permit any
deterministic \(n_d\) with \(\sqrt d=o(n_d)\) and \(n_d=o(d)\),
while keeping \(h_d=\lceil96\log d\rceil\): these conditions make the
score remainders vanish, preserve \(m_d/d\to R\), and imply both
\(d(h_d/n_d)^4\to0\) and
\(p_{\mathrm{den},d}=o(d^{-1})\).
\end{revision}
The expansion in Equation~\eqref{eq:score-taylor-expansion}, clipping, and Cauchy--Schwarz now show that the
scaled risk on \(\mathsf A_{e,d}\) tends to zero.
The endpoint-return and exceptional branches were already bounded, so
\begin{equation}
  \rho_d\to1\quad\Longrightarrow\quad
  d\E_{\rho_d}(\widehat\rho_d-\rho_d)^2\longrightarrow0.
  \label{eq:endpoint-risk-limit}
\end{equation}
On the good events used above the pilot counts are unequal: the near
count is below \(n_d/4\), while the far count is at least \(n_d/4\).
The transformation \((\rho,B)\mapsto(-\rho,-B)\) interchanges these
counts, negates the selected pilot and scores, and preserves the
information ranking.  It therefore preserves squared error on those
events.  Pilot ties lie in the exceptional events bounded above, so their
contribution is negligible under the stated tie rule.  The same bounds thus
prove Equation~\eqref{eq:endpoint-risk-limit} at \(-1\).

To obtain the uniform worst-case limit, for each \(d\) we choose \(\rho_d\)
whose risk is within \(d^{-2}\) of the supremum.
Every subsequence has a further subsequence converging either to an endpoint
or to a point in a compact interior interval.  Equation~\eqref{eq:endpoint-risk-limit} handles
the endpoints, and Equation~\eqref{eq:interior-risk-limit} handles the interior.  This proves the upper
bound in Equation~\eqref{eq:uniform-protocol-risk}.  Proposition~
\ref{prop:dither-oracle} shows that
\(\J_{W,R}\) diverges at the endpoints and has an interior
minimizer.  Applying Equation~\eqref{eq:interior-risk-limit} at one minimizer proves the reverse
inequality for this protocol.
\end{proof}

\subsection{Extension from the Sphere to the Unit Ball}
\label{sec:unit-ball-extension}

\begin{revision}
To extend the directional result to general vectors, we also store the
source norm.
\end{revision}
\textcolor{black}{The norm grid in Section~\ref{sec:unit-ball-protocol}
uses \(\lceil\log_2(d+1)\rceil\) bits and has error at most \(1/(2d)\).
Its contribution to normalized risk vanishes, preserving the directional
constant.}

\begin{proof}[Proof of Theorem~\ref{thm:sharp-dither-reader}]
\begin{revision}
The common implementation in
Section~\ref{sec:unit-ball-protocol} is admissible at every finite
budget and has the stated fixed two-batch boundaries for all large
\(d\).  Its remaining directional budgets satisfy
\(N_d/d\to W\) and \(k_d/d\to R\).
For nonzero inputs, apply Proposition~\ref{prop:dither-uniform} to the
normalized directions and use Equation~\eqref{eq:common-norm-transfer}.
The directional root mean square error is \(O(d^{-1/2})\), whereas
norm rounding contributes at most \(1/(2d)\).  Squaring their sum adds
only \(O(d^{-3/2})\) to the risk, uniformly over inputs.
Consequently,
\end{revision}
\[
  \limsup_{d\to\infty}
  d\,\mathcal E_d^{\mathrm{prot}}(W,R)
  \le
  \frac1{\J_{W,R}^{\min}}.
\]
\begin{revision}
For unit vectors, the source norm is represented exactly.  Moreover,
every \(\rho\in[-1,1]\) is realized by two unit vectors whenever
\(d\ge2\).  Restricting the risk supremum to these vectors and applying
Proposition~\ref{prop:dither-uniform} gives the reverse limit inequality.
This proves the asserted limit for the protocol risk.
\end{revision}
Since the minimax risk is no larger than the
risk of this protocol, Equation~\eqref{eq:adaptive-upper-bound} follows.  Equation~\eqref{eq:sparse-oracle-information} follows from
Proposition~\ref{prop:dither-oracle}.  If either vector is zero,
return zero.
\end{proof}

\subsection{Numerical Illustration of the Sharp Refinement}
\label{sec:refinement-numerics}

Figure~\ref{fig:mse-separation} compares the exact query-independent
optimum with a numerical evaluation of the asymptotic constant for
Protocol~\ref{prot:dithered-reader} at \(R=1/4\) and \(W=\Lambda/4\).
We use the Gaussian upper-tail formula in
Equation~\eqref{eq:oracle-information}.
The rotated coordinates in the proof of
Proposition~\ref{prop:dither-oracle} reduce the calculation to
one-dimensional integrals.
We evaluate them by 256-point
Gauss--Legendre quadrature on \([-10,10]\).
At every plotted ratio, the smallest information on the correlation grid
\(\{0,0.05,\ldots,0.95,0.97,0.99\}\) occurs at zero; symmetry covers
negative correlations.  We use the zero-correlation values for the
orange curve.

These values approximate \(1/\J_{\Lambda/4,1/4}^{\min}\) from
Theorem~\ref{thm:sharp-dither-reader}.
The curves cross near \(\Lambda=5.18\) in this numerical
evaluation of the limiting risks.  Certifying that location would require
minimizing over all correlations and bounding the numerical error.
The proved sparse asymptotic for the orange curve is \(\pi/\log\Lambda\).
 
\section{Information Bounds for Fully Branching Access}
\label{app:adaptive-converse}

We prove Theorem~\ref{thm:finite-adaptive-converse} in four steps.
First, we answer repeated probes from cached values and pad with ignored
probes until the reader has visited exactly \(k\) distinct cells.
The canonical label records these addresses and values.
It determines the returned transcript once the query is known and has at most
\[
  \mathsf C_{N,k}(q)=\binom Nkq^k
\]
possible values.  Each label event also factors as
the set of sources with the recorded values times a compatible query set.
Second, we bound the information in these rectangle-factorized labels
by treating them as pseudotranscripts.
The compensated scalar DSBS inequality in
Equation~\eqref{eq:compensated-dsbs-inequality}, averaged over a
pseudotranscript, gives
\[
  \mathsf N(T)\le\rho^2\mathsf E(T),\qquad
  \mathsf N(T)=I(X;Y)-I(X;Y\mid T),\quad
  \mathsf E(T)=I(T;X,Y).
\]
Two block identities extend this inequality from one coordinate pair
to independent pairs: their remainder is a nonpositive combination of
conditional mutual informations.
A KL identity then bounds the divergence of the observed label and
query from their law at zero correlation.

Third, we represent correlation \(\rho_1\) by fixing a randomly selected
subset of coordinate pairs and using independent DSBS pairs on the
remaining coordinates.  The combined distribution has correlation
\(\rho_0\) when the residual correlation is zero.  We show that conditioning
on the fixed coordinates preserves the source--query factorization in the
remaining coordinates.  Applying the pseudotranscript inequality and the
quadratic expansion of KL divergence yields
\[
  J_\rho(\Omega,\Gamma,Y)
  \le
  \frac{2\log\mathsf C_{N,k}(q)}{(1-|\rho|)^2}.
\]
Finally, a cosine-squared prior supported on \(|\rho|<\varepsilon\)
\textcolor{black}{allows us to apply van Trees' inequality and bound the Bayes
MSE for estimating \(\rho\).}  Minimizing
\(2\log\mathsf C_{N,k}(q)/(1-\varepsilon)^2+\pi^2/\varepsilon^2\)
over the prior width gives the constant \(\mathfrak A(N,k,q)\).
\textcolor{black}{We then pass from correlation estimation to inner-product
estimation.  Conditional on the correlation,
\(\Var_\rho(X^\top Y)\le1/d\).  The \(L_2\) triangle inequality accounts
for this fluctuation and gives Equation~\eqref{eq:finite-branching-converse}.}

\subsection{Canonical Labels for Branching Transcripts}

\begin{revision}
We count partial assignments to the cells visited by a reader.
Different queries can visit the same assignment in different orders.
We first simulate the original reader.  We answer repeated probes from
cached values and make a physical probe only for a previously unseen
address, retaining the original sequence of returned bits.
Once the simulation ends, we probe the least unused addresses until
exactly \(k\) distinct cells have been visited.  The decoder reconstructs
the original transcript and ignores the padding values.
The simulation makes at most \(k\) distinct probes, and \(k\le N\)
guarantees enough unused addresses for padding.  The normalized reader
therefore preserves the original output for every source, query, and seed.
It uses only the unrestricted computation allowed by the model.
\end{revision}

Fix such a normalized deterministic \(q\)-ary protocol.  Its
\emph{canonical label} is the partial assignment
\(\gamma=(S,a)\), where \(|S|=k\) and \(a\in[q]^S\) records the values
at the probed addresses.
There are at most
\[
  \mathsf C_{N,k}(q)
  =
  \binom Nkq^k
\]
possible labels.
\begin{revision}
When \(k=0\), the empty assignment is the unique label, including when
\(N=0\).
\end{revision}

Let
\[
  C_\gamma
  =
  \{x:\operatorname{Enc}(x)_S=a\}
\]
be the set of sources whose stored values agree with
the label.  We define \(A_\gamma\) by replaying the reader
with query \(y\).  If it requests an address in \(S\), we return the
assigned value.  If it requests an address outside \(S\), we reject the
replay.  We accept exactly when the replay terminates having visited all
of \(S\).  Then \(A_\gamma\) is the set of accepted queries.
Several query-dependent orders can share the same label \(\gamma\).

\begin{proposition}[Canonical Rectangle Identity]
\label{prop:canonical-rectangle}
For every normalized deterministic protocol and every canonical label
\(\gamma\),
\begin{equation}
  \{\Gamma=\gamma\}
  =
  C_\gamma\times A_\gamma.
  \label{eq:canonical-rectangle}
\end{equation}
There are at most \(\mathsf C_{N,k}(q)\) such labels.  Given the query and
the deterministic reader, the \textcolor{black}{normalized} returned-value transcript and canonical
label determine one another.
\end{proposition}

\begin{proof}
If the actual execution has label \(\gamma\), its source lies in
\(C_\gamma\), and replay with the same query accepts.  Conversely, suppose
\(x\in C_\gamma\) and \(y\in A_\gamma\).  Induction over probes shows that
the actual execution and replay request the same address and receive the
same value at each round.  They therefore terminate with the same set
\(S\), proving Equation~\eqref{eq:canonical-rectangle}.
\textcolor{black}{The set \(A_\gamma\) is measurable because acceptance is a
finite combination of measurable preimages under the address maps.}

There are \(\binom Nk\) choices of \(S\) and \(q^k\) assignments to it.
\begin{revision}
Given the query and the normalized returned values, the reader
reconstructs its distinct addresses recursively and hence the label.
Given the query and a realized label, the accepting replay reconstructs
those returned values.  The cached-probe simulation then reconstructs
the original transcript and output.
\end{revision}
\end{proof}

{\color{black}
\subsection{Geometric Lower Bound}
\label{sec:geometric-rectangle-bound}

The rectangle identity gives a lower bound using independent source and
query vectors.  Within each rectangle, its source set determines the
uncertainty that remains after the label is observed.  The query set
determines the directions in which the decoder must resolve that
uncertainty.

\begin{proposition}[Geometric Rectangle Bound]
\label{prop:geometric-rectangle-bound}
For integers \(d\ge1\), \(q\ge2\), and \(N\ge k\ge0\),
\begin{equation}
  \D_d^{(q)}(N,k)
  \ge
  \frac{c_d^2}{d}\,\mathsf C_{N,k}(q)^{-2/d},
  \label{eq:geometric-rectangle-bound}
\end{equation}
where, as in Lemma~\ref{lem:ball-shannon},
\[
  v_d=\operatorname{vol}(\Bdd)
      =\frac{\pi^{d/2}}{\Gamma(d/2+1)},
  \qquad
  c_d=\frac{d\,v_d^{2/d}}{2\pi e}\longrightarrow1.
\]
More generally, let \(M\ge1\) be an integer.  Every randomized protocol
whose label events, for each fixed choice of seeds, partition
\(\Bdd\times\Bdd\) into at most \(M\) measurable rectangles has
worst-pair MSE at least
\(c_d^2M^{-2/d}/d\), when the decoder observes the label and the query.
\end{proposition}

\begin{proof}
Draw \(X\) and \(Y\) independently and uniformly from \(\Bdd\),
independently of the collection of protocol seeds.  Reveal all seeds
to the decoder and condition on their values.  The input law is unchanged,
and the original decoder remains available.  Normalize the reader as in
Proposition~\ref{prop:canonical-rectangle}.  Its label events form a
measurable rectangle partition with at most
\(M=\mathsf C_{N,k}(q)\) labels.

For a label \(t\) of positive probability, write its event as
\(C_t\times A_t\).  Its probability is
\[
  p_t=\frac{\operatorname{vol}(C_t)\operatorname{vol}(A_t)}{v_d^2}.
\]
Conditional on this label, \(X\) and \(Y\) are independent and uniform
on \(C_t\) and \(A_t\).  Write their means as
\(\mu_{X,t},\mu_{Y,t}\) and their covariance matrices as
\(\Sigma_{X,t},\Sigma_{Y,t}\).  Since the decoder knows \(Y\), the
conditional mean of \(X^\top Y\) given its observations is
\(\mu_{X,t}^{\top}Y\).  The conditional-variance identity therefore
bounds the MSE of every scalar decoder from below by
\[
  \E[Y^\top\Sigma_{X,t}Y\mid t]
  =
  \operatorname{tr}(\Sigma_{X,t}\Sigma_{Y,t})
  +\mu_{Y,t}^{\top}\Sigma_{X,t}\mu_{Y,t}
  \ge
  \operatorname{tr}(\Sigma_{X,t}\Sigma_{Y,t}).
\]

Both sets have positive volume, so their covariance matrices are
positive definite.  Indeed, a zero-variance direction would confine the
corresponding set, up to a null set, to a hyperplane of zero volume.
For a vector \(V\) uniform on either set \(B\), let \(\mu\) and
\(\Sigma\) be its mean and covariance.  Expanding its relative entropy
with respect to the Gaussian with the same mean and covariance gives
\[
  0\le \KL\!\left(P_V\,\middle\|\,N(\mu,\Sigma)\right)
  =
  -\log\operatorname{vol}(B)
  +\frac12\log\!\left((2\pi e)^d\det\Sigma\right).
\]
Here the Gaussian quadratic term has expectation
\(\operatorname{tr}(\Sigma^{-1}\Sigma)=d\).
Consequently,
\[
  \det\Sigma_{X,t}
  \ge\frac{\operatorname{vol}(C_t)^2}{(2\pi e)^d},
  \qquad
  \det\Sigma_{Y,t}
  \ge\frac{\operatorname{vol}(A_t)^2}{(2\pi e)^d}.
\]
We now apply the arithmetic--geometric mean inequality to the positive
eigenvalues of
\(\Sigma_{X,t}^{1/2}\Sigma_{Y,t}\Sigma_{X,t}^{1/2}\).
Its trace is \(\operatorname{tr}(\Sigma_{X,t}\Sigma_{Y,t})\),
and its determinant is
\(\det\Sigma_{X,t}\det\Sigma_{Y,t}\).  Hence,
\[
  \operatorname{tr}(\Sigma_{X,t}\Sigma_{Y,t})
  \ge
  d(\det\Sigma_{X,t}\det\Sigma_{Y,t})^{1/d}
  \ge
  \frac{c_d^2}{d}\,p_t^{2/d}.
\]

The label probabilities sum to one.  Pad their list with zeros to
length \(M\); convexity of \(u\mapsto u^{1+2/d}\) gives
\[
  \sum_t p_t^{1+2/d}
  \ge M\left(\frac1M\right)^{1+2/d}
  =M^{-2/d}.
\]
Averaging the conditional error bounds over labels therefore gives
MSE at least \(c_d^2M^{-2/d}/d\) for the fixed seeds.  The same input
distribution was used for every seed, so averaging over the seeds
preserves this lower bound.  The resulting average error is at most the
worst-pair risk of the original protocol.  Taking the infimum over
protocols proves Equation~\eqref{eq:geometric-rectangle-bound}.
The argument also proves the stated bound for any measurable rectangle
partition.  Finally, the volume formula and Stirling's formula give
\[
  c_d=\frac{d}{2e}\Gamma(d/2+1)^{-2/d}\longrightarrow1.
\]
\end{proof}
\par}

\subsection{Pseudotranscript Information Bounds}

\begin{revision}
Rectangle factorization lets us treat a canonical label as a
pseudotranscript in the sense of
Definition~\ref{def:pseudotranscript} under a correlated binary
distribution.  The bound compares the reduction in
dependence between the two sources with the information revealed about
the pair.  We prove the bound for one coordinate pair, then use two block
identities to extend it to independent pairs.  This extension is called
tensorization.  A KL identity then expresses the bound in a form that
controls Fisher information.
\end{revision}

\begin{definition}[Pseudotranscript]
\label{def:pseudotranscript}
Let \((X,Y)\) have a finite joint law.
A randomized rule with finite output \(T\)
is a \emph{pseudotranscript}, in the terminology of
\cite{prabhakaran-pseudotranscript}, if
\begin{equation}
  \Prb\{T=t\mid X=x,Y=y\}
  =
  \alpha_t(x)\beta_t(y)
  \label{eq:pseudotranscript-factorization}
\end{equation}
for every pair in
\(\operatorname{supp}(P_X)\times\operatorname{supp}(P_Y)\), for
nonnegative functions \(\alpha_t,\beta_t\), with
\(\sum_t\alpha_t(x)\beta_t(y)=1\) on that Cartesian product.
The kernel is thus defined and rectangle-factorized even at pairs having
zero probability under the joint law.  This condition is needed when the
same kernel is applied to \(P_XP_Y\) below.  Define
\[
  \mathsf E_P(T)=I_P(T;X,Y)
\]
and
\[
  \mathsf N_P(T)
  =
  I_P(X;Y)-I_P(X;Y\mid T)
  =
  I_P(T;X)+I_P(T;Y)-I_P(T;X,Y).
\]
\end{definition}

Here, \(\mathsf E(T)\) is the information that \(T\) reveals about the
pair \((X,Y)\), while \(\mathsf N(T)\) is the reduction in their mutual
information after conditioning on \(T\).

For \(|\rho|<1\), let \(\mathsf B_\rho\) be the law of a doubly symmetric
binary source (DSBS):
\[
  \mathsf B_\rho(x,y)
  =
  \frac{1+\rho xy}{4},
  \qquad
  x,y\in\{-1,1\}.
\]
Both signs are uniform, and their correlation is \(\rho\).  The product
law \(\mathsf B_\rho^{\otimes d}\) consists of \(d\) independent such pairs.

Conditioning on a pseudotranscript value reweights this law by separate
factors in \(x\) and \(y\); we call this a \emph{marginal tilt}.
The extra term in the next inequality has mean zero when averaged over
the pseudotranscript.  This cancellation is the reason for the
compensated form of the bound.

\begin{lemma}[Compensated Scalar DSBS Inequality]
\label{lem:scalar-dsbs}
Fix \(|\rho|<1\), and let \(U\) be uniform on
\(\{-1,1\}\).  Suppose
\[
  \frac{dP}{d\mathsf B_\rho}(x,y)
  =
  \frac{f(x)g(y)}
       {\E_{\mathsf B_\rho}[f(X)g(Y)]}
\]
for nonnegative \(f,g\) with positive denominator.  Then
\begin{equation}
  \KL(P_X\|U)+\KL(P_Y\|U)
  \le
  (1+\rho^2)\KL(P\|\mathsf B_\rho)
  +
  \rho\{\E_P[XY]-\rho\}.
  \label{eq:compensated-dsbs-inequality}
\end{equation}
\end{lemma}

\begin{proof}
At \(\rho=0\), the tilt is a product law, so its joint divergence is the
sum of its marginal divergences and the inequality is equality.
Replacing \(Y\) by \(-Y\) reduces the
general case to \(r=\rho\in(0,1)\).  First, assume \(f,g>0\).
After rescaling, there are \(u,v\in(-1,1)\) such that
\[
  \frac{dP}{d\mathsf B_r}(x,y)
  =
  \frac{(1+ux)(1+vy)}{z},
  \qquad
  z=1+ruv.
\]
Write
\[
  a=\E_PX,
  \qquad
  b=\E_PY,
  \qquad
  m=\E_P[XY].
\]
Direct summation gives
\begin{equation}
  a=\frac{u+rv}{z},
  \qquad
  b=\frac{v+ru}{z},
  \qquad
  m=\frac{r+uv}{z}.
  \label{eq:binary-tilt-means}
\end{equation}

For \(s\in(-1,1)\), put
\[
  F(s)
  =
  s\operatorname{arctanh}s
  +
  \frac12\log(1-s^2),
  \qquad
  D_F(s\|t)
  =
  F(s)-F(t)
  -
  \operatorname{arctanh}(t)(s-t).
\]
\textcolor{black}{Thus, \(F(s)\) gives the divergence from the uniform sign
law for a sign with mean \(s\), while \(D_F\) is the divergence between
two sign laws specified by their means.}
Using
\[
  \log(1+us)
  =
  s\operatorname{arctanh}u
  +
  \frac12\log(1-u^2),
  \qquad s\in\{-1,1\},
\]
we obtain
\[
  C:=\KL(P\|\mathsf B_r)
  =
  a\operatorname{arctanh}u
  +
  b\operatorname{arctanh}v
  +
  \frac12\log(1-u^2)
  +
  \frac12\log(1-v^2)
  -
  \log z.
\]
The difference between the right and left sides of Equation~\eqref{eq:compensated-dsbs-inequality} is
\[
  G
  =
  (1+r^2)C+r(m-r)-F(a)-F(b).
\]

\begin{revision}
The proof of \(G\ge0\) has three parts.
First, we use a curvature comparison to show that the two divergence
gaps \(\mathcal R_1\) and \(\mathcal R_2\), defined below, are nonnegative.
Next, a one-variable estimate bounds the magnitude of the negative
interaction between tilts with the same sign.
Finally, we combine these estimates in an explicit decomposition of \(G\).
This decomposition keeps the dependence on both tilts explicit.
\end{revision}

Define
\[
  \mathcal R_1
  =
  r^2D_F(a\|rv)-D_F(b\|v),
  \qquad
  \mathcal R_2
  =
  r^2D_F(b\|ru)-D_F(a\|u).
\]
\textcolor{black}{With \(v\) fixed, eliminating \(u\) from
Equation~\eqref{eq:binary-tilt-means} gives}
\begin{equation}
  b
  =
  \frac{
    (1-r^2)v+r(1-v^2)a
  }{
    1-r^2v^2
  },
\qquad
  s_v:=\frac{db}{da}
  =
  \frac{r(1-v^2)}{1-r^2v^2}.
  \label{eq:tilted-mean-relation}
\end{equation}
Since \(F''(s)=(1-s^2)^{-1}\),
\[
  \frac{d^2\mathcal R_1}{da^2}
  =
  \frac{r^2}{1-a^2}
  -
  \frac{s_v^2}{1-b^2}.
\]
\textcolor{black}{Substituting Equation~\eqref{eq:tilted-mean-relation} and
combining the two fractions gives the exact identity}
\[
  \frac{d^2\mathcal R_1}{da^2}
  =
  \frac{
    r^2(1-r^2)(1-v^2)
    \left[
      a^2(1-v^2)-2arv+v^2(1+r^2)
    \right]
  }{
    (1-r^2v^2)^2(1-a^2)(1-b^2)
  }.
\]
\textcolor{black}{Substituting \(a=(u+rv)/z\) rewrites the bracket as}
\[
  \frac{1-r^2v^2}{z^2}
  \left[
    u^2+v^2-(1+r^2)u^2v^2
  \right],
\]
\textcolor{black}{where the expression in square brackets is}
\[
  u^2(1-v^2)+v^2(1-u^2)+(1-r^2)u^2v^2\ge0.
\]
\begin{revision}
For fixed \(v\), the map
\(u\mapsto (u+rv)/(1+ruv)\) is strictly increasing and maps
\((-1,1)\) onto \((-1,1)\).  Its derivative is
\((1-r^2v^2)/z^2>0\).  \textcolor{black}{Thus,} the curvature calculation proves that
\(\mathcal R_1\) is convex on the entire interval \(a\in(-1,1)\).
At \(u=0\), equivalently \(a=rv\), both \(\mathcal R_1\) and its derivative
vanish, so \(\mathcal R_1\ge0\).
Interchanging \(u,v\) gives \(\mathcal R_2\ge0\).
\end{revision}

It remains to bound the interaction term.  Define
\[
  L_r(s)=r^2F(s)-F(rs),
\]
\[
  \varphi_r(s)
  =
  r^3(1-s^2)\operatorname{arctanh}s
  -
  (1-r^2s^2)\operatorname{arctanh}(rs)
  +
  rs(1-r^2).
\]
The function \(L_r\) is even, \(L_r(0)=L_r'(0)=0\), and
\[
  L_r''(s)
  =
  \frac{
    r^2s^2(1-r^2)
  }{
    (1-s^2)(1-r^2s^2)
  }
  \ge0,
\]
so \(L_r\ge0\).
\begin{revision}
Also \(L_r'(s)\ge0\) for \(s\ge0\).
The function \(\varphi_r\) is odd, vanishes at zero, and satisfies
\[
  \varphi_r'(s)=-2rsL_r'(s).
\]
Consequently \(\varphi_r(s)\le0\) on \([0,1)\).
\end{revision}

Let
\[
  \Psi_r(s)
  =
  (1+rs)L_r(s)+\varphi_r(s).
\]
Then \(\Psi_r(0)=\Psi_r'(0)=0\) and
\begin{revision}
\[
  \Psi_r'(s)=rL_r(s)+(1-rs)L_r'(s).
\]
Differentiating once more cancels the two first-derivative terms:
\[
  \Psi_r''(s)
  =(1-rs)L_r''(s)
  =
  \frac{
    r^2s^2(1-r^2)
  }{
    (1-s^2)(1+rs)
  }
  \ge0
  \qquad (0\le s<1).
\]
\end{revision}
Therefore,
\begin{equation}
  -\varphi_r(s)
  \le
  (1+rs)L_r(s),
  \qquad 0\le s<1.
  \label{eq:scalar-interaction-bound}
\end{equation}

We finish by decomposing \(G\) into terms controlled by these bounds.
The addition formulas following from Equation~\eqref{eq:binary-tilt-means} are
\[
  \operatorname{arctanh}a
  =
  \operatorname{arctanh}u
  +
  \operatorname{arctanh}(rv),
  \qquad
  \operatorname{arctanh}b
  =
  \operatorname{arctanh}v
  +
  \operatorname{arctanh}(ru).
\]
Together with
\[
  1-a^2
  =
  \frac{
    (1-u^2)(1-r^2v^2)
  }{z^2},
\qquad
  1-b^2
  =
  \frac{
    (1-v^2)(1-r^2u^2)
  }{z^2},
\]
the four \(D_F\) terms reduce to two formulas and their
\(u,v\) interchanges:
\[
\begin{aligned}
  D_F(a\|rv)
    &=a\operatorname{arctanh}u+\tfrac12\log(1-u^2)-\log z,\\
  D_F(b\|v)
    &=b\operatorname{arctanh}(ru)
      +\tfrac12\log(1-r^2u^2)-\log z.
\end{aligned}
\]
For example, substituting the addition formula for
\(\operatorname{arctanh}a\) cancels the
\(a\operatorname{arctanh}(rv)\) terms in the first identity.
\begin{revision}
Substituting these four formulas and the expressions for \(F(a),F(b)\)
into \(2G-\mathcal R_1-\mathcal R_2\) cancels all \(\log z\) terms.
Subtracting \(L_r(u)+L_r(v)\) leaves
\[
\begin{aligned}
 &2G-\mathcal R_1-\mathcal R_2-L_r(u)-L_r(v)\\
 &\quad =
 r^2(a-u)\operatorname{arctanh}u
 +r^2(b-v)\operatorname{arctanh}v\\
 &\qquad\quad
 +(ru-b)\operatorname{arctanh}(ru)
 +(rv-a)\operatorname{arctanh}(rv)
 +2r(m-r).
\end{aligned}
\]
The coefficients simplify to
\[
\begin{aligned}
 a-u&=\frac{rv(1-u^2)}z,
 & b-v&=\frac{ru(1-v^2)}z,\\
 ru-b&=-\frac{v(1-r^2u^2)}z,
 & rv-a&=-\frac{u(1-r^2v^2)}z,\\
 m-r&=\frac{uv(1-r^2)}z.
\end{aligned}
\]
We split the last term equally between the two groups.  The terms
multiplied by \(v/z\) then give \(\varphi_r(u)\), and those multiplied by
\(u/z\) give \(\varphi_r(v)\).  We obtain the decomposition
\end{revision}
\begin{equation}
  G
  =
  \frac{\mathcal R_1+\mathcal R_2}{2}
  +
  \frac12
  \left[
    L_r(u)+L_r(v)
    +
    \frac{
      v\varphi_r(u)+u\varphi_r(v)
    }{z}
  \right].
  \label{eq:scalar-gap-decomposition}
\end{equation}
If \(uv\le0\), oddness and the sign of \(\varphi_r\) make both cross terms
nonnegative.  If \(uv\ge0\), simultaneous sign reversal leaves the bracket
unchanged, so take \(u,v\ge0\).
\begin{revision}
Equation~\eqref{eq:scalar-interaction-bound} gives
\[
  -v\varphi_r(u)
  \le
  v(1+ru)L_r(u)
  \le
  zL_r(u),
\]
where the last inequality uses \(z-v(1+ru)=1-v\ge0\).
Interchanging \(u,v\) gives \(-u\varphi_r(v)\le zL_r(v)\).
\textcolor{black}{Hence,} the bracket in Equation~\eqref{eq:scalar-gap-decomposition}
is nonnegative, proving \(G\ge0\).

For nonnegative \(f,g\), replace them by \(f+\delta,g+\delta\)
and let \(\delta\downarrow0\).
The original denominator is positive, so the tilted law converges.
Since \(\mathsf B_r\) and \(U\) have full support, both joint and
marginal divergences are continuous on these finite probability
simplices, with \(0\log0=0\).  The compensated term is continuous as
well.  This proves the boundary case.
\end{revision}
\end{proof}

\begin{lemma}[Two-Block Tensor Identities]
\label{lem:two-block}
Let \((A,B)\) and \((C,D)\) be independent finite pairs, and let \(T\) be a
pseudotranscript between \((A,C)\) and \((B,D)\).  Put
\[
  U=(T,D),
  \qquad
  V=(T,A),
  \qquad
  K=I(A;D\mid T),
  \qquad
  L=I(B;C\mid A,D,T).
\]
Then \(U\) is a pseudotranscript between \(A,B\), \(V\) is a
pseudotranscript between \(C,D\), and
\begin{equation}
  \mathsf E(U)+\mathsf E(V)-\mathsf E(T)
  =
  K-L,
  \label{eq:tensor-information-identity}
\end{equation}
\begin{equation}
  \mathsf N(U)+\mathsf N(V)-\mathsf N(T)
  =
  K+L.
  \label{eq:tensor-dependence-identity}
\end{equation}
\begin{revision}
For \(U\), the source pair is \((A,B)\); for \(V\), it is \((C,D)\);
for \(T\), it is \(((A,C),(B,D))\).
These pairs apply to both \(\mathsf E\) and \(\mathsf N\).
\end{revision}
\end{lemma}

\begin{proof}
If
\[
  \Prb\{T=t\mid A=a,C=c,B=b,D=d\}
  =
  \alpha_t(a,c)\beta_t(b,d),
\]
then
\[
  \Prb\{U=(t,d)\mid A=a,B=b\}
  =
  \beta_t(b,d)
  \sum_cP_{CD}(c,d)\alpha_t(a,c),
\]
which factors in \(a,b\).  The analogous calculation for \(V\) proves the
second pseudotranscript claim\textcolor{black}{:}
\begin{revision}
\[
  \Prb\{V=(t,a)\mid C=c,D=d\}
  =
  \alpha_t(a,c)\sum_bP_{AB}(a,b)\beta_t(b,d).
\]
Both formulas give normalized conditional laws throughout the full
Cartesian products of the respective marginal supports.  For example,
summing the first formula over \(t,d\) gives
\[
  \sum_{c,d}P_{CD}(c,d)\sum_t\alpha_t(a,c)\beta_t(b,d)=1.
\]
This verifies the support requirement in
Definition~\ref{def:pseudotranscript}, including source pairs of zero
joint probability.
\end{revision}

\begin{revision}
For the first identity, we express each mutual information as the
difference between source entropy and conditional source entropy.
Because \(AB\) and \(CD\) are independent, the unconditional entropies
cancel, giving
\[
 \mathsf E(U)+\mathsf E(V)-\mathsf E(T)
 =H(ABCD\mid T)-H(AB\mid T,D)-H(CD\mid T,A).
\]
Expanding first in \(A,D\) and then in \(B,C\), the right side becomes
\[
\begin{aligned}
 &H(AD\mid T)-H(A\mid D,T)-H(D\mid A,T)\\
 &\quad
 +H(BC\mid A,D,T)-H(B\mid A,D,T)-H(C\mid A,D,T)\\
 &=I(A;D\mid T)-I(B;C\mid A,D,T)=K-L.
\end{aligned}
\]
This proves Equation~\eqref{eq:tensor-information-identity}.
\end{revision}

For the second identity, the same source independence gives
\[
  \mathsf N(U)+\mathsf N(V)-\mathsf N(T)
  =
  I(AC;BD\mid T)
  -
  I(A;B\mid T,D)
  -
  I(C;D\mid T,A).
\]
The chain rule in the order \(A,D,B,C\) reads
\[
  I(AC;BD\mid T)
  =
  I(A;D\mid T)
  +
  I(A;B\mid T,D)
\]
\[
  \quad
  +
  I(C;D\mid T,A)
  +
  I(C;B\mid T,A,D).
\]
The middle two terms cancel, leaving \(K+L\), which proves Equation~\eqref{eq:tensor-dependence-identity}.
\end{proof}

\begin{proposition}[Product-DSBS Pseudotranscript SDPI]
\label{prop:pseudotranscript-sdpi}
If \((X,Y)\sim \mathsf B_\rho^{\otimes d}\), \(|\rho|<1\), every finite
pseudotranscript \(T\) satisfies
\begin{equation}
  \mathsf N(T)
  \le
  \rho^2\mathsf E(T).
  \label{eq:product-dsbs-sdpi}
\end{equation}
\end{proposition}

This inequality applies to every rectangle-factorized pseudotranscript.
It bounds the reduction of the \(X\)--\(Y\) dependence by a \(\rho^2\)
fraction of the information revealed about the pair.

\begin{proof}
For \(d=1\), condition on \(T=t\).  By Equation~\eqref{eq:pseudotranscript-factorization}, the posterior is a
marginal tilt of \(\mathsf B_\rho\), so
Lemma~\ref{lem:scalar-dsbs} applies.  Averaging its left side
gives \(I(T;X)+I(T;Y)\), and averaging
\(\KL(P_{XY\mid T}\|\mathsf B_\rho)\) gives \(I(T;X,Y)\).  The compensated term
cancels:
\[
  \sum_tP_T(t)
  \rho\{\E[XY\mid T=t]-\rho\}
  =
  0.
\]
Thus,
\[
  I(T;X)+I(T;Y)
  \le
  (1+\rho^2)I(T;X,Y),
\]
which is Equation~\eqref{eq:product-dsbs-sdpi}.

Suppose the proposition holds for two independent source blocks.  By
Lemma~\ref{lem:two-block},
\[
  \mathsf N(T)
  =
  \mathsf N(U)+\mathsf N(V)-K-L
\]
\[
  \le
  \rho^2\{\mathsf E(U)+\mathsf E(V)\}
  -
  K-L
\]
\[
  =
  \rho^2\mathsf E(T)
  -
  (1-\rho^2)K
  -
  (1+\rho^2)L
  \le
  \rho^2\mathsf E(T).
\]
Induction over coordinates completes the proof.
\end{proof}

\begin{lemma}[Pseudotranscript KL Identity]
\label{lem:pseudotranscript-kl}
Let \(P_{XY}\) be finite, let
\(\overline P_{XY}=P_XP_Y\), and pass both laws through the same
pseudotranscript kernel.  Then
\begin{equation}
  \mathsf N_P(T)
  =
  \KL(P_{TX}\|\overline P_{TX})
  +
  \KL(P_{TY}\|\overline P_{TY})
  -
  \KL(P_T\|\overline P_T).
  \label{eq:pseudotranscript-kl-identity}
\end{equation}
In particular,
\begin{equation}
  \max\{
    \KL(P_{TX}\|\overline P_{TX}),
    \KL(P_{TY}\|\overline P_{TY})
  \}
  \le
  \mathsf N_P(T).
  \label{eq:pseudotranscript-kl-bound}
\end{equation}
\end{lemma}

\begin{proof}
Under \(\overline P\), the factorization in Equation~\eqref{eq:pseudotranscript-factorization} makes \(X,Y\)
conditionally independent given \(T\).  Also,
\[
  \frac{dP_{TXY}}{d\overline P_{TXY}}(t,x,y)
  =
  \frac{dP_{XY}}{d(P_XP_Y)}(x,y),
\]
so
\[
  \KL(P_{TXY}\|\overline P_{TXY})
  =
  I_P(X;Y).
\]
\begin{revision}
We apply the KL chain rule first to \((T,X)\), then add and subtract
\(\log P(Y\mid T)\) in the conditional term.  Conditional independence
under \(\overline P\) gives
\end{revision}
\[
  I_P(X;Y)
  =
  \KL(P_{TX}\|\overline P_{TX})
  +
  I_P(X;Y\mid T)
\]
\[
  \quad
  +
  \KL(P_{TY}\|\overline P_{TY})
  -
  \KL(P_T\|\overline P_T).
\]
Rearrangement proves Equation~\eqref{eq:pseudotranscript-kl-identity}.  Data processing gives
\[
  \KL(P_{TX}\|\overline P_{TX})
  \ge
  \KL(P_T\|\overline P_T)
\]
and the analogous inequality for \(Y\), proving Equation~\eqref{eq:pseudotranscript-kl-bound}.
\end{proof}

\subsection{Fisher Information of Canonical Labels}

\begin{revision}
To apply van Trees' inequality, we bound the Fisher information carried by
the canonical label.  The next lemma supplies two facts for a finite
hidden source: the observed score is the conditional mean of the
source's score, and the quadratic term in KL divergence is one half
of the Fisher information.
The first fact is Fisher's identity \cite{louis-fisher}; here it follows
by differentiating a finite sum.
We then split a \(\rho_1\)-correlated source into a
\(\rho_0\)-correlated part and a residual DSBS part.
Applying the pseudotranscript inequality to the residual pair bounds the
KL divergence.  Its quadratic expansion then gives the Fisher information
bound.
\end{revision}

\begin{revision}
\begin{lemma}[Score and KL Expansion for a Finite Source]
\label{lem:finite-latent}
Let \(Z\) have a finite alphabet and strictly positive mass functions
\(p_\theta(z)\) that are twice continuously differentiable on an open
interval.  Obtain an observation \(O\) by a fixed randomized rule:
given \(Z=z\), draw \(O\) from a probability law \(K_z\) that does
not depend on \(\theta\).  The observed score is
\[
  \dot\ell_\theta(O)
  =
  \E_\theta[
    \partial_\theta\log p_\theta(Z)
    \mid O
  ],
\]
and
\begin{equation}
  \KL(P_{\theta+h}^{O}\|P_\theta^{O})
  =
  \frac{h^2}{2}
  \E_\theta[
    \dot\ell_\theta(O)^2
  ]
  +
  o(h^2).
  \label{eq:finite-latent-kl-expansion}
\end{equation}
Here, \(h\to0\) with \(\theta\) and the finite source family fixed.
\end{lemma}

\begin{proof}
We use the finite measure \(\nu=\sum_zK_z\) to write all output laws
as densities.  If \(k_z=dK_z/d\nu\), the density of \(O\) is
\[
  q_\theta(o)=\sum_zp_\theta(z)k_z(o).
\]
Since \(\sum_zk_z=1\) almost everywhere under \(\nu\), strict positivity
of the finite source law gives
\(q_\theta\ge\min_zp_\theta(z)>0\) on this common support.
This holds even when the observation contains continuous random seeds.
Differentiating the finite sum gives
\[
  \frac{\partial_\theta q_\theta(o)}{q_\theta(o)}
  =
  \sum_z
  \frac{p_\theta(z)k_z(o)}{q_\theta(o)}
  \frac{p_\theta'(z)}{p_\theta(z)}
  =
  \E_\theta[
    \partial_\theta\log p_\theta(Z)
    \mid O=o
  ].
\]
Moreover,
\[
  \frac{q_{\theta+h}(o)}{q_\theta(o)}
  =
  \sum_z
  \Prb_\theta\{Z=z\mid O=o\}
  \frac{p_{\theta+h}(z)}{p_\theta(z)}.
\]
Fix a compact neighborhood \(K\) of \(\theta\).  The finite source
alphabet and strict positivity give the following bound, uniformly in \(z\):
\[
  \frac{p_{\theta+h}(z)}{p_\theta(z)}
  =
  1+h s_\theta(z)+h^2R_h(z),
  \qquad
  |R_h(z)|\le C,
  \qquad
  C=\frac{\max_{z,\,t\in K}|p_t''(z)|}
          {2\min_{z,\,t\in K}p_t(z)},
\]
where \(s_\theta(z)=\partial_\theta\log p_\theta(z)\).
These finitely many source scores are bounded, so their conditional
mean \(\dot\ell_\theta\) is bounded as well.  Taking the
posterior average in the preceding display yields
\[
  r_h(o):=
  \frac{q_{\theta+h}(o)}{q_\theta(o)}
  =
  1+h\dot\ell_\theta(o)+h^2\overline R_h(o),
  \qquad
  |\overline R_h(o)|\le C.
\]
In particular, \(r_h\) is uniformly bounded away from zero for all
sufficiently small \(h\).  The elementary expansion
\[
  r\log r
  =
  (r-1)+\frac12(r-1)^2+O(|r-1|^3)
\]
holds uniformly near one.  Normalization cancels its linear term:
\(\int q_\theta(r_h-1)\,d\nu=0\).
The uniform likelihood-ratio expansion then gives
\[
\begin{aligned}
  \KL(P_{\theta+h}^{O}\|P_\theta^{O})
  &=\frac12\int q_\theta(r_h-1)^2\,d\nu+O(|h|^3)\\
  &=\frac{h^2}{2}\E_\theta[\dot\ell_\theta(O)^2]+o(h^2),
\end{aligned}
\]
proving Equation~\eqref{eq:finite-latent-kl-expansion}.
\end{proof}
\end{revision}

\begin{proposition}[Canonical-Label Fisher Bound]
\label{prop:canonical-fisher}
\begin{revision}
Let \(\Omega\) be a seed with a parameter-independent law,
independent of \((X,Y)\).  Let \(\Gamma\) be a jointly measurable,
parameter-independent function of \((X,Y,\Omega)\) taking values in a
common finite alphabet of size at most \(M\ge1\).
For every seed value, suppose the preimage of each label is a rectangle in
\[
  \{-1,1\}^d\times\{-1,1\}^d.
\]
Under \(\mathsf B_\rho^{\otimes d}\), put
\end{revision}
\[
  O_\rho=(\Omega,\Gamma,Y).
\]
For \(|\rho_0|<1\) and
\[
  \xi
  =
  \frac{\rho_1-\rho_0}{1-|\rho_0|}
  \in[-1,1],
\]
\begin{equation}
  \KL(P_{\rho_1}^{O}\|P_{\rho_0}^{O})
  \le
  \xi^2\log M.
  \label{eq:masked-kl-bound}
\end{equation}
\begin{revision}
For every \(|\rho|<1\), its Fisher information satisfies
\end{revision}
\begin{equation}
  J_\rho(O_\rho)
  \le
  \frac{
    2H_\rho(\Gamma\mid\Omega)
  }{
    (1-|\rho|)^2
  }
  \le
  \frac{2\log M}{(1-|\rho|)^2}.
  \label{eq:canonical-fisher-bound}
\end{equation}
\end{proposition}

\begin{revision}
The bound depends on the branching partition only through its label
count \(M\).  The observation's Fisher information about the correlation
is therefore at most \(2\log M/(1-|\rho|)^2\).
\end{revision}

\begin{proof}
Put \(\alpha=|\rho_0|\) and choose
\(\sigma\in\{-1,1\}\) with \(\sigma\alpha=\rho_0\).
\begin{revision}
We keep the reference value \(\rho_0\) fixed throughout this comparison.
\textcolor{black}{Thus,} the mask distribution below stays fixed as
\(\rho_1\), and hence \(\xi\), varies.
\end{revision}
Independently across coordinates, draw
\[
  \beta_i\sim\operatorname{Bernoulli}(\alpha),
\]
a uniform sign \(Z_i\), and
\[
  (U_i,V_i)\sim \mathsf B_\xi.
\]
\begin{revision}
For each coordinate, the mask, the frozen sign, and the residual pair
are mutually independent; they are also independent of \(\Omega\).
When \(\rho_0=0\), take \(\sigma=1\); all masks are then zero.
\end{revision}
Define
\begin{equation}
  X_i'=\beta_iZ_i+(1-\beta_i)U_i,
\qquad
  Y_i'=\sigma\beta_iZ_i+(1-\beta_i)V_i.
  \label{eq:dsbs-masking}
\end{equation}
\begin{revision}
The two signs in each pair in Equation~\eqref{eq:dsbs-masking} are
uniform.  Its correlation is
\[
  \alpha\sigma+(1-\alpha)\xi
  =
  \rho_1.
\]
Uniform binary marginals and the correlation determine the four joint
probabilities, so \((X',Y')\sim\mathsf B_{\rho_1}^{\otimes d}\).
At \(\xi=0\), its law is \(\mathsf B_{\rho_0}^{\otimes d}\).
\end{revision}

\begin{revision}
Conditioning on \(S=(\beta,Z,\Omega)\) splits the coordinates into the
frozen set \(F_1=\{i:\beta_i=1\}\) and the free set
\(F_0=\{i:\beta_i=0\}\).  On \(F_1\), both \(X_i'=Z_i\) and
\(Y_i'=\sigma Z_i\) are determined by \(S\).  On \(F_0\), we have
\((X_i',Y_i')=(U_i,V_i)\).
\end{revision}

Every label event has the form
\(\mathbf1_{C_\gamma}(X')\mathbf1_{A_\gamma}(Y')\).
Proposition~\ref{prop:canonical-rectangle} gives this factorization for
canonical labels.  \textcolor{black}{Once \(S\) is fixed, substituting the
frozen coordinates into each indicator leaves}
\[
  c_{\gamma,S}(U_{F_0})\,a_{\gamma,S}(V_{F_0}),
\]
which is a rectangle in the free residual coordinates.  Equivalently, it
is a rectangle in the full vectors \((U,V)\) that ignores \(F_1\).
The argument uses only the rectangle factorization between the two sources.
Definition~\ref{def:pseudotranscript} therefore applies: \(\Gamma\) is a
pseudotranscript between \(U\) and \(V\) conditional on \(S\).  Proposition~
\ref{prop:pseudotranscript-sdpi} and
Lemma~\ref{lem:pseudotranscript-kl}, comparing correlation
\(\xi\) with independence, give
\[
  \E_S\!\left[
    \KL(
      P_\xi^{\Gamma,V\mid S}
      \|
      P_0^{\Gamma,V\mid S}
    )
  \right]
  \le
  \xi^2
  I_\xi(\Gamma;U,V\mid S)
  \le
  \xi^2\log M.
\]
\begin{revision}
Averaging over the parameter-independent \(S\) and applying data
processing from \((S,\Gamma,V)\) to \((\Omega,\Gamma,Y')\) proves
Equation~\eqref{eq:masked-kl-bound} for \(|\xi|<1\).
\end{revision}
The endpoints \(\xi=\pm1\) follow by continuity of the finite input laws
through the same fixed \textcolor{black}{randomized rules}.  The reference law at \(\xi=0\) has
positive mass on every input pair.
\begin{revision}
To justify this endpoint limit, observe that
\[
  \frac{d\mathsf B_\xi^{\otimes d}}
       {d\mathsf B_0^{\otimes d}}(u,v)
  =\prod_{i=1}^d(1+\xi u_iv_i)\le2^d,
  \qquad |\xi|\le1.
\]
An output likelihood ratio is a conditional average of this input
ratio, so it has the same uniform bound.  The function \(r\log r\),
extended by zero at \(r=0\), is bounded on \([0,2^d]\).
Dominated convergence therefore justifies the endpoint limit even when
the seed distribution is continuous.
\end{revision}
\textcolor{black}{The same argument also gives the sharper bound}
\[
  \KL(P_{\rho_1}^{O}\|P_{\rho_0}^{O})
  \le
  \xi^2 I_\xi(\Gamma;U,V\mid S).
\]

Because \(\Gamma\) is deterministic given \((U,V,S)\),
\[
  I_\xi(\Gamma;U,V\mid S)
  =
  H_\xi(\Gamma\mid S).
\]
As \(\xi\to0\), finite-alphabet
continuity and dominated convergence over \(\Omega\) give
\[
  I_\xi(\Gamma;U,V\mid S)
  \to
  H_0(\Gamma\mid S)
  \le
  H_{\rho_0}(\Gamma\mid\Omega).
\]
\begin{revision}
The entropy integrands are bounded by \(\log M\), so dominated convergence
applies.  The last inequality removes the conditioning on \((\beta,Z)\).
At \(\xi=0\), the remaining law of \((\Omega,\Gamma)\) is exactly its law
at correlation \(\rho_0\).
\end{revision}
\begin{revision}
Lemma~\ref{lem:finite-latent} applies to the finite DSBS source followed
by the protocol's fixed randomized rule.  Use
\end{revision}
\(\rho_0=\rho\), \(\rho_1=\rho+h\), and
\(\xi=h/(1-|\rho|)\).  Divide the sharper bound by \(h^2/2\) and let
\(h\to0\).  Equation~\eqref{eq:finite-latent-kl-expansion} gives
\[
  J_\rho(O_\rho)
  \le
  \frac{2H_\rho(\Gamma\mid\Omega)}
       {(1-|\rho|)^2},
\]
and \(H_\rho(\Gamma\mid\Omega)\le\log M\) completes Equation~\eqref{eq:canonical-fisher-bound}.
\end{proof}

\subsection{Fisher Information and Minimax Risk}
\label{sec:branching-risk-proof}

\begin{revision}
For the final step, we choose a DSBS submodel whose target inner product
is an empirical correlation.  The canonical-label bound controls the
observed Fisher information on every interval
\(|\rho|<\varepsilon<1\).  We apply van Trees' inequality
\cite{van-trees} with a cosine-squared prior to obtain a lower bound on the
risk of estimating \(\rho\), then optimize the prior width.
Minkowski's inequality accounts for the \(d^{-1/2}\) fluctuation of the
empirical inner product around its mean.
\end{revision}

\begin{revision}
\begin{lemma}[Cosine-Prior Information-to-Risk Bound]
\label{lem:cosine-van-trees}
Let \(\varepsilon>0\).  Suppose a finite hidden source \(Z\) has
strictly positive mass functions \(p_\rho(z)\) that are twice
continuously differentiable on an open interval containing
\([-\varepsilon,\varepsilon]\).  Obtain \(O\) from \(Z\) by a fixed
randomized rule, as in Lemma~\ref{lem:finite-latent}, and suppose its
Fisher information satisfies
\[
  \sup_{|\rho|<\varepsilon}J_\rho\le J_\star<\infty.
\]
Define
\begin{equation}
  \pi_\varepsilon(\rho)
  =
  \varepsilon^{-1}
  \cos^2\left(
    \frac{\pi\rho}{2\varepsilon}
  \right)
  \mathbf1_{\{|\rho|<\varepsilon\}}.
  \label{eq:cosine-prior}
\end{equation}
The prior information is
\begin{equation*}
  I(\pi_\varepsilon)
  =
  \frac{\pi^2}{\varepsilon^2},
\end{equation*}
and every bounded estimator \(\widehat\rho=\widehat\rho(O)\) that
does not depend on the unknown parameter satisfies
\begin{equation}
  \int
  \E_\rho[
    (\widehat\rho-\rho)^2
  ]
  \pi_\varepsilon(\rho)\,d\rho
  \ge
  \frac1{
    J_\star+\pi^2/\varepsilon^2
  }.
  \label{eq:cosine-prior-risk}
\end{equation}
\end{lemma}

\begin{proof}
The density in Equation~\eqref{eq:cosine-prior} integrates to one and
vanishes at both endpoints.
Its information is
\[
  \int\frac{\pi_\varepsilon'(\rho)^2}{\pi_\varepsilon(\rho)}\,d\rho
  =\frac{\pi^2}{\varepsilon^3}
    \int_{-\varepsilon}^{\varepsilon}
      \sin^2\!\left(\frac{\pi\rho}{2\varepsilon}\right)d\rho
  =\frac{\pi^2}{\varepsilon^2}.
\]
Let \(K_z\) be the fixed observation law given \(Z=z\), and let
\(\dot\ell_\rho\) be the observed score.
The estimator's mean is the finite sum
\[
  m(\rho)=\E_\rho\widehat\rho
  =\sum_zp_\rho(z)\int\widehat\rho(o)\,K_z(do).
\]
Each integral is finite and independent of \(\rho\).
Differentiating this sum and using the score formula in
Lemma~\ref{lem:finite-latent} gives
\[
  m'(\rho)
  =\sum_zp_\rho'(z)\int\widehat\rho(o)\,K_z(do)
  =\E_\rho[\widehat\rho\,\dot\ell_\rho].
\]
The same calculation with the constant estimator one gives
\(\E_\rho\dot\ell_\rho=\sum_zp_\rho'(z)=0\).
Because \(m\) is continuously differentiable on the prior interval,
integration by parts yields
\[
\begin{aligned}
  0
  &=\bigl[(m(\rho)-\rho)\pi_\varepsilon(\rho)
       \bigr]_{-\varepsilon}^{\varepsilon}\\
  &=\int_{-\varepsilon}^{\varepsilon}
      \{(m'(\rho)-1)\pi_\varepsilon(\rho)
        +(m(\rho)-\rho)\pi_\varepsilon'(\rho)\}\,d\rho.
\end{aligned}
\]
The boundary term vanishes since \(m(\rho)-\rho\) remains bounded.
Substitution of the mean derivative and the zero-mean score gives
\[
  1
  =
  \int
  \E_\rho
  \left[
    (\widehat\rho-\rho)
    \left\{
      \dot\ell_\rho
      +
      \frac{\pi_\varepsilon'(\rho)}
           {\pi_\varepsilon(\rho)}
    \right\}
  \right]
  \pi_\varepsilon(\rho)\,d\rho.
\]
Under the joint law of the prior and observation, the two score terms
have zero cross moment.  Their sum therefore has second moment
\[
  \int_{-\varepsilon}^{\varepsilon}
    J_\rho\pi_\varepsilon(\rho)\,d\rho
  +I(\pi_\varepsilon)
  \le J_\star+\frac{\pi^2}{\varepsilon^2}.
\]
Applying Cauchy--Schwarz to the score identity shows that the Bayes
mean-squared error times \(J_\star+\pi^2/\varepsilon^2\) is at least one.
Rearrangement proves Equation~\eqref{eq:cosine-prior-risk}.
\end{proof}
\end{revision}

\begin{proof}[Proof of Theorem~\ref{thm:finite-adaptive-converse}]
\begin{revision}
Fix an arbitrary randomized protocol and write \(D\) for its worst-case
mean-squared error.  If \(D=\infty\), the claimed inequality is immediate.
Otherwise, we normalize the reader as above and clip its output to
\([-1,1]\).  Since every target \(x^\top y\) lies in this interval,
clipping cannot increase the error, so the resulting risk is at most
\(D\).
\end{revision}
Restrict source and query to
\[
  X=\frac{\widetilde X}{\sqrt d},
  \qquad
  Y=\frac{\widetilde Y}{\sqrt d},
\]
where
\[
  (\widetilde X,\widetilde Y)
  \sim
  \mathsf B_\rho^{\otimes d}.
\]
\begin{revision}
Every such pair lies in \(\Bdd\times\Bdd\).
The worst-case criterion permits this correlated source--query law;
only the protocol seeds must be independent of the inputs.
\end{revision}
The empirical target
\[
  \tau_d=X^\top Y
  =
  \frac1d
  \sum_{i=1}^d
  \widetilde X_i\widetilde Y_i
\]
satisfies
\begin{equation}
  \E_\rho \tau_d=\rho,
  \qquad
  \Var_\rho(\tau_d)
  =
  \frac{1-\rho^2}{d}
  \le
  \frac1d.
  \label{eq:empirical-correlation-variance}
\end{equation}

\begin{revision}
Include every protocol seed in \(\Omega\), whose law is independent of
the source pair and \(\rho\), and let \(\Gamma\) be the canonical label.
Revealing encoder-private and decoder-private seeds gives the
lower-bound experiment more information.  The original estimator is still
admissible in this experiment.
Given \((Y,\Omega)\), Proposition~\ref{prop:canonical-rectangle}
recovers the normalized reader's sequence of returned
bits from \(\Gamma\),
and hence the original protocol output.
The clipped output is therefore measurable with respect to
\end{revision}
\[
  O_\rho=(\Omega,\Gamma,\widetilde Y),
\]
and for each seed there are at most
\[
  M=\mathsf C_{N,k}(q)
\]
labels.

For any \(0<\varepsilon<1\),
Proposition~\ref{prop:canonical-fisher} gives
\begin{equation*}
  \sup_{|\rho|<\varepsilon}J_\rho(O_\rho)
  \le \frac{2\log M}{(1-\varepsilon)^2}.
\end{equation*}
\begin{revision}
Apply Lemma~\ref{lem:cosine-van-trees} to this clipped estimator.
The finite hidden source is \((\widetilde X,\widetilde Y)\).
Its DSBS product probabilities are strictly positive and twice
continuously differentiable on \((-1,1)\), which contains the closed
prior interval because \(\varepsilon<1\).
All protocol randomness is included in the fixed observation rule,
and clipping makes the estimator bounded.
Consequently,
\end{revision}
\[
  \E_{\rho\sim\pi_\varepsilon}\E_\rho(\widehat s-\rho)^2
  \ge
  \left\{\frac{2\log M}{(1-\varepsilon)^2}
              +\frac{\pi^2}{\varepsilon^2}\right\}^{-1}.
\]
For \(L=\log M>0\), the denominator is minimized at
\[
  \varepsilon_\star
  =\frac{\pi^{2/3}}{(2L)^{1/3}+\pi^{2/3}},
  \qquad
  \inf_{0<\varepsilon<1}
  \left\{\frac{2L}{(1-\varepsilon)^2}
              +\frac{\pi^2}{\varepsilon^2}\right\}
  =\{(2L)^{1/3}+\pi^{2/3}\}^{3}.
\]
Indeed, setting the derivative to zero gives
\(2L\varepsilon^3=\pi^2(1-\varepsilon)^3\), and the denominator is
strictly convex.  If \(L=0\), let \(\varepsilon\uparrow1\) to obtain
the same infimum, \(\pi^2\).

\begin{revision}
Under the joint law of the prior, the DSBS source, and the protocol
seeds, the risk for the empirical target is at most \(D\).
The pointwise decomposition
\(\widehat s-\rho=(\widehat s-\tau_d)+(\tau_d-\rho)\),
Minkowski's inequality, and
Equation~\eqref{eq:empirical-correlation-variance} give
\end{revision}
\[
  \left\{
    \E[
      (\widehat s-\rho)^2
    ]
  \right\}^{1/2}
  \le
  \sqrt D+\frac1{\sqrt d}.
\]
\begin{revision}
Combining this inequality with the Bayes lower bound, optimizing the prior
width as above, and taking the positive part gives
Equation~\eqref{eq:finite-branching-converse}, since \(M=\mathsf C_{N,k}(q)\).
\end{revision}
\begin{revision}
The bound is uniform over protocols, so taking their infimum proves the
minimax statement.
\end{revision}
\end{proof}

\subsection{Rate Consequences for Finite Alphabets}
\label{sec:finite-alphabet-rates}

The rate calculation in Corollary~\ref{cor:adaptive-rate-bounds} extends
to any fixed alphabet size \(q\ge2\).  Suppose the integer budgets
\(N_d\ge k_d\ge0\) satisfy
\[
  \frac1d\log\mathsf C_{N_d,k_d}(q)\longrightarrow\ell\in[0,\infty).
\]
The optimized denominator obeys
\[
  \frac{\mathfrak A(N_d,k_d,q)}d
  =
  \left\{
    \left(\frac{2\log\mathsf C_{N_d,k_d}(q)}d\right)^{1/3}
    +\pi^{2/3}d^{-1/3}
  \right\}^{3}
  \longrightarrow2\ell.
\]
\begin{revision}
Multiplying the finite converse by \(d\) gives
\[
  d\,\D_d^{(q)}(N_d,k_d)
  \ge
  \left[\sqrt{\frac d{\mathfrak A(N_d,k_d,q)}}-1\right]_+^2.
\]
Consequently,
\[
  \liminf_{d\to\infty}d\,\D_d^{(q)}(N_d,k_d)
  \ge\left[(2\ell)^{-1/2}-1\right]_+^2,
\]
where the right side is \(+\infty\) when \(\ell=0\).
Indeed, in that case \(\mathfrak A(N_d,k_d,q)/d\to0\), so the square
root in the finite bound diverges.
The same expression therefore covers both exponential and subexponential
label growth.
\end{revision}

For packets of a fixed integer width \(b\ge1\), take \(q=2^b\) and
fixed bit rates \(0<R<W\).  The packet budgets and label exponent are
\[
  N_d=\left\lfloor\frac{Wd}{b}\right\rfloor,\qquad
  k_d=\left\lfloor\frac{Rd}{b}\right\rfloor,\qquad
  \ell_b(W,R)=\frac Wb h(R/W)+R\log2,
\]
where \(h\) is the binary entropy in Equation~\eqref{eq:binary-entropy}.
The binomial type bounds used in the corollary give
\[
  \frac1d\log\mathsf C_{N_d,k_d}(2^b)
  =
  \frac1d\log\binom{N_d}{k_d}+\frac{k_db}{d}\log2
  \longrightarrow\ell_b(W,R).
\]
Substituting this positive exponent gives
\begin{equation*}
  \liminf_{d\to\infty}d\,\D_d^{(2^b)}(N_d,k_d)
  \ge
  \left[(2\ell_b(W,R))^{-1/2}-1\right]_+^2.
\end{equation*}
The right-hand side is positive exactly when \(\ell_b(W,R)<1/2\).
At the binary full-read endpoint \(k_d=N_d=\lfloor Wd\rfloor\), the
label count is \(2^{N_d}\) and its exponent is \(W\log2\).
\begin{revision}
After taking the dimension limit at fixed \(W>0\), the resulting lower
bound is asymptotic to \(1/(W\log4)\) as \(W\downarrow0\).
This agrees with the leading term of the exact full-read constant
\((4^W-1)^{-1}\).
\end{revision}

\end{document}